\documentclass[a4paper,12pt]{article}
\usepackage{graphicx,rotating,colortbl,xcolor,wrapfig,hyperref,slashed,cite}
\hypersetup{colorlinks,bookmarksopen,bookmarksnumbered,
linkcolor=blus,pdfstartview=FitH,urlcolor=rossos,citecolor=verde}

\newcommand{\figs}{figsjpg}

\def\bma#1{\mbox{\boldmath{$#1$}}}
\def\to{\rightarrow}
\def\bea{\begin{eqnarray}}
\def\eea{\end{eqnarray}}
\def\eq#1{eq.~(\ref{#1})}
\def\meq#1{~(\ref{eq:#1})}
\def\mp{M_{\rm Pl}}
\definecolor{Gray}{gray}{0.95}
\newcommand{\bbox}[1]{\fcolorbox{gray}{Gray}{~$\displaystyle #1$~}}

\newcommand{\Mcr}{M_{\rm cr}}
\newcommand{\Rcr}{R_{\rm cr}}
\newcommand{\that}{\hat{t}}
\newcommand{\Hub}{H}
\newcommand{\Rt}{\tilde{R}}
\newcommand{\Hig}{\Phi_H }
\newcommand{\Mtexp}{173.34}

\newcommand{\asdiff}{\, \frac{\alpha_3(M_Z)-0.1184}{0.0007} }

\definecolor{rosso}{cmyk}{0,1,1,0.4}
\definecolor{rossos}{cmyk}{0,1,1,0.55}
\definecolor{rossoc}{cmyk}{0,1,1,0.2}
\definecolor{blu}{cmyk}{1,1,0,0.3}
\definecolor{blus}{cmyk}{1,1,0,0.6}
\definecolor{bluc}{cmyk}{1,1,0,0.1}
\definecolor{verde}{cmyk}{0.92,0,0.59,0.25}
\definecolor{verdec}{cmyk}{0.92,0,0.59,0.15}
\definecolor{verdes}{cmyk}{0.92,0,0.59,0.4}

 \def\be   {\begin{equation}}   \def\ee   {\end{equation}}
 \def\ba   {\begin{array}}      \def\ea   {\end{array}}

\font\tenrsfs=rsfs10 at 12pt
\font\sevenrsfs=rsfs7
\font\fiversfs=rsfs5
\newfam\rsfsfam
\textfont\rsfsfam=\tenrsfs
\scriptfont\rsfsfam=\sevenrsfs
\scriptscriptfont\rsfsfam=\fiversfs
\def\mathscr#1{{\fam\rsfsfam\relax#1}}

\def\Lag{\mathscr{L}}

\newcommand{\GeV}{\,{\rm GeV}}

\def\circa#1{\,\raise.3ex\hbox{$#1$\kern-.75em\lower1ex\hbox{$\sim$}}\,}

\newcommand{\beq}{\begin{equation}}
\newcommand{\eeq}{\end{equation}}

\def\hhref#1{\href{http://arxiv.org/abs/#1}{arXiv:#1}} % in bibliography
\def\arXiv#1{\href{http://arxiv.org/abs/#1}{arXiv:#1}} % in bibliography

 \def\ex{\epsilon}
 
 \def\kx{\kappa}

\def\ttau{{\tilde{\tau}}}

\def \gta {\mathrel{\vcenter
     {\hbox{$>$}\nointerlineskip\hbox{$\sim$}}}}

\begin{document}
CERN-PH-TH-2015/119   \hfill IFUP-TH/2015

\thispagestyle{empty}
\vspace{0.1cm}
\begin{center}
{\huge \bf \color{rossos} 
The cosmological Higgstory\\[1ex] of the vacuum instability}  \\[6mm]

{\bf\large Jos\'e R. Espinosa$^{a,b}$, Gian F. Giudice$^c$,  \\[1mm]
Enrico Morgante$^d$, Antonio Riotto$^d$, Leonardo Senatore$^e$, \\[2mm]
Alessandro Strumia$^{f,g}$, Nikolaos Tetradis$^h$}  \\[5mm]

{\it $^a$ IFAE, Universitat Aut\'onoma de Barcelona, 08193 Bellaterra, Barcelona}\\[0mm]
{\it $^b$ ICREA, Instituci\'o Catalana de Recerca i Estudis Avan\c{c}ats, Barcelona, Spain}\\[0mm]
{\it $^c$ CERN, Theory Division, Geneva, Switzerland}\\[0mm]
{\it $^d$ D\'epartement de Physique Th\'eorique and Centre for Astroparticle Physics (CAP),\\[-1mm]
Universit\'e de Gen\`eve, Geneva, Switzerland}\\[0mm]
{\it $^e$ Stanford Institute for Theoretical Physics and Kavli Institute for Particle Astrophysics\\[-1mm] and Cosmology, 
Physics Department and SLAC, Stanford, CA 94025, USA}\\[0mm]
{\it $^f$ Dipartimento di Fisica dell'Universit{\`a} di Pisa and INFN, Italy}\\[0mm]
{\it $^g$ National Institute of Chemical Physics and Biophysics, Tallinn, Estonia}\\[0mm]
{\it $^h$ Department of Physics, University of Athens, Zographou 157 84, Greece}

\vspace{8mm}
{\large\bf\color{blus} Abstract}
\begin{quote}\large
The Standard Model Higgs potential becomes unstable at large field values.
After clarifying the issue of gauge dependence of the effective potential,
we study the cosmological evolution of the Higgs field in presence of this instability
throughout inflation, reheating and the present epoch.
We conclude that anti-de Sitter patches in which the Higgs field lies at its true vacuum are
lethal for our universe. From this result, we derive upper bounds on the Hubble constant during inflation, which
depend on the reheating temperature and on the Higgs coupling to the scalar curvature or to the inflaton. Finally we study how a speculative link between Higgs meta-stability and consistence of quantum gravity leads to a sharp prediction for the Higgs and top masses, which is consistent with measured values.

\end{quote}
\end{center}

\pagebreak\large
\tableofcontents
\newpage\normalsize

%\xxx{New notation, maybe not yet consistently employed.
%Hubble constant = $\Hub$.
%Higgs doublet = $\Hig$.
%Higgs = $h$.}

\section{Introduction}

The measurements of the Higgs-boson and top-quark masses imply the surprising fact that, in the context of the Standard Model (SM) with no additional physics, our universe lies at the edge between stability and instability of the electroweak vacuum~\cite{instab} (see~\cite{instab2} for earlier analyses). For the present best fit values of the SM parameters, the Higgs potential develops an instability well below the Planck scale, but the proximity to the stability region insures that the electroweak vacuum lifetime can be exceedingly longer than the age of the universe. 
 
This intriguing result offers a testing ground for phenomena occurring in the early universe. Indeed, the presence of a minimum of the SM potential deeper than the electroweak vacuum raises many cosmological issues: how did the Higgs field end up today in the false vacuum?
Why didn't the primordial dynamics destabilise the Higgs field? 
How did patches of the universe with large Higgs values evolve in time without swallowing all space? 
Addressing these questions leads to interesting constraints on early-time phenomena and inflationary dynamics. These constraints are the subject of this paper. Several aspects about electroweak-vacuum decay from thermal or inflationary
Higgs fluctuations have already been studied in the literature~\cite{Espinosa:2007qp,cosmo2,herranen,zurek}, but here we give a comprehensive description of the phenomenon and reach new conclusions.

\smallskip

The effects of the thermal bath during the radiation-dominated phase of the universe are twofold. On one side, thermal fluctuations can trigger nucleation of bubbles that probe Higgs-field values beyond the instability barrier. On the other side, thermal corrections to the Higgs potential tend to stabilise low field values, creating an effective barrier. For the observed values of the SM parameters, the latter effect is dominant and thermal corrections do not destabilise the electroweak vacuum, even when the reheating temperature is close to the Planck scale~\cite{Espinosa:2007qp}.

More subtle is the issue of the Higgs-field fluctuations generated during inflation. In the case in which the Higgs has no direct coupling to the inflaton and is minimally coupled to gravity (and hence is effectively massless during inflation), the field develops fluctuations with amplitude proportional to $H$, the Hubble rate during inflation. These fluctuations pose a threat to vacuum stability. For values of $H$ smaller than the height of the potential barrier, the Higgs field can tunnel into anti-de Sitter (AdS), according to the Coleman-de Luccia bubble nucleation process~\cite{deluccia}. When $H$ becomes comparable to the barrier height, the transition is well described by the Hawking-Moss instanton~\cite{hawkingmoss}, which corresponds to a thermal overcoming of the barrier due to the effective Gibbons-Hawking temperature $T=H/2\pi$~\cite{gibbons} associated with the causal horizon of de Sitter (dS) space. However, a more convenient way to compute the evolution of the Higgs fluctuations during inflation is through a stochastic approach based on a Fokker-Planck equation that describes the probability to find the Higgs field at a given value $h$    
and time $t$~\cite{fokker}. This approach was followed in~\cite{Espinosa:2007qp,zurek} to derive the probability distribution of Higgs patches in the universe. In this paper, we describe the long-wavelength modes of the Higgs field using a Langevin equation sourced by a Gaussian random noise that mimics quantum fluctuations during inflation. This method has the advantage of bypassing the problem of choosing boundary conditions and it is shown to agree with the results from the Fokker-Planck approach with appropriate boundary conditions.

Quantifying the probability for the existence of a patch of the Higgs field in the SM vacuum sufficiently large to encompass our observable universe is a subtle issue, which requires an understanding of how AdS bubbles (with large Higgs-field configurations) evolve in a dS background, during inflation, and in a Minkowski background, after inflation. As correctly pointed out in~\cite{zurek}, patches in which the Higgs probes field values beyond the barrier do not necessarily end up in the AdS vacuum, as long as their evolution is driven by the stochastic quantum term. Only when classical evolution takes over, the field falls into its deep minimum.
 In \cite{zurek} it was assumed that these AdS patches rapidly evolve into relic defects that are not necessarily dangerous, hence arguing that large Higgs fluctuations do not pose a cosmological threat. In our analysis, we reach opposite conclusions. 

The evolution of AdS bubbles in an inflationary dS background depends on their size, internal energy, surface tension, and initial wall velocity. Depending on the characteristics of the bubbles, we find a variety of possible evolutions. Bubbles shrink, if they start with small radius and low velocity; expand but remain hidden inside the Schwarzschild horizon, if the gravitational self-energy of their surface overwhelms the difference between the vacuum energy in the exterior and interior; and expand at the expense of exterior space, otherwise. The seemingly paradoxical situation of an expanding bubble of crunching AdS space is resolved by understanding the difference in space-time coordinates on the two sides of the wall.  While an observer inside the bubbles will experience space contracting because of the negative cosmological constant, an external observer will see the surface of large bubbles expand. Although we expect that the process of inflation with large $H$ will generate a distribution of AdS expanding bubbles, we conclude that such bubbles will never take over all dS space. The inflationary space expansion always beats the causal expansion of bubbles, efficiently diluting them. 

At this stage, it may seem that the remnant AdS bubbles can be compatible with the presently observed universe. The problem starts when we consider post-inflationary evolution of the AdS patches in a flat background. The bubble wall keeps on expanding at the speed of light and an AdS patch eventually engulfs all space. 
This means that a necessary requirement for our present universe to exist is that the probability to find an expanding AdS bubble in our past light-cone must be negligible. Unfortunately, we cannot make firm statements about the formation of expanding AdS Higgs bubbles during inflation because the answer depends on energy considerations based on the Higgs potential in the Planckian region. However, our study suggests that it would be very difficult to imagine a situation in which all large-field Higgs patches shrink and none expands. Therefore, barring the presence of AdS Higgs bubbles in our past light-cone is a well-justified requirement for a viable cosmology. This line of reasoning leads to an interesting bound on $H$, the Hubble constant during inflation, which we compute not only in the case of a minimally-coupled Higgs, but also in the presence of a gravitational interaction between the Higgs bilinear and the scalar curvature. 

\smallskip

Having established the dangers of patches in which the Higgs field falls  into the trans-Planckian region, we consider the fate of patches in which, at the end of inflation, the Higgs field has fluctuated beyond the potential barrier, but has not yet experienced the classical evolution that wants to drive it towards very large values. The eventual fate of such bubbles is determined by the subsequent thermal evolution of the universe. Thermal effects can rescue such patches of the universe by effectively pushing the potential barrier to larger field values, allowing the Higgs field to relax into its SM vacuum. We study this phenomenon during the preheating and reheating stages of the universe, when the energy stored in the inflaton oscillating around its minimum is released into thermal energy carried by SM particles. In this way we can express the constraint on $H$ as a function of the reheating temperature after inflation.

\medskip

Finally, in a more speculative vein, we explore the consequences of a conjecture put forward in the context of quantum theories of gravity. It has been argued that no formulation of quantum mechanics in dS spaces can be consistent. On the other hand, we observe today a positive cosmological constant. The resolution of this conflict between a conceptual obstruction and an empirical fact can be found by assuming that the asymptotic state of our universe is not dS. In other words, we are only living in a transitory situation and today's dS space will soon terminate. Of course, there are many ways in which the universe could escape the allegedly dreadful dS condition, but it is tempting to speculate that the instability of the electroweak vacuum is the emergency exit chosen by nature. If we take this hypothesis seriously, we obtain a rather precise prediction for a combination of the Higgs and top masses in the SM, in good agreement with experimental measurements. The predicted strip in parameter space can be narrowed further by the hypothesis that the universe must have been sufficiently hot in the past (for instance, for allowing some high-temperature mechanism of baryogenesis).

\bigskip

The paper is organised as follows. In section~\ref{Vxi} we address the preliminary technical issue of the gauge dependence of the effective potential. The generation and evolution of Higgs fluctuations during inflation is studied in section~\ref{HI}, while the subsequent evolution after inflation is the subject of section~\ref{sec4}. Our speculations on the quantum-gravity prediction of the Higgs mass are discussed in section~\ref{QG}, and our results are summarised in section~\ref{concl}. The details of the general-relativity calculation of the evolution of AdS bubbles in dS or Minkowski backgrounds are contained in the appendix.

\section{Gauge dependence of the SM effective potential}\label{Vxi}

The critical Higgs mass below which the SM Higgs potential becomes unstable is gauge-independent; however the instability scale of the SM potential, {\it e.g.} $h_{\rm max}$,  defined as the Higgs field value at which $V_{\rm eff}(h)$ is maximal, is gauge-dependent (as recently emphasised  in~\cite{diLuzio,Schwartz}). Therefore, one has to be cautious in extracting from the potential a physically meaningful scale associated
to the instability. 

%This issue is really bypassed: the instability scale can be alternatively defined in a gauge-independent way as
%$V_{\rm max} \equiv V_{\rm eff}(h_{\rm max})$, with the numerical result~\cite{instab}
%\beq
%\log_{10}\frac{V_{\rm max}^{1/4}}{\GeV} = 9.5 + 0.7\Mhdiff-1.0\Mtdiff + 0.3 \asdiff .
%\label{eq:lambdai}
%\eeq
%Indeed, the Nielsen theorem~\cite{Nielsen} guarantees that the value of the effective potential at any one of its extrema (maximum or minimum) is gauge-independent.

There is a number of ways in which one can try to identify scales that
track the potential instability and are gauge-invariant because expressed in terms of extrema.\footnote{For example, this could be done through the scale of a higher-dimensional operator 
$h^n$ (with $n>4$) that, added to the SM, cures its instability.} 
However, our cosmological computations will employ the full SM effective potential also away from its extrema, so 
we are confronted with the issue of the gauge-dependence of the effective potential shape, an old topic much debated in the literature.\footnote{We summarise some of the main approaches here.
Nielsen~\cite{Nielsen2014} proved that the gauge dependence of the effective potential can be reabsorbed by a re-definition of the fields.
Tye~\cite{Tye} found that the effective potential and the effective kinetic terms are separately gauge-invariant
if the perturbative expansion is performed by decomposing
the Higgs doublet into the physical Higgs field $h$ and the 3 angular coordinates $\pi$, such that the Goldstone fields $\pi$ are massless at any value of $h$, not only at the extrema of the potential.
The results then agree with the unitary gauge.
Buchmuller et al.~\cite{Buch} computed the effective potential in terms of the gauge-invariant combination $\Hig^\dagger \Hig$
claiming a gauge-invariant effective potential; again this selects the radial mode of the Higgs doublet such that Goldstone are always massless.
Schwartz et al.~\cite{Schwartz} argue that finding a  gauge-invariant definition of the effective potential is a misguided enterprise
and that the contribution of Goldstone bosons should be neglected (at leading order) in a consistent perturbative expansion
around the field value at which $\lambda=0$.
}
In dealing with this issue we follow a pragmatic approach. First, we insist on calculating physical quantities, that can be proven to be gauge independent. Second, we make sure that the approximations we use in those calculations are consistent, in the sense that any residual gauge dependence is smaller than the precision of our approximations.

The gauge-independence of our results is ultimately based on the 
Nielsen identity that describes how the effective action depends on the gauge-fixing parameters and how to extract out of it gauge independent quantities. Let us briefly discuss how this works.

The fact that the Higgs effective potential $V(h)$ depends on the gauge parameters (generically denoted as $\xi$) follows from the fact that the effective action $S_{\rm eff}$ itself is a gauge-dependent object. In spite of this, as is well known, both the potential and the effective action are extremely useful and physical quantities extracted from them (like particle masses, $S$-matrix elements, the vacuum energy density, tunnelling rates in the case of metastable vacua, etc.) turn out to be gauge-independent, as they should.

\subsection{Gauge (in)dependence of the effective action}
The Nielsen identities~\cite{Nielsen} tell us that the gauge dependence of the effective action can be compensated by a local field redefinition.
In other words, different gauges describe the same physics in terms of different coordinates in field space (leading to different potentials but also to different kinetic terms).
Particularising to cases with Higgs background only, one has
\be
\label{NIS}
\xi \frac{\partial S_{\rm eff}}{\partial\xi}=-\int d^4 x \ K[h(x)] \frac{\delta S_{\rm eff}}{\delta h(x)}\ ,
\ee
where $K[h(x)]$ is a functional of $h$ that can be found in~\cite{Nielsen}. 

One immediate consequence of the Nielsen identity is that the action evaluated on a solution of the equation of motion for $h$, $\delta S_{\rm eff}/\delta h=0$, is gauge-independent. We also see that the gauge-independence of the extremal values of the effective potential follows directly by applying the previous general fact to constant field configurations. 

Writing the effective action in a derivative expansion
%\footnote{From a Wilsonian perspective the first two terms in this expansion capture the dominant effects in the infrared. We are implicitly assuming that possible effects from Planckian physics are negligible even at the large field values we will consider later on (as long as they are sub-Planckian).}
\be
S_{\rm eff}[h] = \int  d^4 x \left[-V(h)+ \frac{1}{2}Z(h)(\partial_\mu h)^2 + {\cal O}(\partial^4)\right] \ ,
\ee
we can find a series of Nielsen identities  for the coefficients of this expansion \cite{MetaWe,GarKon} by
expanding in the same way $K[h]$ and $\delta S_{\rm eff}/\delta h$ in (\ref{NIS}), as
\be
K[h]= C(h) + D(h) (\partial_\mu h)^2 - \partial^\mu [\tilde D(h)\partial_\mu h] + {\cal O}(\partial^4) \ ,
\ee
\be\label{EoM}
\frac{\delta S_{\rm eff}}{\delta h}= -V' + \frac{1}{2} Z'  (\partial_\mu h)^2-
 \partial^\mu [Z(h)\partial_\mu h] + {\cal O}(\partial^4) \ ,
\ee
where primes denote $h$-derivatives. The higher-order derivative terms are expected to be suppressed by an energy scale, which can be as low as the value of the Higgs field, and by a one-loop factor, which is the same for any order in derivatives. Therefore, our derivative expansion is valid only when the gradient of the Higgs field is smaller than the homogeneous value of the field under consideration.

At the lowest order in the derivative expansion, we find the expression for the gauge-dependence of the effective potential
\be\label{NIV}
\xi\frac{\partial V}{\partial  \xi} + C(h)  V'=0
\ ,
\ee
which, as anticipated, ensures the gauge-independence of the values of the potential at its extremal points. This 
Nielsen identity also tells us that the explicit $\xi$-dependence of the potential  can be compensated by an implicit
$\xi$-dependence of the field as:
\be\label{NIh}
\xi \frac{d h}{d  \xi} = C(h)\ ,
\ee
so that $d V/d \xi=0$.
At order ${\cal O}(\partial^2)$ we get
\be\label{NIZ}
\xi\frac{\partial Z}{\partial  \xi} = - C Z'- 2 Z C' + 2 D V' + 2 \tilde D V''
\ ,
\ee
where we suppressed the $h$ dependence of all functions.

It is useful to consider the order in weak gauge couplings (denoted generically by $g$ in this paragraph) of the different functions that appear in the previous identities \cite{MetaWe}. As we will be interested in the potential region where the Higgs quartic coupling 
gets negative, eventually inducing a new minimum radiatively, we 
will use the counting $\lambda\sim g^4$. The function $C(h)$
starts at one loop and is ${\cal O}(g^2)$. The Nielsen identity  (\ref{NIV})
then implies that the $\xi$ dependence of $V$ starts at ${\cal O}(g^6)$. On the other hand, the Nielsen identity (\ref{NIZ}) implies that
the $\xi$ dependence of $Z$ starts at ${\cal O}(g^2)$, with the
terms involving $D$ and $\tilde D$ being of higher order in $g$. As we will see in the next subsection, it will be sufficient for our purposes to
deal with the dominant $\xi$ dependence of the potential so that we will neglect the effect of the subleading $D$ and $\tilde D$ terms in what follows, as in \cite{MetaWe}.

\bigskip

Let us next consider the $\xi$-dependence of the equation of motion for $h$
which we write using
\be
{\rm EoM}[h] \equiv \sqrt{Z} \partial^2 h + 
\frac{1}{2\sqrt{Z}}Z'  (\partial_\mu h)^2 
+\frac{1}{\sqrt{Z}}V' \ ,
\ee
and its solutions $\bar h(x)$, which satisfy ${\rm EoM}[\bar h]=0$. 
It is straightforward to show that
\be
\xi \left.\frac{d}{d\xi}  {\rm EoM}[h]\right|_{h=\bar h}=0\ ,
\ee
up to ${\cal O}(\partial^4)$ corrections, provided we use $\xi d\bar h/d\xi = C$. In principle, one can continue the check of the gauge invariance of the equations of motion iteratively up to infinite order in the number of derivatives.\footnote{Using the previous identities {one can also check, to all orders}, the $\xi$-independence of the scalar physical mass $M_h^2\equiv \left. V''/Z\right|_{\rm min}$,
evaluated at the minimum of the potential, as indicated.}
This means that, if some $\bar h_\xi(x)$ solves the equation of motion for some choice of $\xi$ and we shift $\xi \rightarrow \xi +d\xi$, the shifted solution is $\bar h_\xi(x)+d\bar h_\xi (x)$ with $d \bar h_\xi(x)= C(\bar h_\xi) d\xi /\xi$. In other words, the field rescaling that 
can balance the effect of changing $\xi$ in the effective potential
is the same field rescaling that applies to the solutions of the equation of motion for different $\xi$. 

\bigskip

The same rescaling works for the Fokker-Planck and Langevin equations that we will use later on to describe the Higgs fluctuations during inflation. These equations take the form, $\hbox{\sc Langevin}[h_L]=0$ and $\hbox{\sc FokkerPlanck}[P(h,t)]=0$, with
\be\label{Lang}
\hbox{\sc Langevin}[h]\equiv \sqrt{Z}\frac{d h}{dt} +\frac{1}{3H\sqrt{Z}} V' -\eta  \ ,
\ee
and 
\be\label{FP}
\hbox{\sc FokkerPlanck}[P(h,t)]\equiv \frac{1}{\sqrt{Z}}\frac{\partial}{\partial h}\left\{\frac{1}{\sqrt{Z}}\left[
\frac{\partial}{\partial h}\left(\frac{H^3}{8\pi^2}\frac{P}{\sqrt{Z}}\right)+\frac{1}{3H} \frac{PV' }{\sqrt{Z}}
\right]\right\}-\frac{1}{\sqrt{Z}}\frac{\partial P}{\partial t}\ .
\ee
Here $P(h,t)\, dh$ is the probability for finding the Higgs field in the infinitesimal interval between $h$ and $h+dh$ at time $t$ during inflation. The fact that $P$ is a probability density explains why $P$ enters in \eq{FP} through the ratio $P/\sqrt{Z}$.

Using these expressions and the $\xi$-dependence of $V(h)$, $Z(h)$, and $h$ as described in eqs.~(\ref{NIV})--(\ref{NIZ}), we get
\be
\left.\frac{ d}{ d\xi}  \hbox{\sc Langevin}[h]\right|_{h=h_L}= 0\ ,\qquad
\left.\frac{ d}{ d\xi}  \hbox{\sc FokkerPlanck}[P(h,t)]\right|_{P=\bar P}= 0\ ,
\ee
(where $h_L$ and $\bar P$
are solutions of the Langevin and Fokker-Planck equations, respectively)
up to corrections that can be shown to be subleading.\footnote{Here we are explicitly using the derivative expansion previously introduced to derive the gauge transformation properties of $V,\,Z$ and $h$. Indeed, the Langevin and Fokker-Planck formalism represent a truncation of the theory at the lowest order in derivatives, where the approximation is justified by the smallness of the gradient of the field with respect to the Hubble parameter.  In the rest of the paper, we will be using these equations to describe evolutions of the Higgs field for values of the Hubble parameters even quite larger than the Higgs vev itself. For this reason, we cannot naively apply the zeroth order truncation in derivatives of the effective action, because, as we discussed, the derivative expansion is suppressed only by the Higgs vev. However, since the higher derivative corrections are suppressed by at least a one-loop factor and are not log-enhanced (at one-loop), a consistent truncation is to use the Langevin and Fokker-Planck equations as derived from an effective action where the only corrections that are included are the non-derivative, leading log-enhanced, ones. This will be how we will use the Langevin and Fokker-Planck equations in the rest of the paper.
%In the physics problem at hand, the fluctuations in the Higgs field must be small enough 
%in order to avoid falling in the true minimum, and
%therefore neglecting higher derivatives is  as good an approximation as using the Langevin and Fokker-Planck formalism.
}$^,$\footnote{
Note also that, concerning the dependence on the renormalisation scale $\mu$, one can show the $\mu$-independence
of $M_h^2$,  $ {\rm EoM}[h]$, $\hbox{\sc Langevin}[h]$ and $\hbox{\sc FokkerPlanck}[P(h,t)]$ just making use of $dV/d\mu=0$, $dh/d\log\mu=\gamma h$,
$d(\partial V/\partial h)/d\log\mu = -\gamma (\partial V/\partial h)$, $dZ/d\log\mu = -2\gamma Z$, etc.
}
This shows once again that if we have a solution of the Langevin  equation for a given value of $\xi$, we automatically obtain a solution for $\xi+ d\xi$ by the shift $h(\xi)+ C(h(\xi))  d\xi /\xi$. 
For the Fokker-Planck equation, a solution for general $Z$ is formally related to a solution for $Z=1$ again by a field rescaling, with
$P(h,t)/\sqrt{Z(h)}= P_c(h_c(h))$, where the relation between
$h$ and the canonical field $h_c$ follows from $dh_c/dh=\sqrt{Z(h)}$.
 As a result, the integrated probability is independent of the field rescaling: 
\be
\int_{h_{c,i}}^{h_{c,f}}P_c(h_c) dh_c = 
\int_{h_i}^{h_f}P_c(h_c(h)) \sqrt{Z(h)} dh =
 \int_{h_i}^{h_f}P(h) dh \ .
\ee
This implies that the 
probability of finding the field beyond $h_{\rm max}$ after a given number of e-folds is a gauge invariant quantity: although 
the value of $h_{\rm max}$ depends on $\xi$, the $\xi$-change of the ratio $P(h,t)/\sqrt{Z(h)}$ corresponds to the same field-rescaling 
$h(\xi)\rightarrow h(\xi)+ C(h(\xi))  d\xi /\xi$ and leaves the integrated probabilities unchanged.

To sum up, the key idea is that a change in a given gauge parameter $\xi$ is equivalent to a redefinition of the Higgs field, which should leave physics invariant. The effective potential, the equations of motion for the Higgs field and the Fokker-Planck and Langevin equations enjoy a sort of ``covariance" under changes of the gauge parameters. The equations are changed in such a way that the change induced in their solutions is just a common field redefinition dictated by the Nielsen identity.
% As an example relevant for our discussion, the probability $P(h>h_{\rm max})$ of finding the Higgs at a given time during inflation with a value larger than the instability scale $h_{\rm max}(\xi)$ (gauge-dependent, as indicated) can be obtained by solving a Langevin equation that is explicitly dependent on $\xi$ also. Changing $\xi$ is equivalent to a rescaling of $h_{\rm max}$ and the same rescaling for the solution of the Langevin equation. In the integral that determines $P(h>h_{\rm max})$ the change of $\xi$ is equivalent to a redefinition of the integration variable, with no effect on the final result, in such a way that $P(h>h_{\rm max})$ is in fact a gauge-independent quantity.

\subsection{Effective potential including only log-enhanced corrections}

For the previous appealing properties to hold, the interplay between the effective potential and the kinetic term in the effective action is crucial. For the SM case at very large field values we write
\beq \Lag_{\rm eff} = Z(h,\xi)  \frac{(\partial_\mu h)^2}{2}-
\lambda_{\rm eff}(h,\xi)\frac{h^4}{4} + \cdots\eeq
where the ellipsis denotes higher derivative terms and both $ Z(h,\xi)$ and $\lambda_{\rm eff}(h,\xi)$ include radiative corrections and depend on $\xi$. As usual, it proves convenient to 
use a canonically normalised Higgs field $h_{\rm can}(h,\xi)$ as $dh_{\rm can}/dh = Z^{1/2} $ and to re-express the
effective Lagrangian in terms of $h_{\rm can}$, obtaining
\beq \Lag_{\rm eff} = \frac{(\partial_\mu h_{\rm can})^2}{2}- \lambda_{\rm can}(h_{\rm can},\xi) \frac{h_{\rm can}^4}{4} + \cdots\eeq

\begin{figure}[t]
$$\includegraphics[height=5.7cm]{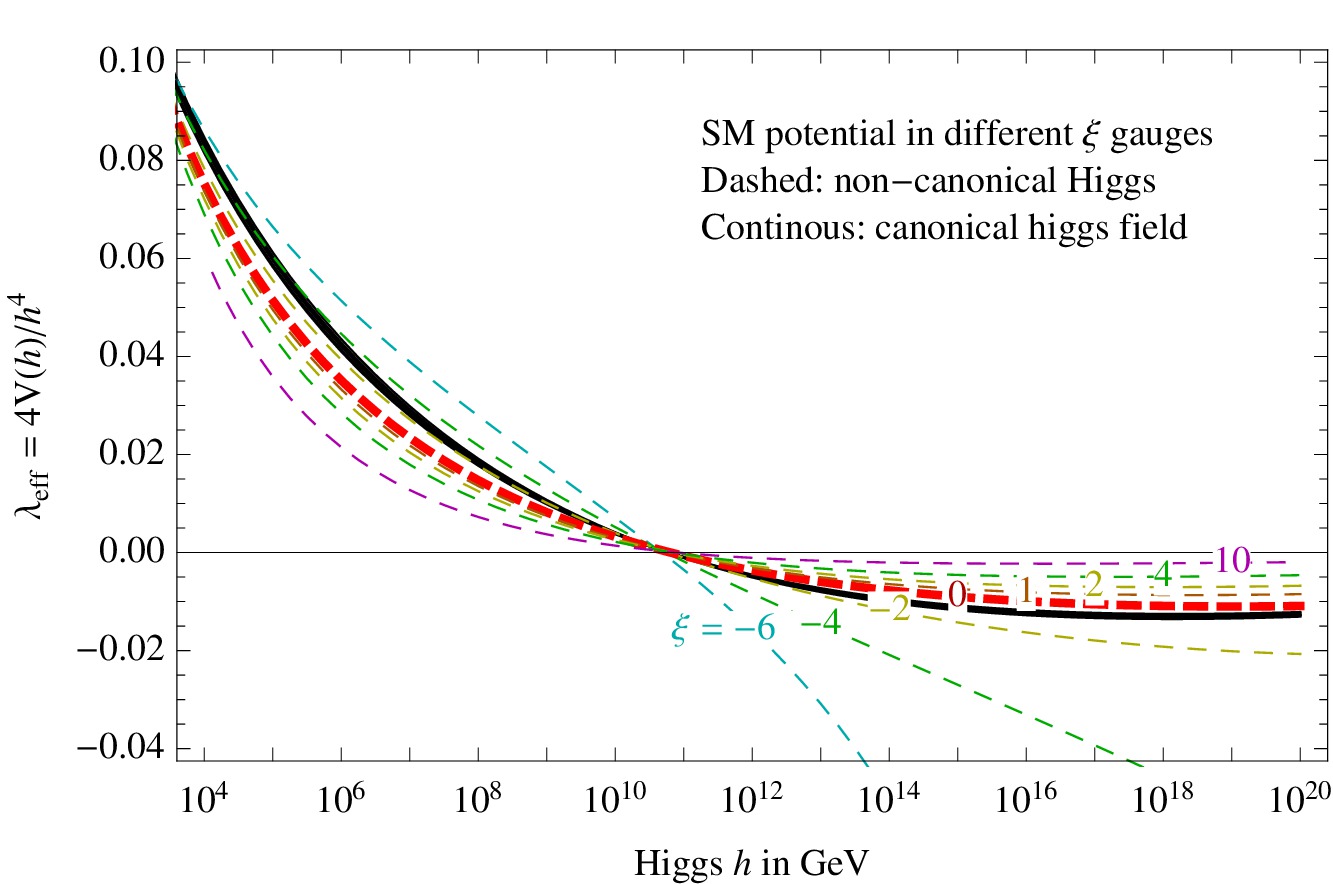}\qquad
\includegraphics[height=5.5cm]{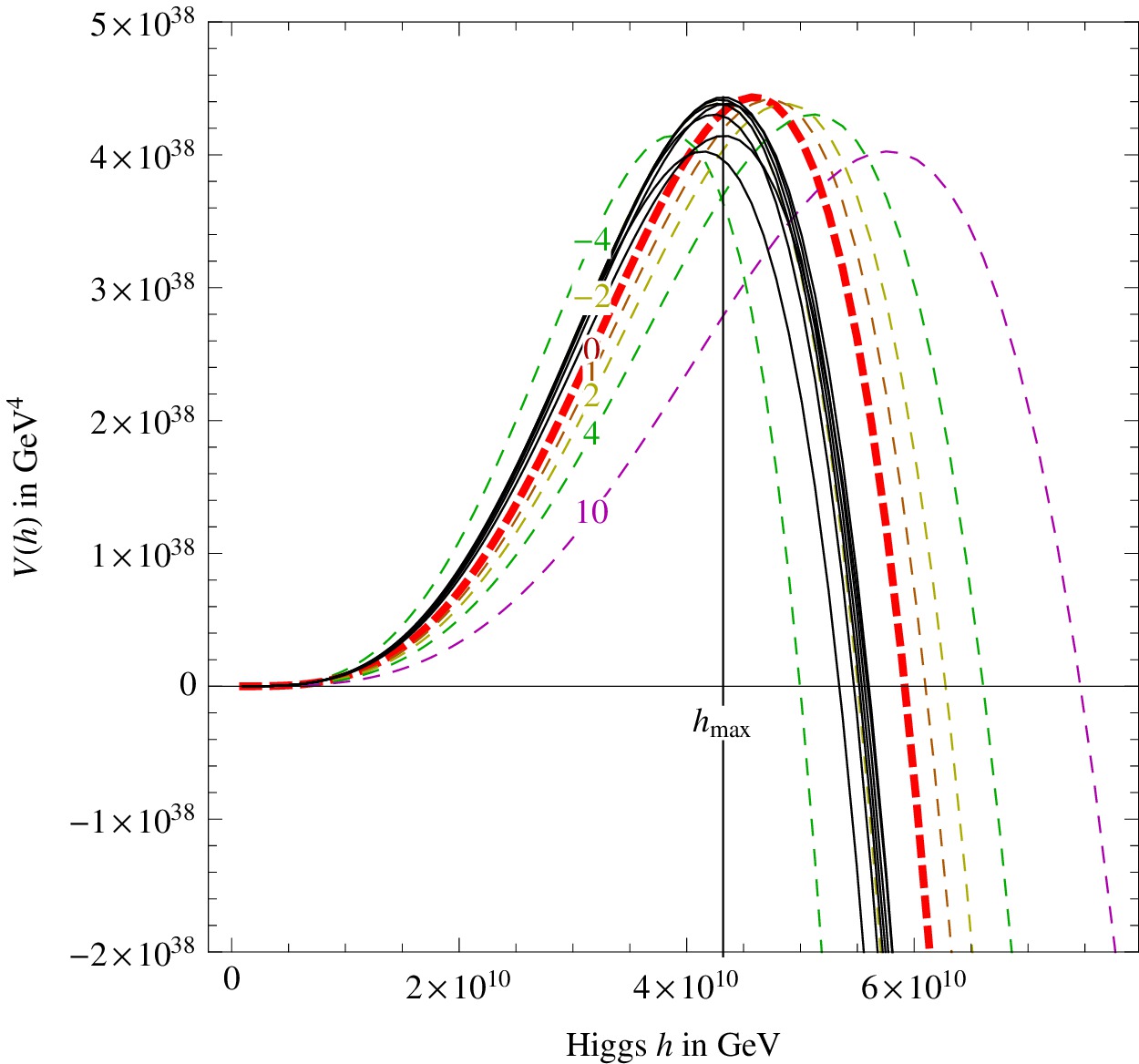}$$
\caption{\em 
\label{fig:VSMxi} The dashed curves show the effective quartic coupling (left) and effective SM potential (right) computed at next-to-leading order in a generic Fermi $\xi$-gauge.
The thick red dashed curve corresponds to the Landau gauge, $\xi=0$.
The right handed panel shows that the height of the potential barrier is only approximately gauge-independent (a measure of the residual gauge dependence).
The  black continuous curves show the same potential expressed
in terms of the canonical field $h_{\rm can}$: the gauge dependence in the potential gets compensated
by the gauge-dependence of the kinetic term, such that the continuous curves nearly overlap.}
\end{figure}

In terms of the canonical field all the equations become simpler as we do not have to drag the $Z$ factor around. An additional bonus is that 
the residual $\xi$ dependence in our approximations will be significantly reduced. Let us see how this works examining the gauge dependence of the effective potential. The coloured dashed curves in fig.~\ref{fig:VSMxi} show  $\lambda_{\rm eff}(h,\xi)$,  which tracks the large field behaviour of the SM effective potential, as computed at next-to-leading order (NLO) accuracy in the Fermi $\xi$ gauges~\cite{diLuzio}\footnote{At this level of approximation the potential has a residual dependence on the RG scale $\bar\mu$  comparable to the gauge dependence.
We here adopted the choice $\bar\mu = h e^{\Gamma}$ that minimises the error.}
further improved by performing a resummation of IR-divergent Goldstone loops~\cite{IRresum}\footnote{Around the minimum of the potential, this resummation becomes equivalent to the expansion of~\cite{Schwartz}, which makes the truncation of the potential compatible with the Nielsen identity.}. We take into account the running of $\xi_1$ and $\xi_2$ (gauge-fixing parameters for hypercharge and SU(2)$_L$),
assuming a common value $\xi$ renormalised at $M_t$.
We confirm that $\lambda_{\rm eff}(h,\xi)$ significantly depends on the gauge-parameter $\xi$.
The black curves  in fig.~\ref{fig:VSMxi} show $\lambda_{\rm can}(h_{\rm can},\xi)$,
again computed at NLO in Fermi $\xi$ gauges:
we see that the dependence on $\xi$ almost completely disappeared --- all black curves almost merged into a single curve.

One can explain analytically why the gauge dependence approximately cancels out by looking at the dominant corrections enhanced by large logarithms, which are
resumed by solving the RG equations and setting the  RG scale  $\bar\mu$
around the field value of interest:
\beq \lambda_{\rm eff}(h,\xi) \approx e^{4\Gamma(\bar\mu\approx h,\xi)}
\lambda(\bar\mu\approx h),\qquad
Z_{\rm eff}(h,\xi) \approx e^{2\Gamma(\bar\mu\approx h,\xi)}\ ,
\eeq
where $\Gamma=\int_{M_t}^{\bar \mu} \gamma \, d\ln\bar\mu$, $\gamma$ is the gauge-dependent
anomalous dimension of the Higgs field,
and $\lambda(\bar \mu)=\int_{M_t}^{\bar \mu} \beta_\lambda \, d\ln\bar\mu$  is the running quartic coupling, where $\beta_\lambda$ is gauge-independent.
In this leading-order (LO) approximation one has
\beq 
h_{\rm can} \approx h e^{\Gamma(\bar\mu \approx h)},\qquad 
\lambda_{\rm can}(h_{\rm can})\approx \lambda(\bar\mu\approx h)
\ ,
\eeq
which is gauge-independent because the RGE for $\lambda$ and all other couplings of the theory are gauge-independent. 
The order-of-magnitude gauge dependence of $h_{\rm max}$ found in~\cite{diLuzio} disappears because it is almost entirely due to the RG factor $\Gamma$.

Such LO cancellation has been noticed before, see {\it e.g.} \cite{Frere,Sher}. Note however that the field redefinition dictated by the Nielsen identity we discussed earlier and the field redefinition
required to make the field canonical are the same only at LO. Moreover, the field
redefinition from the Nielsen identity becomes considerably more complicated at NLO~\cite{Nielsen2014}. Its use to define a ``gauge-independent" potential is in fact equivalent to choosing a particular gauge and therefore does not solve the problem of how to extract gauge-invariant quantities out of the effective action. For this reason
we refrain from attempting to use it as a way of defining a gauge-invariant potential and simply use the canonical field as a way of reducing the residual gauge dependence of our results.

The previous discussion has been carried out in Fermi gauge at NLO to help us clarify the issues related to gauge invariance. Having understood them, we can now use the state-of-the-art computation of the effective potential in the Landau gauge $(\xi=0$) with NNLO accuracy (2 loop finite corrections and 3 loop RGE corrections) \cite{instab} combined with the use of a field redefinition to make the field canonical (taking into account the effect of potentially large logarithms in $Z$). 
Using that canonically normalised Higgs field $h$,
in the region around  the top of the barrier, the SM Higgs potential can be analytically approximated as
\be
 V_{\rm eff}( h) \approx - b \ln \left(\frac{ h^2}{h_{\rm max}^2 \sqrt{e}}\right) \frac{ h^4}{4}\ ,
\label{StrV}
\ee
where $h_{\rm max}$ is the field value at which  $V_{\rm eff}(h)$ 
takes its maximal gauge-invariant value $V_{\rm eff}(h_{\rm max})= b h_{\rm max}^4/8$.
Using the value $b \approx 0.16/(4\pi)^2$ for the $\beta$ function of $\lambda$ around $h_{\rm max}$,
we find  $h_{\rm max}=5\times 10^{10}\GeV$ for the present best-fit values of $M_t$, $M_h$ and $\alpha_3$. Although this value of $h_{\rm max}$ is computed in Landau gauge and it would be slightly different in other gauges, the reader should keep in mind that the results we present in the following sections are gauge-invariant
even if for convenience we express them in terms of $h_{\rm max}$.\footnote{Alternatively, we could choose other scales associated (more indirectly) to the instability which are explicitly gauge invariant. 
One could be the renormalisation scale $\mu_0$ at which the quartic Higgs coupling $\lambda$ crosses zero; another choice is the scale  $\mu_X$ at which the one-loop radiatively corrected Goldstone mass is zero (as used in \cite{Schwartz}). For the same central values above we get $\mu_0=1\times 10^{10}$ GeV and 
$\mu_X=4.5\times 10^{10}$~GeV.}

\section{Higgs fluctuations during inflation}\label{HI}
The instability of the Higgs potential leads to an interesting dynamics during inflation.
We focus on the relevant radial mode $h = \sqrt{2  |\Phi_H|^2}$ of the Higgs doublet.
If the Hubble constant ${\Hub }$ is large enough, 
{$h$} fluctuates beyond the potential barrier.
If  the true vacuum is deep enough, inflation stops in the regions where the Higgs falls, while inflation continues in the (possibly rare) regions where accidentally {$h< h_{\rm max}$}.
In this section we compute the probability of the possible outcomes at the end of inflation, while
in the next section we will discuss what happens after inflation.

%This leads, after inflation, to a large universe containing regions of true vacuum.
%If these would behave a stable masses, their presence would not be a problem,
%and they might even contribute to the Dark Matter density.
%However in section~\ref{AdS} we show that, under generic conditions, these bubbles expand
%at the speed of light, such that their presence is unacceptable.

In the absence of a large Higgs mass term, the evolution of the long wavelength modes of the $h$ field is controlled by the Langevin equation~\cite{Staro}
\be
\frac{d h}{d t} + \frac{1}{3\Hub} \frac{d V(h)}{d h} = \eta(t),
\label{eq:Langevin}
\ee
where $\eta$ is a Gaussian random noise with
\be
\langle \eta(t) \eta(t') \rangle = \frac{\Hub^3}{4\pi^2} \delta(t-t').
\ee
It is important to realise that eq.~(\ref{eq:Langevin}) is valid only if the positive effective mass squared $V''(h)$ of the Higgs field is light enough
compared to  $\Hub^2$. Only under these circumstances the long wavelength super-Hubble fluctuations
of the Higgs field are generated. On the contrary, if $V''(h)>9\Hub ^2/4$, the resulting power spectrum
of Higgs fluctuations is both  suppressed by  ${\rm exp}(-2V''(h)/\Hub ^2)$ and by the fact that the spectrum
is strongly tilted on the blue side \cite{sup}\footnote{Indeed, the solution of the Klein-Gordon equation for a spin 0 
 particle with mass $m$ in de Sitter goes like ${\rm exp}(-\pi\mu/2)(-\tau)^{3/2}H_{i\mu}(-k\tau)$, where
$\mu=\sqrt{m^2/\Hub^2-9/4}$ and $\tau$ is the conformal time. }.

\subsection{Higgs fluctuations during inflation for ${\xi_H=0}$}\label{xisec}
%\subsection{Higgs fluctuations during inflation}

It is convenient to rewrite \eq{eq:Langevin} replacing time $t$ with the number of $e$-folds $N=Ht$,
and to normalise the Higgs field and its potential in units of the 
Higgs value
$h=h_{\rm max}$ at which $V(h_{\rm max})=V_{\rm max}$ is maximal,
\be
\overline{h} = \frac{h}{h_{\rm max}} \qquad {\rm and} \qquad \overline{V}(\bar h) \equiv  \frac{V}{h_{\rm max}^4}\approx
 - b \ln\left(\frac{\bar h^2}{\sqrt{e}}\right) \frac{\bar h^4}{4}.
\label{StrV2}
\ee
%having used the approximation in \eq{StrV}.
%%where $b \approx 0.15/(4\pi)^2$  is a typical SM value for $h_{\rm max} \approx 10^{10}$.
%The potential $\bar V$ is maximal at $\bar h= 1$.

After these redefinitions, the Langevin equation in eq.~(\ref{eq:Langevin}) becomes
\be
\frac{d \overline{h}}{d N} + \frac{h_{\rm max}^2}{3H^2} 
\frac{d \overline V (\overline h)}{d \overline h}  = \overline\eta(N) 
\ee
where the noise
$\overline\eta(N) $ obeys
\be
\langle \overline\eta(N) \overline\eta(N') \rangle
%= \frac{1}{\Lambda_{\rm max}^2} \langle \widetilde\eta(N) \widetilde\eta(N') \rangle
= \left( \frac{\Hub}{2\pi h_{\rm max}}\right)^2
%\frac{1}{4\pi^2} \frac{\Hub^2}{h_{\rm max}^2} 
\delta(N-N').
\ee

%{\cred Leonardo: ````I must be stupid, but shouldn't this be $\bar h=1$?""}
%\as{just impose $\bar V'=0$...} {\cred[LS: sorry, how $\Lambda_{\rm max} $ is defined? I thought $V'(h=\Lambda_{\rm max})=0$. It is not even true that $V_{\rm max}=\Lambda_{\rm max}^4$. I know I am stupid, but I am confused.]}
%\as{I merged notes written by me and others, without noticing that there is an inconsistency,
%$\bar h \neq 1$.  To fix this we should first choose how we define $\Lambda_{\rm max}$,
%the best would be telling that $V$ is maximal at $h=\Lambda_{\rm max}$ 
%and that this is gauge invariant.}

\begin{figure}[t]
\begin{center}
$$\includegraphics[width=0.6\textwidth]{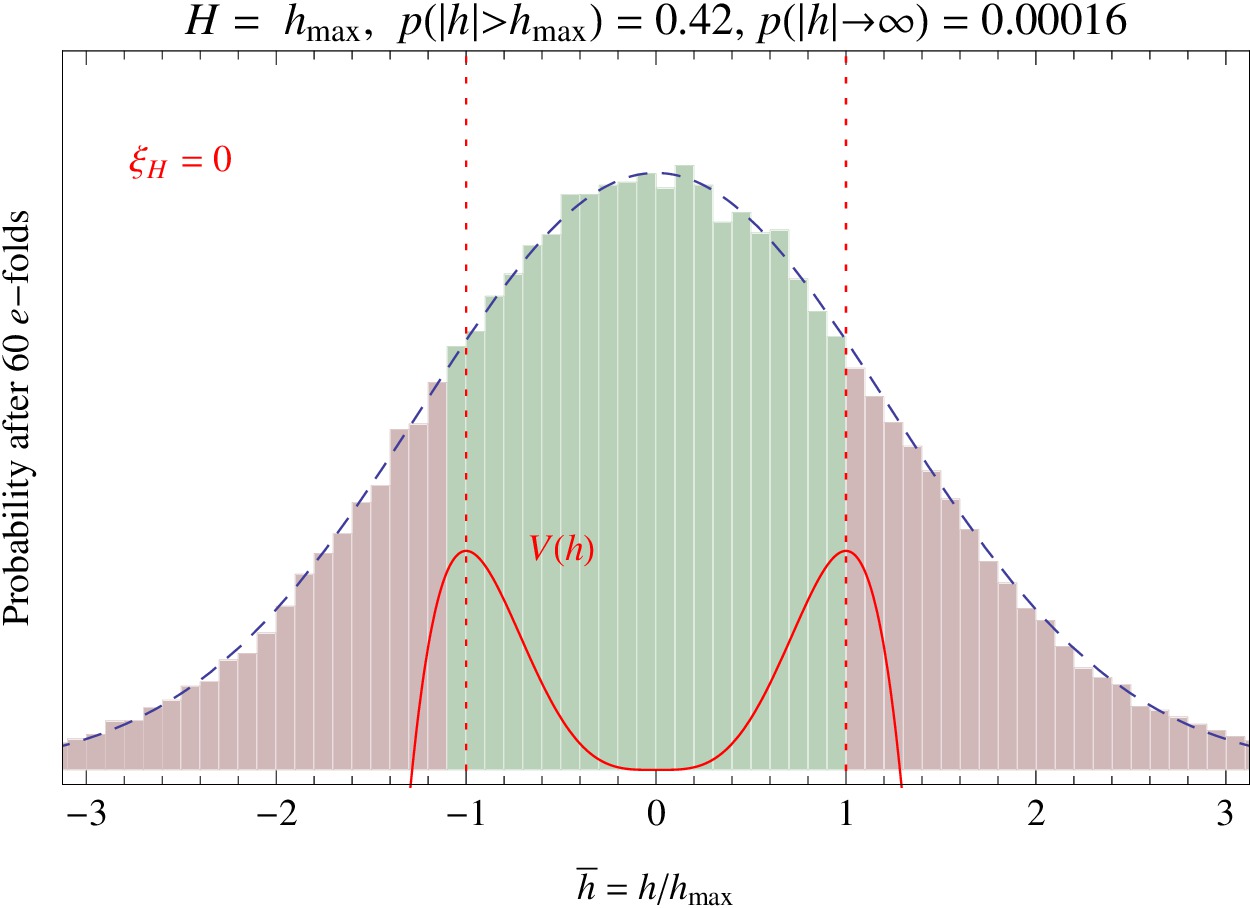}$$
\caption{\em Random distribution of the Higgs field $\bar h = h/h_{\rm max}$
after $N=60$ e-folds of inflation with Hubble constant equal to the Higgs instability scale, $\Hub  = h_{\rm max}$.
The blue dashed curve is the $V=0$ Gaussian approximation of eq.\meq{hGauss}.
The red curve is the SM Higgs potential $\bar V(\bar h)$, in arbitrary units.
\label{ran}
}
\end{center}
\end{figure}

\begin{figure}[t]
\begin{center}
$$\includegraphics[width=0.7\textwidth]{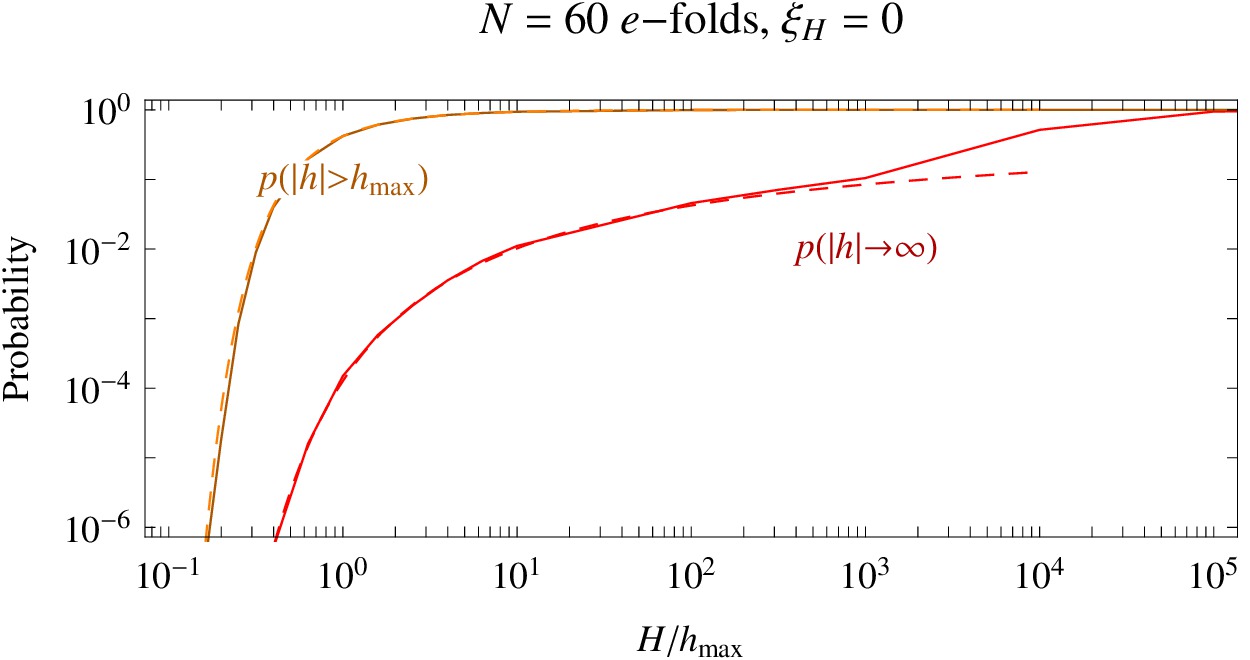}$$
\caption{\em Minimal probability that, after $N=60$ e-folds of inflation,
the Higgs fluctuated above the SM potential barrier (orange curve),
or fall down to the true minimum (red curve).  The continuous curves are the numerical results;
the dashed curves are the analytical approximations presented in the text. 
\label{prob}
}
\end{center}
\end{figure}

\bigskip

One can now numerically generate random realisations of the Higgs evolution in $N$, in steps of $dN$, as
\be \bar h(N+dN) = \bar h(N)  - \frac{h_{\rm max}^2}{3\Hub^2} \bar V' (\bar h) \, dN+ r
\label{LangStr}
\ee
where $r$ are random numbers extracted from a Gaussian distribution
with zero mean and standard deviation $ \sigma = \Hub \sqrt{dN}/(2\pi h_{\rm max})$.

Indeed, for $h\gg \sigma$, the same result is reproduced by the
analytic solution to the Fokker-Planck equation for the probability $P(h,N)$ of finding the Higgs field at the value $h$ after $N$ e-folds of inflation,
\beq
 \displaystyle \frac{\partial P}{\partial N} = \frac{\partial^2 }{\partial h^2} \bigg(  \frac{\Hub^2}{8\pi^2}  P\bigg)+
 \frac{\partial }{\partial h}  \bigg(\frac{V'}{3\Hub^2} P\bigg) ,
 \label{Fokker}
 \eeq
taking $V=0$ and boundary conditions at $h=\pm\infty$.

%\beq \displaystyle \frac{\partial P}{\partial N} = \frac{\partial P}{\partial h} \bigg[  \frac{H^2}{8\pi^2} \frac{\partial P}{\partial h}+
%\frac{V'}{3H^2} P\bigg]\eeq.

Figure~\ref{ran} shows the resulting probability density of the value of $h$ after $N=60$ e-foldings, 
starting from $\bar h=0$ at the beginning of inflation.
The result, a quasi-Gaussian distribution, has a simple interpretation.
Given that the quartic Higgs coupling vanishes around the instability scale, for a large range of Higgs values around the instability
scale the classical evolution (sourced by the gradient of the potential) is negligible with respect to the quantum evolution (sourced by the random noise $\eta$).
As a consequence, even assuming that the Higgs starts from $h=0$, the field
$h$ acquires a Gaussian distribution with zero mean and variance that grows with $N$:
\beq\label{eq:hGauss}
P(h,N) = \frac{1}{\sqrt{2\pi \langle  h^2\rangle}}\exp \left( -\frac{h^2}{2\langle  h^2\rangle}\right) \, , ~~~~~
\sqrt{ \langle  h^2\rangle} = \frac{\Hub}{2\pi} \sqrt{N}.\eeq
The distribution shown in fig.~\ref{ran} maintains its quasi-Gaussian shape also for values of the Higgs field well above $h_{\rm max}$. Therefore, during inflation, the Higgs field can fluctuate above the barrier without being sucked into the negative-energy (AdS) true vacuum.
%During inflation the Higgs vev mostly fluctuates below, around and {\em above} the barrier;
The only regions where the Higgs falls into the true minimum are those where $h$
fluctuates to field values so large that the potential slope can no longer be neglected.
%because $\lambda(h)$ is large and negative.

\subsubsection*{Regions that fluctuate above the potential barrier}
The minimal probability that the Higgs ends up beyond the top of its potential barrier after $N$ e-folds is 
\beq \label{phmax}
p(|h|> h_{\rm max}) \approx1- \hbox{erf}\bigg(\frac{\sqrt{2}\pi  h_{\rm max}}{\sqrt{N} \Hub}\bigg).\eeq
This probability, obtained by integrating the Gaussian distribution for $|h|> h_{\rm max}$, is minimal because it corresponds to the initial condition $h=0$. Shifting the peak of the distribution to a non-vanishing value of $h$ will only increase $p(|h|> h_{\rm max})$.
The solid orange curve in fig.~\ref{prob} shows our numerical result for this probability as a function of the Hubble constant during inflation,
in units of the Higgs instability scale $\Hub/h_{\rm max}$. The dashed orange curve corresponds to the analytic expression in \eq{phmax}, which is evidently an excellent approximation.
%\footnote{Eq.~(\ref{Fokker}) solves for the probability that a given region of space has value of the Higgs field equal to $h$ by the end of inflation. We impose this probability to be very small for $h> h_{\rm max}$. This same question can be addressed also using a different formalism aimed at deriving the expectation value of the fraction of reheating volume created by inflation where the Higgs field is at or above a certain value of the higgs field. It is interesting to verify that we obtain compatible results. This formalism has been developed in~\cite{Creminelli:2008es,Dubovsky:2008rf,Dubovsky:2011uy,Lewandowski:2013aka}, and here we quote simply the relevant result. Using eq.~(54) of~\cite{Dubovsky:2011uy}, the expectation value of reheating volume with the Higgs field $h$  in the infinitesimal interval $[h_{\rm max},h_{\rm max}+dh]$ is given by
%\be
%\left\langle\begin{array}{cc}
%{\rm reheating\; volume\; with\;} h\in[h_{\rm max},h_{\rm max}+dh]\\
%{\rm\; in\; units\; of\; the\;inflationary\; Hubble\;volume}
%\end{array}\right\rangle \simeq \frac{dh}{H}\; e^{- \frac{2\pi^2}{N} \frac{h_{\rm max}^2}{H^2}}\ ,
%\ee 
%having neglected pre-exponential factors. 
%Imposing the overall reheating volume with $h\geq  h_{\rm max}$ to be less that one Hubble volume, gives a result compatible with the main text.
%}

% 1 - erf(mu/Sqrt[2]/sigma)
No constraints arise if, after inflation, the regions with $|h|> h_{\rm max}$ fall back to the SM minimum, pushed by thermal effects (see section~\ref{sec4}).
If instead, after inflation, the regions with $|h|> h_{\rm max}$ fall down into the true AdS minimum, then their probability
should be smaller than  $e^{-3 N}$, so that it is unlikely to find the Higgs away from its EW vacuum in any of the $\sim e^{3N}$
causally independent regions that are formed during inflation and that constitute the observable universe today.
Using $1-\hbox{erf}(x) \simeq e^{-x^2}/\sqrt{\pi}x$ for large $x$, this condition implies
\beq\bbox{ \frac{\Hub}{h_{\rm max}} < \sqrt{\frac23} \frac{\pi}{N}  \approx 0.04} .
\label{box1}
\eeq

\subsubsection*{Regions that fall to the true minimum during inflation}

The approximation of neglecting the scalar potential $V$, which led to the quasi-Gaussian distribution of the Higgs field values, breaks down at large $h$. There, the gradient of the potential 
dominates over quantum fluctuations, and  $h$ falls down to its true minimum already during inflation.
The solid red curve in fig.~\ref{prob} shows our numerical result for such probability.

%In such regions one possibly has $h\sim M_{\rm Pl}$  and $V \sim -M_{\rm Pl}^2$. 

We can analytically estimate the probability for $h$ to fall into its true vacuum after $N$ e-folds of inflation.
We first consider a potential  $V = \lambda h^4/4$  with constant $\lambda$
and assume that the bulk of the Higgs field probability distribution is still given by the Gaussian in \eq{eq:hGauss}, cut at large field values.
The location of the cut is estimated by demanding that the classical evolution becomes more important than the quantum fluctuations \cite{zurek}. This can be quantified by requiring that the second term in the right-hand side of \eq{Fokker} dominates over the first one,
\beq
 \left| \frac{\partial }{\partial h}  \bigg(\frac{V'}{3\Hub^2} P\bigg) \right| > k
\left| \frac{\partial^2 }{\partial h^2} \bigg(  \frac{\Hub^2}{8\pi^2}  P\bigg) \right| \, ,
\label{FokFok}
\eeq
where $k$ is a fudge factor and $P$ is given in  \eq{eq:hGauss}. Equation~(\ref{FokFok}) implies that the Gaussian distribution must be cut for $h^2 >3kH^2/(2|\lambda | N)$, and values of $h$ that satisfy this inequality are sucked into the true minimum.
 
%are the ones where $h^2  > 3 \Hub^4/8\pi^2|\lambda|\langle h^2\rangle$.
%\xxx{Brandenberger 1504.00867 says that the stochastic boundary is instead $\frac{3}{2\pi} H^3 > |V'|$; this gives $h> H (2\pi /3|\lambda|)^{1/3}$ (different dependence on $N$ and on $\lambda$) and so a different probability. It fits worse.  AS: for me Brandenberger is oversimplified i.e. wrong.  }
Therefore, the probability of falling to infinity is  exponentially suppressed for small $|\lambda|$:
\beq p(|h|\to \infty) \approx 1- \hbox{erf}\,\left(\frac{\pi\sqrt{3k}}{N  \sqrt{|\lambda|}} \right).\eeq
Such probability satisfies $p(|h|\to\infty)<e^{-3N}$ for $|\lambda|<k\pi^2/N^3$.

Considering now the more realistic case of the SM potential with a running coupling $\lambda (h)= -b \ln (h^2/h_{\rm max}^2\sqrt{e})$,
we find
\beq p(|h|\to \infty) \approx 1- \hbox{erf}\,\left(\frac{\pi\sqrt{3k}}{N  \sqrt{b B}} \right),\qquad
\hbox{where}\qquad B = \hbox{PL}\left( \frac{3k\Hub^2}{2b N h_{\rm max}^2}\right) \,
\eeq
where PL is the ProductLog function. This analytic approximation of $p(|h|\to \infty)$ is shown in
fig.~\ref{prob} as the dashed red line and agrees  well  with the numerical computation, once we fit the fudge factor to be $k=2.6$.
Next, we need to extrapolate the analytic approximation to probabilities much smaller than those that can be computed numerically.
The probability $p(|h|\to \infty)$ is smaller than $e^{-3N}$ for
\beq  \bbox{\frac{\Hub}{h_{\rm max} }< \frac{\pi }{N}  \sqrt{2\over 3}e^{\pi^2 k/2b N^3} \approx 0.045}.
\label{box2}
\eeq

\begin{figure}[t]
\begin{center}
\includegraphics[width=0.6\textwidth]{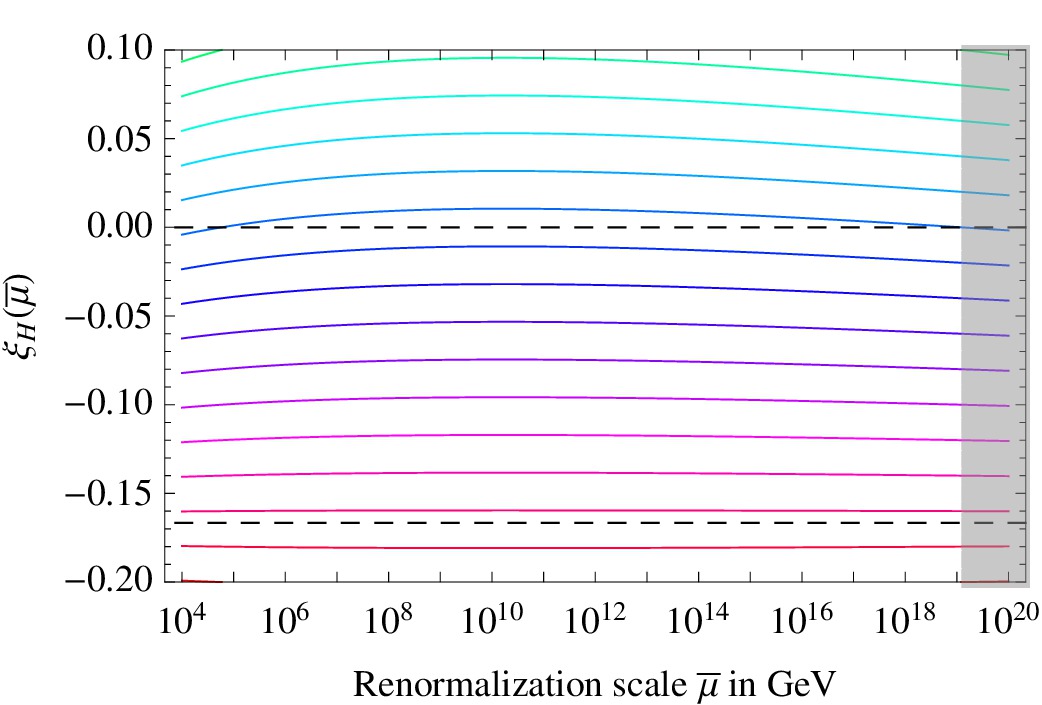}
\end{center}
\caption{\em Running of  the Higgs  coupling to gravity $\xi_H$ as a function of the renormalisation scale in the SM, for different initial conditions at the Planck scale.
The dashed horizontal lines correspond to the special values $\xi_H=-1/6$ and $\xi_H=0$.}
\label{fig:xiRGE}
\end{figure}

\begin{figure}[t]
\begin{center}
$$\includegraphics[width=0.6\textwidth]{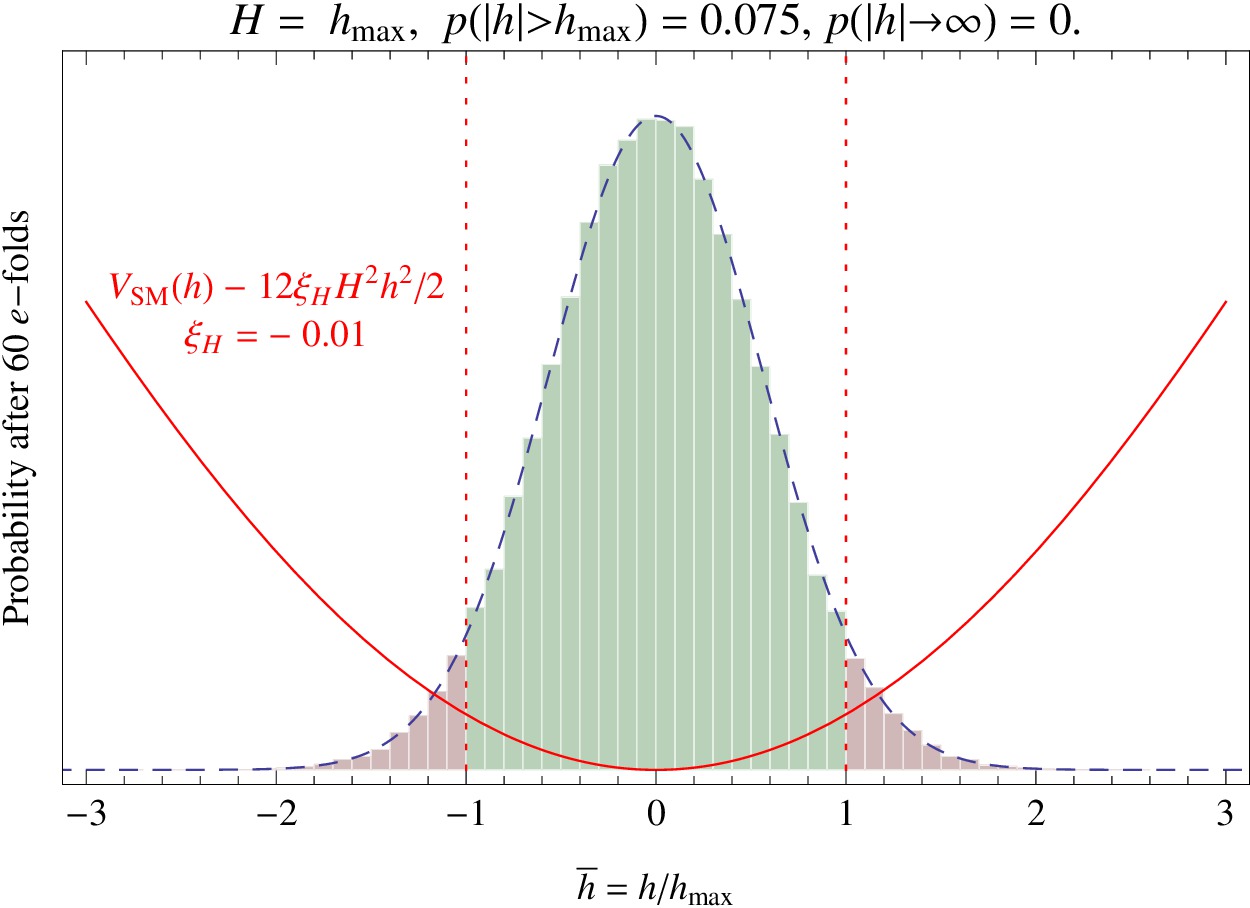}$$
\caption{\em Random distribution of the Higgs field $\bar h = h/h_{\rm max}$
after $N=60$ e-folds of inflation with Hubble constant equal to the Higgs instability scale, $\Hub  = h_{\rm max}$,
and for $\xi_H = -0.01$.
The blue dashed line is the Gaussian approximation of eq.\meq{hGaussxi}.
The red curve is the Higgs potential $V_{\rm SM}(h) - 12\xi_H \Hub^2 h^2/2$, in arbitrary units.
\label{ran2}
}
\end{center}
\end{figure}

\begin{figure}[t]
\begin{center}
$$\includegraphics[width=0.8\textwidth]{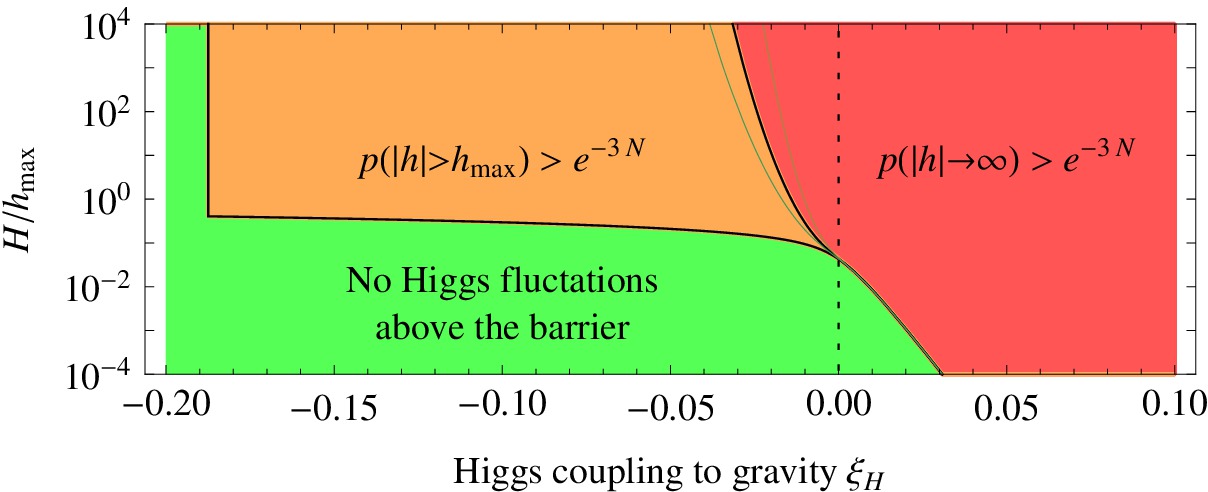}$$
\caption{\em As a function of $\xi_H$ and the Hubble constant in units of the instability scale $h_{\rm max}$ (and for $N=60$ $e$-folds of inflation), we show the three regions where: the probability for the Higgs field to end up in the negative-energy true minimum is larger than $e^{-3N}$ (red); the probability for the Higgs field to fluctuate beyond the potential barrier is larger than $e^{-3N}$ (orange); the latter probability is smaller than $e^{-3N}$ (green).  Higgs fluctuations are damped for $\xi_H < -3/16$.
The uncertainty on the orange/red boundary corresponds to a fudge factor $1/3<k<3$.
\label{RHbounds}
}
\end{center}
\end{figure}

\subsection{Higgs fluctuations during inflation for ${\xi_H\neq0}$}\label{xi}
%\xxx{HIGGS-INFLATON}
Higgs fluctuations during inflation can get damped if the Higgs doublet $\Phi_H$ during inflation
acquires an effective mass $m$.
Various effects can contribute to such mass: 
\begin{enumerate}
\item  a quartic term $\lambda_{h\phi} |\Phi_H|^2 \phi^2$ in the potential, which describes a coupling between the Higgs $\Phi_H$ and the inflaton $\phi$, generates during inflation an extra  contribution $m^2 = \lambda_{h\phi} \phi^2$ 
to the Higgs mass;

\item  a decay of the inflaton into SM particles can generate a non-vanishing temperature during inflation.
Such decay are kinematically blocked when SM particles acquire a thermal mass larger than the inflaton mass, of order $H$.
Thereby the Higgs can acquire a mass $m^2 \approx H^2$;

\item  a non-minimal Higgs coupling to gravity, $- \xi_H |\Phi_H|^2 R$ contributes as $m^2 = \xi_H R =-12 \xi_H H^2$.
\end{enumerate}
These contributions to $m^2$ would have qualitatively similar effects. 
In our quantitative analysis we focus on the latter effect because
%  $\phi$, $\lambda_{HS}|\Phi_H |^2 \phi^2/2$,
the presence of the $\xi_H$  term is unavoidable:
even if $\xi_H = 0$ at some energy scale, SM quantum corrections  generate a non-vanishing value 
of $\xi_H$ at any other energy scale.
Indeed, ignoring gravity, the one-loop running of $\xi_H$ is given by
%%
%\be
%\frac{\de \xi_H}{\de t} = \kappa \left( \xi_H + \frac{1}{6} \right) \left( 12\lambda + 6h_t^2 -\frac{3}{2}g'^2 -\frac{9}{2}g^2 \right).
%\label{eq:RGExi}
%\ee
%\xxx{AS: what is $\kappa$ and $h_t$?.
\beq
\frac{d\xi_H}{d\ln\bar\mu} =
\frac{\xi_H+1/6}{m^2}\frac{d m^2}{d\ln\bar\mu}=
\frac{\xi_H+1/6}{(4\pi)^2 }\left(6 y_t^2-\frac{9}{2} g_2^2 - \frac{9}{10} g_1^2+12\lambda_H\right) + \cdots
\eeq
where $\bar\mu$ is the renormalisation scale.
The RGE for the Higgs mass parameter $m^2$ is known up to 3 loops in the $\overline{\rm MS}$ scheme
(as summarised in~\cite{instab}), and we have shown here only the leading term. The SM couplings are such that $d\ln |m^2|/d\ln\bar\mu$ is positive (negative) at energy roughy below (above) $10^{10}\GeV$.
The evolution of $\xi_H$ for different boundary conditions at $M_{\rm Pl}$ is shown in fig.~\ref{fig:xiRGE};
it has a fixed point at the conformal value $\xi_H=-1/6$.
Notice that this value is not special for our analysis because it does not recover conformal invariance, which is broken at the level of the SM Higgs effective potential.

\bigskip

We consider the following action
\beq\label{eq:LadimSM}
S=  \int  d^4 x \sqrt{g} \bigg[
 -\frac{\bar M_{\rm Pl}^2}{2} R- \xi_H |\Hig |^2 R + |D_\mu \Hig |^2 - V+\cdots \bigg]
\eeq
where $ V \simeq V(\phi) + \lambda |\Hig |^4$ is the scalar potential of the Higgs and of the inflation $\phi$ and $\bar M_{\rm Pl}$ is the reduced Planck mass. We use the approximation that, during inflation, the inflaton potential is constant $V(\phi)\simeq V_I$.%
%In these notations $\xi_H=-1/6$ represents the  the conformal value and $\bar M_{\rm Pl} = M_{\rm Pl}/\sqrt{8\pi}$ is the reduced
%Planck mass.
\footnote{Of course, one could also envisage other operators coupling the Higgs field with gravity, {\it e.g.} 
$ |\Hig |^2 R^2/M_{\rm Pl}^2$. However, in most models of inflation, the Hubble parameter {squared}
decreases linearly with the number of $e$-folds $N_e$ till the end of inflation. Therefore, for $\xi_H\gta (H_I^2/M_{\rm Pl}^2)(N_e/N_I)$, where $H_I$ is the initial value of the Hubble rate when inflation starts and $N_I$ is the total number of $e$-folds, 
the higher-order operator becomes negligible. This condition becomes easier and easier to satisfy as inflation proceeds.
}

The $\xi_H$ coupling of the Higgs to gravity affects the scalar potential during inflation 
by inducing
an effective Higgs mass term $m^2 = \xi_H R =-12 \xi_H H^2\simeq - 4\xi_H V_I/\bar M_{\rm Pl}^2$ which can stabilise the Higgs potential
and suppress Higgs fluctuations. As explained after  eq.~(\ref{eq:Langevin}), Higgs fluctuations are damped if $\xi_H < -3/16$.
For $-3/16<\xi_H<0$, Higgs fluctuations are still present, but become less dangerous than in the case of vanishing $\xi_H$.

\medskip

Neglecting the small Higgs quartic coupling, adding the effective Higgs
mass term $m^2  = - 12\xi_H\Hub^2$, and assuming 
the ansatz of a Gaussian distribution with variance $\langle  h^2\rangle$,
\beq  P(h,N) = \frac{1}{\sqrt{2\pi \langle  h^2\rangle}}\exp \left( -\frac{h^2}{2\langle  h^2\rangle}\right),\eeq
the  evolution of $\langle  h^2\rangle$ is obtained from the Fokker-Planck equation~(\ref{Fokker}) and becomes, at $h^2\gg \langle h^2\rangle$,
\beq \label{eq:evoN}
\frac{\partial \langle  h^2\rangle}{\partial N}=-\frac{2m^2}{3H^2}  \langle  h^2\rangle+
\frac{H^2}{4\pi^2}\qquad\Rightarrow\qquad
\sqrt{ \langle h^2\rangle} =\sqrt{\frac32}\frac{\Hub^2}{2\pi m} \sqrt{1-\exp\left(-\frac{2 m^2 N}{3H^2}\right)}.\eeq
If $m^2<0$, the variance grows exponentially with $N$.
If $m^2>0$,
%Neglecting the small Higgs quartic coupling and adding the effective Higgs
%mass term $m^2  = - 12\xi_H\Hub^2$,
the Higgs probability distribution approaches, after a few $e$-folds,
the limiting distribution given by\footnote{This is larger than what is obtained by naively assuming a Hawking temperature $T=H/2\pi$.} 
\beq \label{eq:hGaussxi}
\sqrt{ \langle h^2\rangle} =  \frac{\Hub}{4\pi\sqrt{-2\xi_H}} \, .
\label{eq:FluctInflxineq0}\eeq
This is to be compared with \eq{eq:hGauss}, which holds for $\xi_H=0$.
Figure~\ref{ran2} shows a numerical example: already for $\xi_H=-0.01$
the variance is significantly reduced.

Using \eq{eq:hGaussxi}, we obtain the following bounds on $H$ from the request that the probabilities of the Higgs fluctuating beyond the barrier ($|h|>h_{\rm max}$) or falling into the true minimum
($|h|\to\infty $) are less than $e^{-3N}$:
\begin{eqnarray}\label{eq:h<hmax-xiH}
p(|h|>h_{\rm max})<e^{-3N}  & \Rightarrow & \frac{\Hub}{h_{\rm max}} < 4\pi \sqrt{\frac{-\xi_H}{3N}}, \\
p(|h|\to\infty )<e^{-3N}  & \Rightarrow & \frac{\Hub}{h_{\rm max}} < 4\pi \sqrt{\frac{-\xi_H}{3N}} e^{32\pi^2\xi_H^2/bN}.
\label{bis}
\end{eqnarray}
These bounds are the analogues of eqs.~(\ref{box1}) and (\ref{box2}), which are valid for $\xi_H =0$.
An order-one fudge factor $k$ can be similarly introduced such that eq.~(\ref{bis}) closely agrees with the numerical result; 
however $k$ depends on $\xi_H$ and thereby differs from what we previously discussed in the limit $\xi_H =0$.
Taking into account how $\xi_H$ stabilises the potential results in a more complicated, but numerically similar, analytic expression.
Furthermore these approximations needs to be extrapolated down to probabilities smaller than those that can be compared to the numerical result.
Conservatively estimating the uncertainty by varying $1/3<k<3$,
in fig.~\ref{RHbounds} we summarise the situation by showing the regions of $\xi_H$ and $H/h_{\rm max}$ where the bounds in eq.s~(\ref{eq:h<hmax-xiH}) and (\ref{bis}) are satisfied for $N=60$.

\bigskip

%\xxx{HIGGS-INFLATON}
In the presence of a $\lambda_{h\phi} |\Phi_H|^2 \phi^2$ potential coupling between the Higgs $\Phi_H$ and the inflaton $\phi$,
during inflation one has an extra  contribution
to the Higgs mass, $m^2 = \lambda_{h\phi} \phi^2$.
This term has a similar effect as the inflationary mass discussed above. 
The Higgs $h$ has no inflationary fluctuations  as long as $m>  3 H/2$.
However, in general $m^2$ changes during inflation in a model-dependent way.
Considering, for example, large-field inflation with a quadratic potential, one has $\phi = 2\bar M_{\rm Pl} \sqrt{N_I-N}$
 during inflation, where $N_I \circa{>} 60$ is the total number of $e$-folds.
Inserting $m^2 = 4\lambda_{h\phi} \bar M_{\rm Pl}^2 (N_I - N)$ into eq.~(\ref{eq:evoN}) one finds
that the maximal Higgs fluctuation is achieved at the end of inflation and is Planck suppressed:
\beq \langle h^2 \rangle  =  \sqrt{\frac{3}{\pi \lambda_{h\phi}}} \frac{H^3}{16\pi \bar M_{\rm Pl}  }.  \label{hhinflm}\eeq

% for $m^2\propto N^2$ one has again $ \langle h^2 \rangle\sim H^4/m^2$.

\subsection{Bubble evolution in de Sitter spacetime}\label{bubble}
After having computed the probability for inflationary fluctuations to form regions where the Higgs field lies at its  true minimum, the next question we have to address is how these regions evolve. In the literature one finds conflicting statements about the evolution of AdS regions in an inflationary background. One point of view, based on
flat-space intuition, is that AdS regions should expand because their interior has  lower energy than the exterior.
A different point of view is that, since AdS space eventually contracts, regions in which the Higgs lies at its true minimum will shrink, possibly leaving some almost point-like relics, which are nevertheless efficiently diluted, and thus made harmless, by the inflationary expansion of space. We will show that addressing the question about the fate of AdS regions involves a number of non-trivial and counter-intuitive issues raised by general relativity.

First, gravitational energy contributes to the total energy budget.
Second, an AdS region might expand, while remaining hidden behind a black-hole horizon.
Third, the interior AdS space is dynamically unstable~\cite{deluccia}:
when described in cosmological FRW coordinates 
it reaches a `big-crunch' singularity in a finite amount of internal time
of order $(G V_{\rm in})^{-1/2}$, where $-V_{\rm in}<0$ is the internal cosmological constant.
If space is empty, this is just a coordinate singularity (AdS can be continued using better coordinates);
if space is filled by a background field (for example the Higgs field), its fluctuations grow until
the energy density becomes infinite and a physical singularity appears.
Furthermore, the AdS geometry has a timelike boundary, such that the evolution cannot 
be predicted after the bubble wall reaches the boundary, unless additional
boundary conditions are imposed there (in other words, 
information must flow in from infinity). 
As a result, a Cauchy horizon appears in the interior of the
bubble. It is expected that, within the full theory beyond the thin-wall limit, 
a physical spacelike singularity must develop before the Cauchy horizon \cite{freivogel}.  This confirms the expectation that the 
AdS bubble is unstable.

\medskip

In order to clarify all these issues we performed a careful (and somewhat lengthy) general-relativistic computation, described in appendix~\ref{AdS}.
Here we summarise the main points.

\medskip

\begin{figure}[t]
$$\hspace{-1cm}\includegraphics[width=70mm,height=60mm]{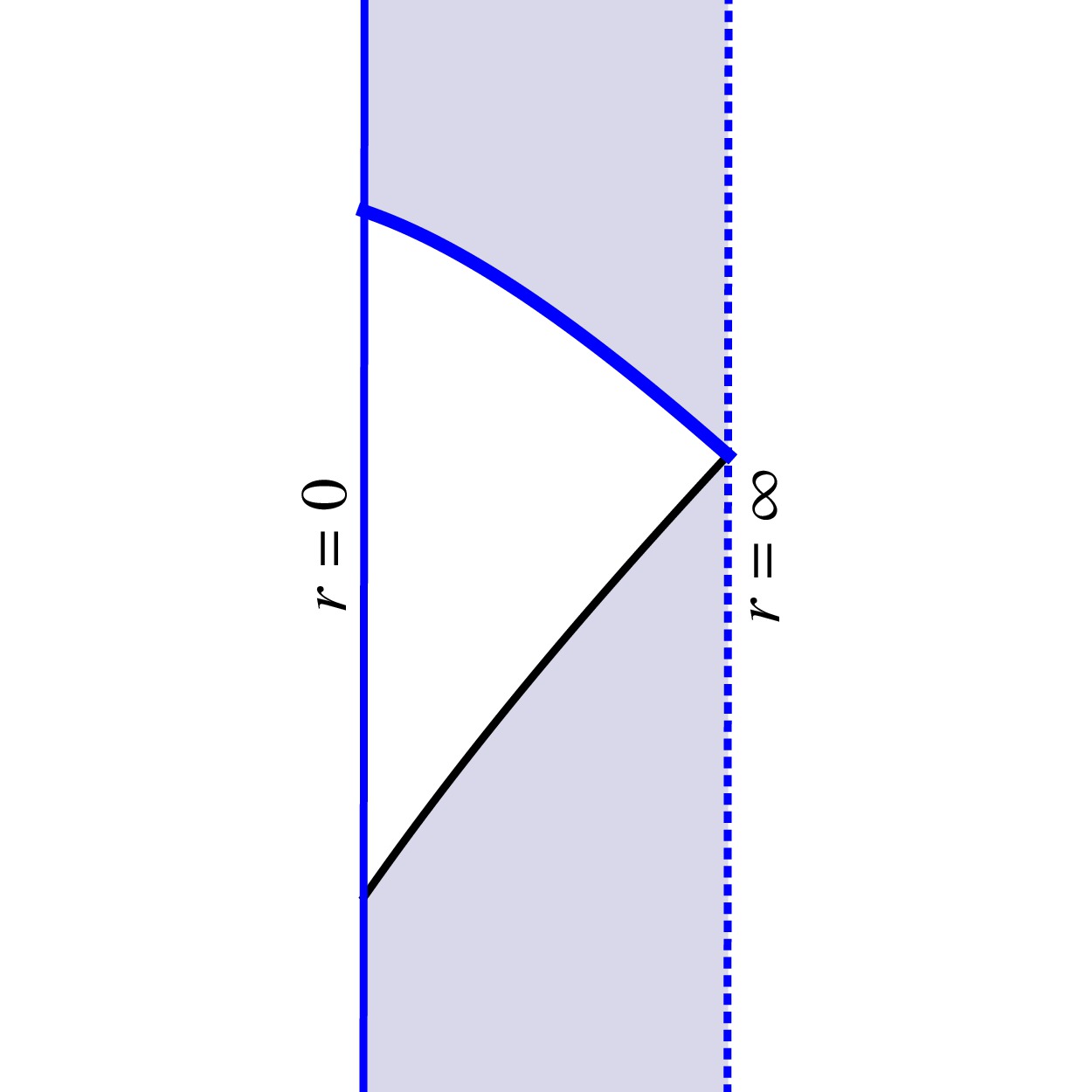}
\includegraphics[width=100mm,height=60mm]{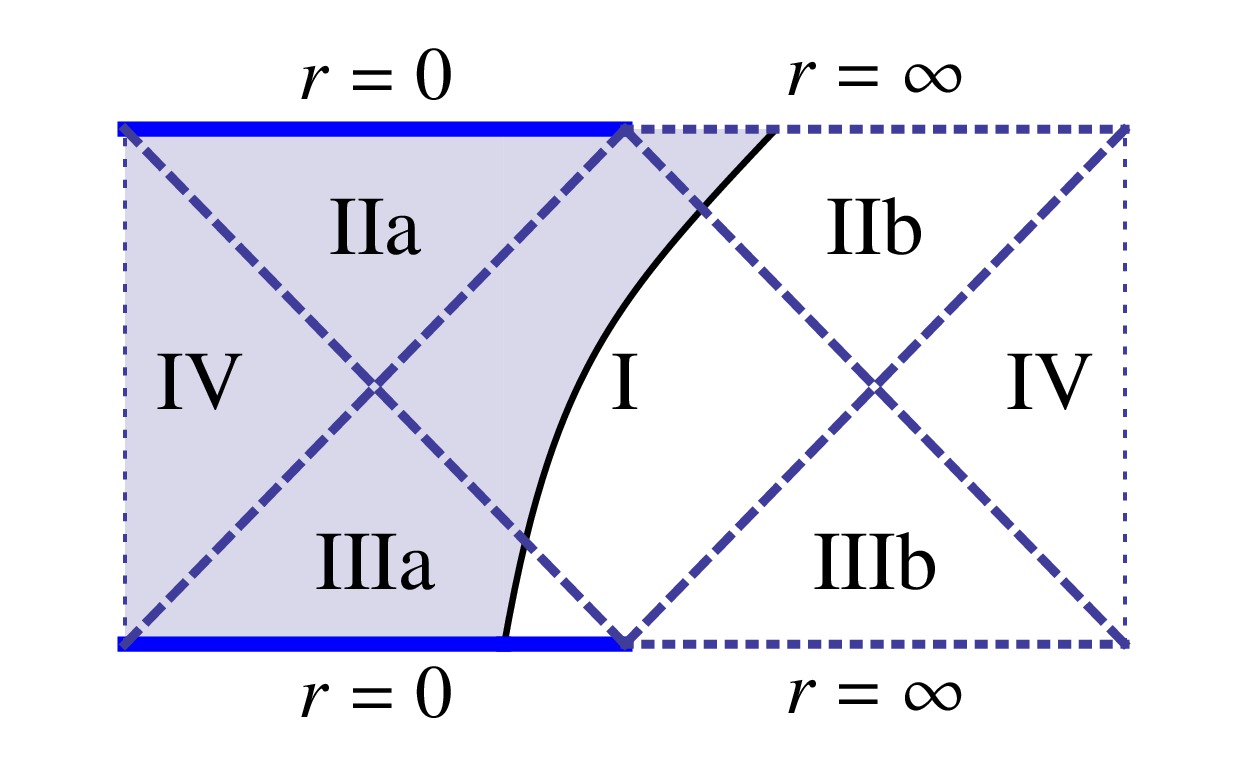} $$
\caption{{\bf}  \em {\bf Penrose diagram describing an AdS bubble that expands in a dS space}.
The  black curve denotes the thin wall separating the two phases; the true space is obtained by patching the 
region in white in the left panel (AdS interior of the bubble) with the region in white in the right panel (Schwarzschild-dS
exterior of the bubble).
%; the shading covers the full unphysical AdS and dS spaces used in the computation.
The dashed lines denote the various horizons, while the thick blue line in the 
left panel denotes the AdS singularity (`crunch').}
\label{W1}
\end{figure}

In order to make the problem tractable analytically, we assume a spherical AdS region (that we thereby call `bubble'),
separated from the outside space by a thin wall with constant surface tension $\sigma$.
The matching of the external and internal geometries requires the presence of such a wall with nonzero energy density.
The fate of the AdS bubble is then determined by computing the motion of the wall separating the AdS interior from the external
space (de Sitter during inflation and Minkowski after inflation).
In the thin-wall approximation,  the motion of the wall is determined by junction conditions that relate the extrinsic curvature
on each of its sides~\cite{israel}. 
The bubbles that we now compute are more general than those
that arise from vacuum decay with zero total energy, already studied in~\cite{deluccia}.
The basic elements of our calculation are the following.
\begin{enumerate}
\item  The space inside the bubble is assumed to be an empty  spherical region of AdS space with metric 
\be
 ds^2=-f_{\rm in}(r)\,  d\eta^2+ \frac{ dr^2}{f_{\rm in}(r)}+r^2  d\Omega^2_2,\qquad r< R,
\label{ads-metric} \ee
expressed in global coordinates.
Here $f_{\rm in}(r)=1+r^2/\ell_{\rm in}^2$ is the usual AdS solution, with vacuum energy $-V_{\rm in}$
corresponding to the length scale $1/\ell^2_{\rm in}=8\pi G V_{\rm in}/3$. 

\item The space outside the bubble is described by the  metric 
\be
 ds^2=-f_{\rm out}(r)\,  dt^2+\frac{ dr^2}{f_{\rm out}(r)}+r^2  d\Omega^2_2,\qquad r >R,
\label{schw-metric} \ee
where $f_{\rm out}(r)=1-r^2/\ell_{\rm out}^2-{2GM}/{r}$ describes 
a Schwarzschild-de Sitter (SdS)
spacetime, with $G=1/(8\pi\bar M_{\rm Pl}^2)$.
Here $M$ is the mass of the bubble as seen by an outside observer, living in an
asymptotically de Sitter space described
by the length scale $1/\ell^2_{\rm out}=8\pi G V_{\rm out}/3 = H^2$.
As discussed later, the metric in \eq{schw-metric} also describes the case of the
asymptotically flat spacetime produced after inflation, which is obtained 
 in the limit $V_{\rm out}\to 0$, 
so that $f_{\rm out}(r)=1-{2GM}/{r}$.

Note that, for $\ell_{\rm out} \gg GM$, the SdS spacetime contains  two horizons, corresponding to the zeros of $f_{\rm out}(r)$:
the inner (Schwarzschild) horizon at $r\approx 2GM$ and the outer (de Sitter) horizon at $r\approx \ell_{\rm out}$.
The corresponding Penrose diagram is depicted in the right panel of 
fig.~\ref{W1}. It is a combination of the 
diagrams for the Schwarzschild and de Sitter spacetimes \cite{gibbons}. 
Thick blue lines denote curvature singularities, 
the dashed lines horizons and the dotted lines conformal infinities.
The two thin vertical lines at the ends of the
diagram indicate that the pattern is repeated indefinitely on either side.

%The choice $f_{\rm out}(r)=1-{2GM}/{r}$ gives the usual Schwarzschild metric
%and $M$ is the mass attribute to the bubble by an outside observer.
%The choice $f_{\rm out}(r)=1-r^2/\ell_{\rm out}^2-{2GM}/{r}$ corresponds to a dS-Schwarzschild
%spacetime. We concentrate on the first choice in the following. However, 
%we shall comment on the bubble evolution in an asymptotically de Sitter 
%spacetime in the last subsection.

\item
The two regions are separated by a domain wall with constant surface tension $\sigma$.
The metric on the domain wall can be written as
\be
 ds^2=- d\tau^2+R^2(\tau)  d\Omega^2_2,
\label{wall-metric} \ee
where $R(\tau)$ denotes the location of the wall in both coordinate systems (\ref{ads-metric}) and (\ref{schw-metric}).
The evolution is expressed in terms of the proper time $\tau$ on the wall. 
In the full problem, $\sigma$ is given by the
kinetic and potential energy of the Higgs field, 
and is different for each Higgs configuration.
Within our approximation, all energy stored in the Higgs potential goes into 
the motion of the wall, leaving the AdS interior empty.
\end{enumerate}
The detailed calculation described in appendix~\ref{AdS} shows that 
the (naively positive) difference between the energy in the exterior ($\ell_{\rm out}^{-2}$) and the interior ($-\ell_{\rm in}^{-2}$) of the bubble  that controls whether the bubble expands or contracts receives 
a gravitational correction $-\kappa^2$, so that the relevant parameter is the quantity:
\beq \Delta  = \frac{1}{\ell^2_{\rm in}}+\frac{1}{\ell_{\rm out}^2}-\kappa^2,\qquad  \kappa\equiv 4\pi G \sigma. \eeq
As discussed after eq.~(\ref{square}), the contribution $\sim \kappa^2$ 
can be interpreted,
from a  Newtonian point of view, as the gravitational self-energy of the wall.
The motion of the wall can be described as
the Newtonian motion of a point particle
\be
\left(\frac{ d\tilde R}{ d\ttau} \right)^2+V(\tilde R)=E,
\label{eomm} \ee
in an effective `potential' given by
\be
V(\tilde{R})=-\left( \frac{1+\ex \tilde{R}^3}{\tilde{R}^2}\right)^2-\frac{\gamma^2}{\tilde{R}}-\delta^2 \tilde{R}^2,
\label{potds}\ee
where $\tilde R=\rho R$ is a rescaled dimensionless coordinate that describes the position of the wall
as a function of a rescaled dimensionless proper time $\ttau={2\kappa}\tau/\gamma^2$.
The various constants are given by
\beq
\delta^2=\frac{4\kappa^2}{\ell_{\rm out}^2 \Delta^2},
\qquad
\rho^3 = \frac{|\Delta|}{2GM},
\qquad
\ex \equiv  {\rm sign} \Delta,\qquad
\gamma=\frac{2\kappa}{\sqrt{\left| \Delta\right|}},\qquad
E=-\frac{\kappa^2}{G^2M^2\rho^4}.
\label{rrhods} \ee

\bigskip

The possible types of bubble evolution are discussed in detail in appendix~\ref{AdS}.
The complete analysis is performed for an asymptotically flat exterior
spacetime, for which there are fewer cases. The evolution within an asymptotically
dS spacetime does not display any novel characteristics, and is discussed
more briefly. 
The study of the `potential' shows that there are two cases in which bubbles do not expand:
either they start small enough, or their expansion is hidden behind a black-hole horizon
(this possibility corresponds to $\Delta <0$).
We will discuss in section~\ref{bubblem}
if any of these possibilities is realised in the simpler case of Higgs bubbles in an external flat spacetime, after the end of inflation.

The standard evolution of sufficiently large bubbles is characterised by expansion, with 
their wall crossing
the outer (dS) horizon. A typical example is presented in the
Penrose diagram of fig.~\ref{W1}: the wall starts below the inner
horizon and subsequently expands, passing through both horizons and eventually
reaching a 
speed close to that of light. The bubble grows in size and takes over part of
the dS spacetime. 
The crucial question is whether the bubble can engulf the total exterior spacetime, 
thus ending inflation.   
It is apparent from fig.~\ref{W1} that this does not happen. Asymptotically the
AdS bubble replaces only part of the spacelike surface $r=\infty$, with the remaining
dS space remaining unaffected. 
In other words, 
expanding bubbles are inflated away.
Inflation precisely has the purpose of splitting the expanding universe into causally disjoint regions,
and this limits the effect of the 
bubble growth: bubbles expand, but the dS space between them also grows.
In the limit of infinite inflation, both the bubbles and the exterior 
de Sitter phase acquire infinite extent.
 
We can estimate the asymptotic bubble size through the use of 
dS planar coordinates, commonly employed in the study of inflation. 
The metric has the form 
\be
 ds^2=- dt_p^2+e^{2Ht_p}\left( dr_p^2+r^2_p  d\Omega^2_2 \right).
\label{planar} \ee
Assuming that the wall follows an almost null trajectory, we find that its
location is given by
\be
r_p=r_{p0}+\frac{1}{H}\left(1-e^{-Ht_p} \right).
\label{locds} \ee
Bubbles are created within the causally connected region, which extends 
up to $1/H$ at $t_{p}=0$. This means that its typical radius $r_{p0}$ 
is of order $1/H$. Its subsequent growth extends this radius by $1/H$.\footnote{Note, however, that the physical bubble radius is obtained after multiplication
by the divergent factor $\exp(Ht_p)$. }
It is reasonable then to expect that during inflation a typical bubble can be created
with a certain probability within a causally connected region and then will roughly 
follow the general expansion of this region outside the horizon. It cannot,
however, engulf the whole spacetime. The picture is completely different for 
an asymptotically flat exterior. As we shall see in the next section, in
that case an expanding bubble can take over the whole spacetime.

Another important question concerns 
the consequences for an outside observer of the AdS `crunch' in the bubble interior.
We discuss this issue in detail in appendix~\ref{AdS5}.
From the point of view of an observer deep inside the bubble, 
the coordinates in which the bubble appears as homogenous are those 
of an expanding and subsequently contracting open FRW universe with constant
negative energy density.
The bubble wall can be roughly identified with the $\hat{t}=0$ surface in this
slicing (see fig.~\ref{wallcrunch} in the appendix).  
After a finite (and short) time $\hat{t}$, of order the AdS radius, 
a singularity forms in the bubble interior. However, this singularity never 
reaches the wall, as the latter expands with the speed of light. 
On the other hand, from the point of view of an external observer the bubble just expands forever
(within either de Sitter or Minkowski spacetime).

\medskip

\begin{figure}[t]
\begin{center}
$$\includegraphics[width=0.45\textwidth]{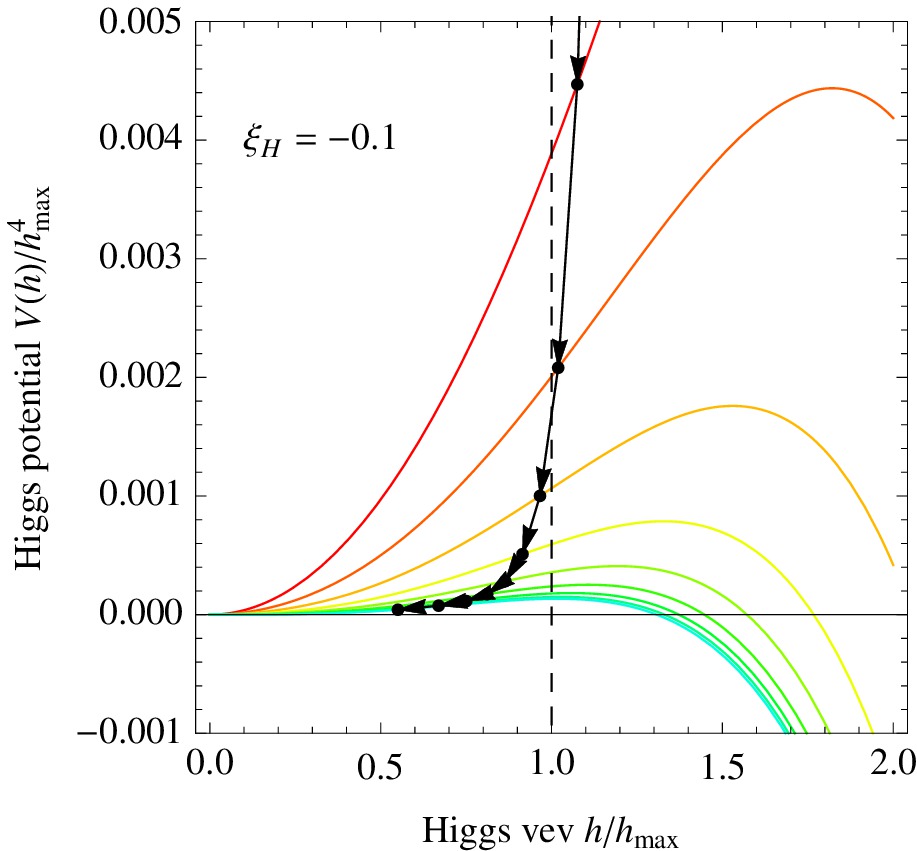}\qquad
\includegraphics[width=0.43\textwidth]{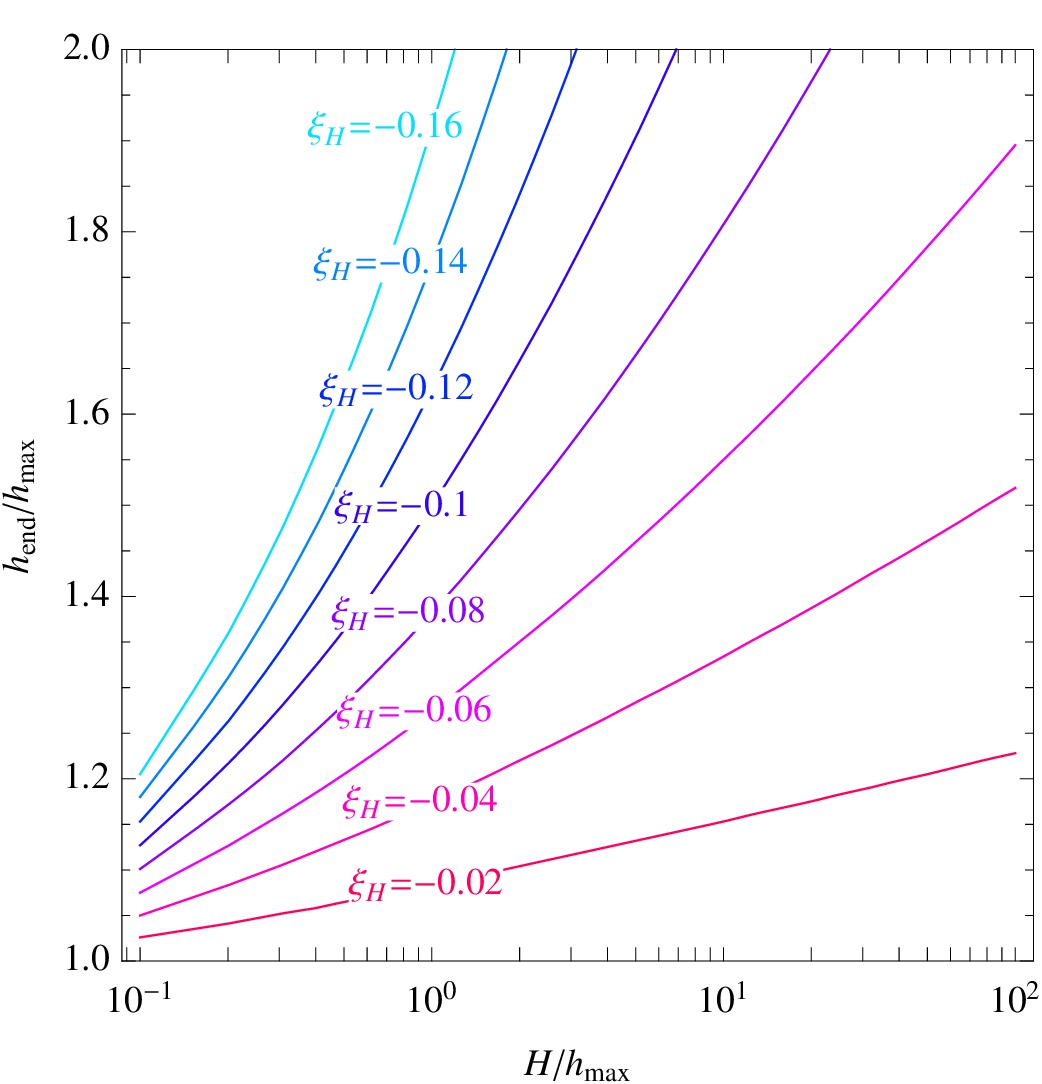}
$$
\caption{\em {\bf Left}: An example of how the dynamical evolution during pre-heating can
bring back the Higgs field into the stable region before the instability takes over. The coloured lines show the potential at successive intervals of time, and the black line shows the trajectory of the Higgs field.
{\bf Right}: Maximal value of the Higgs field at the end of inflation ($h_{\rm end}$)
that is brought back by a non-minimal gravitational coupling $\xi_H$ into the stable region, $h<h_{\rm max}$, shown as a function of
the Hubble constant during inflation and for different values of $\xi_H$.  
\label{Vpostinfl}}
\end{center}
\end{figure}

\section{Higgs evolution after inflation}\label{sec4}
In this section we study the evolution of the Higgs field after inflation, considering that inflation ends with
a matter-dominated phase, characterised by inflaton oscillations, followed by the reheating process, which ignites the usual thermal phase characterised by a gas of SM particles.

\subsection{Higgs  evolution during pre-heating}
We start by considering the pre-heating phase, during which the energy density of the universe is dominated by the inflaton oscillations around its minimum. The interest of this phase lies in the case in which the Higgs potential has an extra mass term $\frac12 m^2 h^2$ induced either by a non-minimal coupling to gravity ($m^2 =-12\xi_H \Hub ^2$) or by a coupling to the inflaton $\phi$
($m^2 =\lambda_{h\phi} \phi^2$). In either case, the mass term rapidly shuts off after inflation. However, we will show that the induced $m^2$ can still have an important stabilising effect during the pre-heating phase.
 
Let us suppose that, at the end of inflation, the universe enters a matter-dominated phase, where the equation of state is that of a pressure-less gas, which is a good approximation when the inflaton field is oscillating before reheating. 
In this case the effective mass of the Higgs field is $m^2 =-3\xi_H \Hub _{\rm m}^2$ and the Hubble rate scales
as $ \Hub _{\rm m}=\Hub/a^{3/2}$
where we have set the value of the scale factor $a$ at the end of inflation to unity.\footnote{We recall that the Ricci scalar is given by $R=-6(\ddot{a}/a+\dot{a}^2/a^2)$ for a spatially flat universe. 
During inflation $a=\exp(Ht)$ with $H$ constant, hence $R=-12H^2$. In a matter-dominated phase ($a\propto t^{2/3}$) we have $R=-3H_{\rm m}^2$ with $H_m=\dot{a}/a\propto a^{-3/2}$, while $R=0$
in the radiation-dominated phase ($a\propto t^{1/2}$).}

We consider a region in which, once inflation ends, the Higgs field has the value $h_{\rm end}$. If the Higgs mass term during inflation ($m^2 =-12\xi_H \Hub ^2$) is larger than $(9/4) \Hub ^2$ ({\it i.e.} if $\xi_H<-3/16$), then the Higgs field is anchored
at the origin and  $h_{\rm end}=0$. We are interested here in the opposite regime, when $-3/16<\xi_H<0$ and quantum fluctuations of the Higgs are generated during inflation; in this case, $h_{\rm end}$ is generally not zero. The subsequent evolution of the Higgs field is governed by the equation
\beq
\label{eqafter}
\frac{d^2 h}{d t^2}+3\Hub_{\rm m}(t) \frac{d h}{d t}+% \frac{4}{3t^2}\xi_H h +
\frac{\partial V}{\partial h}=0
\qquad \Rightarrow \qquad
\frac{d^2 h}{d a^2}+
\frac{5}{2a} \frac{d h}{d a}+ \frac{a}{\Hub^2} \frac{\partial V}{\partial h}=0.
\eeq
Keeping only the mass term in $V$ and neglecting the quartic term, we find $(a/H^2)\partial V /\partial h = -3\xi_H h/a^2$ and then the solution of eq.~(\ref{eqafter}) is
\beq
\label{exact}
h(a) = h_{\rm end}\   a^{-\frac{3}{4}\left(1-\sqrt{1+\frac{16}{3}\xi_H}\right)}\quad
\stackrel{|\xi_H|\to 0}{\approx} \quad h_{\rm end}\ a^{2 \xi_H},
\eeq
where we have neglected the solution with $h\propto a^{-3/2}$, which is rapidly damped with respect to \eq{exact}, and where the last approximation is valid only for  $|\xi_H| \ll 3/16$. 

As the  amplitude of the Higgs field and the contribution to the potential from the  
$m^2 $ term are both decreasing in time, we need to investigate if the Higgs field has time enough to roll down to the safe region $h<h_{\rm max}$ before the instability starts to become more important than the fading $m^2$ term, reverting the evolution of $h$.
An example of the Higgs field behaviour is shown in fig.~\ref{Vpostinfl}a. The time at which the instability starts driving the Higgs dynamics can be estimated by requiring that the quartic term in the scalar potential ($\lambda h^4/4$ with $\lambda=-b \ln h^2/h_{\rm max}^2\sqrt{e}$) is comparable with the mass term ($m^2 h^2/2$ with $m^2 = -3\xi_H H^2/a^3$) at $h_{\rm max}$, which corresponds to
\beq a^3 = a^3_{\rm max} \approx-\frac{12\xi_H\Hub^2}{b h_{\rm max}^2}.\eeq
Thereby, the instability is avoided if
\beq h_{\rm end} \circa{<} h_{\rm max} a_{\rm max}^{-2\xi_H}.\eeq
This estimate is confirmed by the result of a numerical computation illustrated in fig.~\ref{Vpostinfl}b,
in which \eq{eqafter} is solved using the full SM potential.

The qualitative conclusion, whenever $\xi_H\neq 0$, is the following. Higgs field values that, at the end of inflation, are even a factor of ${\cal O}(2)$
above the instability scale $h_{\rm max}$ are brought back into the metastability region ($h<h_{\rm max}$) and saved from collapse into the AdS vacuum by the non-minimal gravitational coupling $\xi_H$ during pre-heating dynamics. The bound on the Hubble scale during inflation in eqs.~(\ref{eq:h<hmax-xiH}) and (\ref{bis}) are correspondingly 
weakened by an ${\cal O}(2)$ factor. 
%The Higgs field reaches the value $h_{\rm max}$ at 
%\beq
%a = a_{\rm max}\sim \left(\frac{h_{\rm max}}{h_{\rm end}}\right)^{\frac{3}{4\xi_H}},
%\eeq
%We then impose the condition
%\be
%-3\xi_H \Hub _{\rm m}^2(a_{\rm max})   h_{\rm max}^2=-3\xi_H \frac{\Hub ^2}{a_{\rm max}^3}h_{\rm max}^2
%>\lambda(h_{\rm max})h_{\rm max}^4.
%\eeq
%As a typical value for $h_{\rm end}$ we can take $(\Hub/4\pi\sqrt{2|\xi_H|})$. The result is shown in Fig. (), where we have retained the exact scaling in Eq. (\ref{exact}).
On the other hand, in regions where the Higgs remains in the instability region $h(a_{\rm max})\circa{>}h_{\rm max}$ during pre-heating,
the field quickly falls down into its deep minimum, in a time $t\sim 4\pi /h_{\rm max}$,
unless a large enough temperature prevents the collapse, as we discuss in the following section.

\bigskip

%\xxx{HIGGS-INFLATON}
The dynamics  discussed above is similar to the one in which the Higgs field is coupled to the inflaton field by a coupling of the form $\lambda_{\phi h}\phi^2h^2/2$  that generates a contribution  $m^2 = \lambda_{\phi h}\phi^2$ to the Higgs mass.
As discussed in section~\ref{xisec}, if during inflation $m>3H/2$ Higgs inflationary perturbations are suppressed and the Higgs is efficiently anchored at $h=0$.
If instead $m<3H/2$,  Higgs fluctuations are generated as in eq.~(\ref{hhinflm}) and they may pose a threat. 
After the end of inflation, when the inflaton field 
oscillates and its amplitude is redshifted away as $\phi\sim a^{-3/2}$, the effective mass squared of the Higgs,
$m^2 = \lambda_{h\phi}\phi^2$, 
decreases as $a^{-3}$, exactly as the $m^2$ induced by the $\xi_H$ coupling ($m^2=-3\xi_H H_{\rm m}^2 $). One can therefore deduce  the dynamics upon identifying the two $m^2$.
% $\lambda_{\phi h}\phi_{I}^2$ with $-12\xi_H H^2$, where $\phi_I$ is the initial amplitude of the inflation oscillations. 
Of course, if the coupling of the inflaton field with the Higgs field is of a different nature, e.g. a non-renormalisable coupling of the form $\phi^4 h^2/M_{\rm Pl}^2$, one needs to account for the different behaviour of the Higgs effective mass.

\subsection{Higgs evolution during reheating}\label{postI}
%Below the result is that the temperature suddenly appears as long as $a\sim{\cal O}(1)$ and
%a small $\Gamma_\phi$ just means a small $T/\Hub \sim [ M_{\rm Pl}^{2} \Gamma_\phi/\Hub^3]^{1/4}$.

In this section we study how the reheating process affects the bounds on the Hubble constant $H$ during inflation. Indeed, the dynamical evolution during the thermal phase can bring back the Higgs field towards the EW vacuum, even in regions where $h$ has fluctuated beyond the instability barrier ($h\circa{>} h_{\rm max}$) at the end of inflation. As a result, the bounds on $H$ derived in section~\ref{HI} are effectively relaxed.
%We now show that the subsequent thermal phase 
%`burns' the regions where the Higgs field fluctuates around its instability scale $h\circa{>} h_{\rm max}$,
%bringing $h$ back to its origin such that these regions are not dangerous.

At the end of inflation, the energy
density of the universe is dominated by the coherent
oscillations of the
inflaton field $\phi$ with energy density $\rho_\phi(t)$.
The oscillations of $\phi$,
started at time $t\sim1/\Hub$,
give a matter-dominated stage that gradually ends at $t\sim1/\Gamma_\phi$, where $\Gamma_\phi$ is the inflaton decay width. The decay  of the inflaton field  into light degrees of
freedom, which quickly thermalise via SM interactions giving rise to an energy density $\rho_R(t)$, initiates the radiation-dominated era of the universe.
The process is described by the equations
\beq\left\{\begin{array}{rcl}\displaystyle
\frac{d \rho_{\phi}}{dt}&=&-3 H_r
\rho_{\phi}-\Gamma_{\phi}\rho_{\phi} \, ,
\label{eq:rho_phi}\\[3mm]  \displaystyle
\frac{d \rho_{R}}{dt}&=& -4 H_r \rho_R + \Gamma_{\phi}
\rho_{\phi} 
 \, , \label{eq:rho_R}
 \end{array}\right.\eeq
where $H_r = \dot a/a = \sqrt{8\pi (\rho_\phi  +
\rho_R)/(3M_{\rm Pl}^2) }$ is the time-dependent
Hubble constant during reheating.
%\xxx{AS:  if I solve the equation for $\rho_R$ during inflation I find that inflaton decays generate $\rho_R = 3 H_I \Gamma_\phi M_{\rm Pl}^2/32\pi$ which means $T\sim T_{\rm max}$ during inflation. Is there some reason why $\Gamma_\phi=0$ during inflation?}

\begin{figure}[t]
\begin{center}
$$\includegraphics[width=0.45\textwidth]{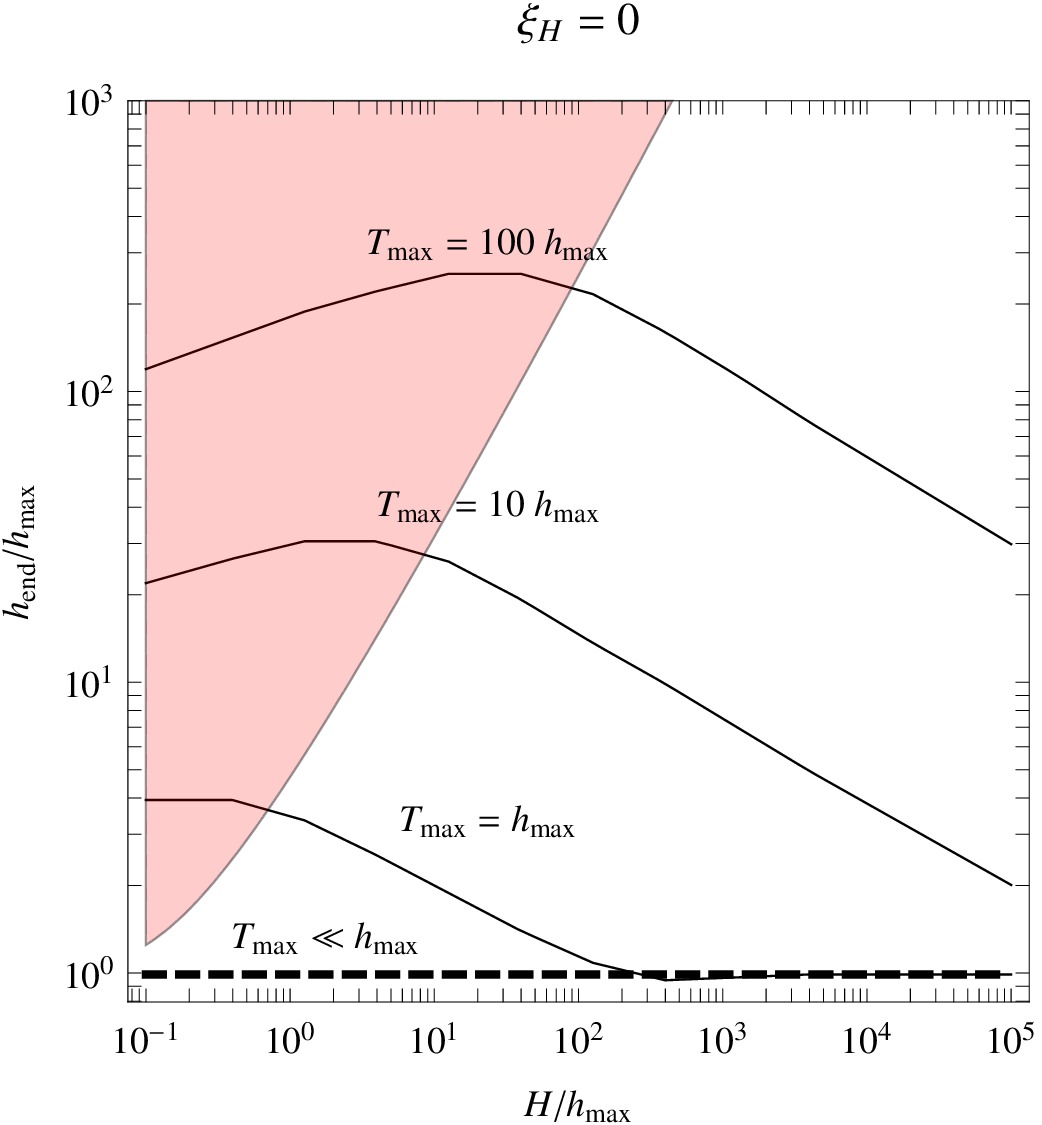}\quad
\includegraphics[width=0.45\textwidth]{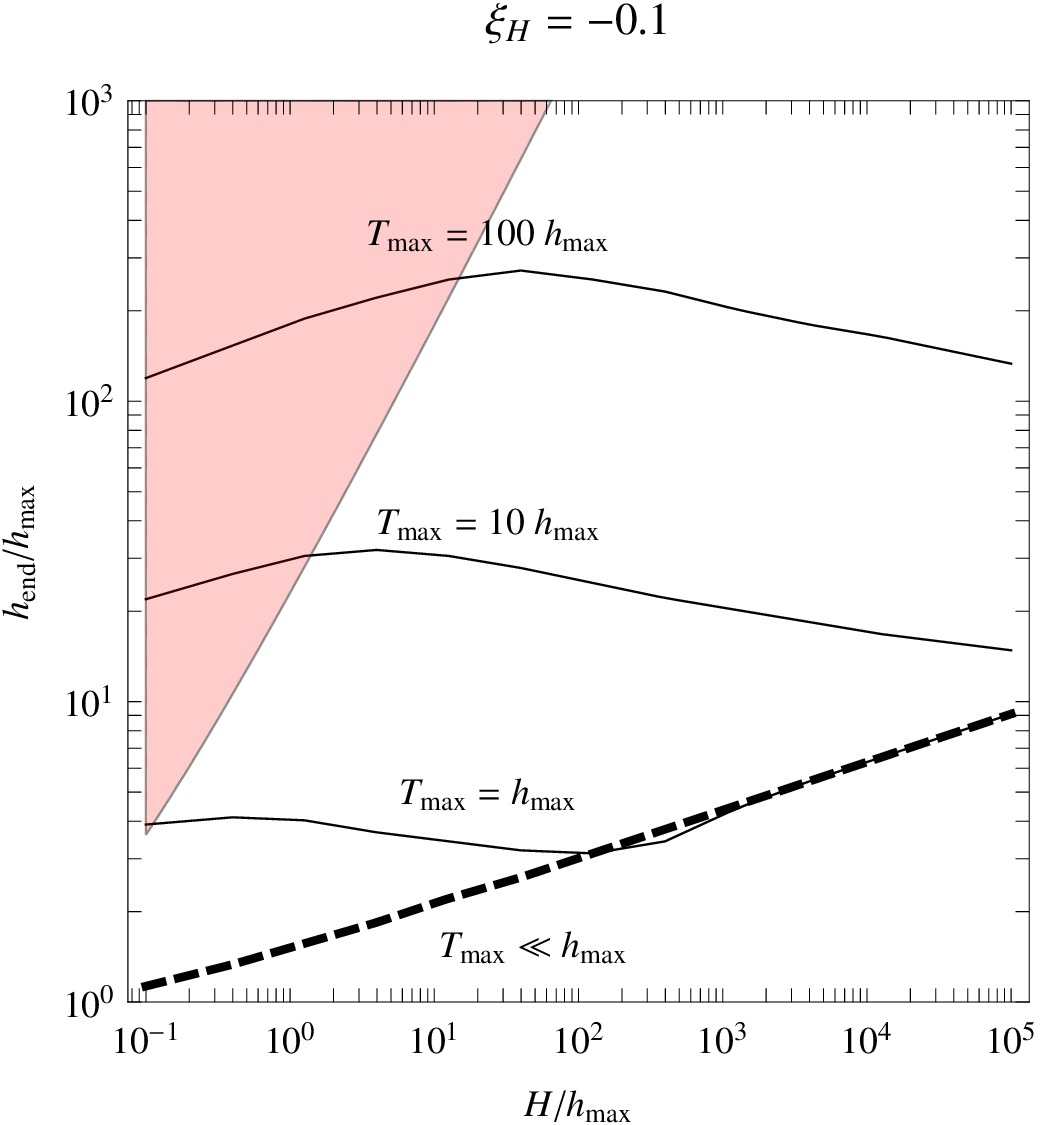}
$$
\caption{\em Maximal value of the Higgs field at the end of inflation $h_{\rm end}$
that is brought back to the stable region, $h<h_{\rm max}$, as a function of
the Hubble constant during inflation, for different values of $T_{\rm max}$ and $\xi_H$.
This result, presented only in terms of ratios, negligibly depends on the absolute value of the instability scale $h_{\rm max}$, which suffers from large uncertainties mainly due to the top quark mass.
In the red region the inflationary fluctuations typically drive the Higgs to its negative-energy minimum, 
and therefore the corresponding values of $h_{\rm end}$ and $H$ are a highly unlikely outcome of inflation.
In the parameter region below the dashed line, the field $h$ rolls towards the SM vacuum, even in the absence of any thermal effect.
\label{hendvsTRH}}
\end{center}
\end{figure}

The solution for the time evolution of $\rho_\phi$ is
\beq 
\rho_\phi(t) = \frac{\rho_{\phi}(0)}{a^3(t)}
e^{-\Gamma_\phi t}, \quad \rho_\phi(0)= \frac{3H^2M_{\rm Pl}^2}{8\pi} 
\label{eq:rhoR}
\eeq
where the initial condition $\rho_\phi(0)$ is given by the total energy density at the end of inflation.
The second equation in the system (\ref{eq:rho_R}) can be more conveniently written as
\beq
\frac{dR}{da} =\frac{\gamma a^{3/2} \Phi}{\sqrt{\Phi + R/a}}
\ , \quad \quad \quad
R\equiv \rho_R a^4 \ , \quad
\Phi \equiv \rho_\phi a^3 \ , \quad
\gamma \equiv \sqrt{\frac{\pi^2 g_*}{30}} T_{\rm RH}^2 \ ,
\label{evR}
\eeq
where $g_*$ is the number of degrees of freedom in the thermal bath ($g_*=106.75$ in the SM) and $T_{\rm RH}$ is the temperature of the system once all the inflaton energy is converted into thermal energy at the decay time,
\beq
T_{\rm RH} =  \left( \frac{45}{4\pi^3 g_*}\right)^{1/4}   M_{\rm Pl}^{1/2}   \Gamma_\phi^{1/2}\ .
\label{TRH}
\eeq
Equation~(\ref{evR}) can be approximately solved at the early stage of reheating ($t\ll \Gamma_\phi^{-1}$), by taking $e^{-\Gamma_\phi t}\approx 1$ in \eq{eq:rhoR} and neglecting the thermal-energy contribution to $H_r$ ($R/a \ll \Phi$). Once we express $\rho_R$ in terms of the effective temperature $T$,
\beq
 \rho_R(t)\equiv  \frac{\pi^2g_{*}}{30} T^4(t) ,
\eeq
the solution of \eq{evR}, at early times, gives the evolution of the temperature $T$ (valid till the universe enters the radiation-dominated phase)
\beq
T\approx k_1\ T_{\rm max}\ a^{-3/8} (1-a^{-5/2})^{1/4}\ , \quad 
T_{\rm max} = k_2 \left( \frac{H M_{\rm Pl} T_{\rm RH}^2}{g_*^{1/2}}\right)^{1/4}\ ,
\label{Tmax}
\eeq
where $k_1=2^{6/5}3^{-3/20}5^{-1/4}=1.3$ and $k_2=(3/8)^{2/5}(5/\pi^3)^{1/8}=0.54$.

The temperature 
$T$ of the SM-particle gas
raises from $0$ to the maximum value $T_{\rm max}$
as long as, soon after inflation, the scale factor of the universe
$a$ grows by an order-one factor in a time $t\sim 1/\Hub$.
After reaching $T_{\rm max}$, the temperature decreases as $a^{-3/8}$, 
signalling the continuous
release of entropy from the decay of the inflaton field. 
When this energy release ends,  at time $t\sim 1/\Gamma_\phi$, the temperature is equal to $T_{\rm RH}$, which is called the reheating temperature, and then radiation cools in the standard way, $T\propto 1/a$, due to space expansion. Note that the entire reheating process can be described by only two parameters, which we choose to be $\Hub$ and $T_{\rm RH}$. The decay of the Higgs condensate at the end of inflation has been discussed in ref.~\cite{Bellido}.
%
%
%reaches a maximum $T_{\rm max} \sim (H_I
%M_{\rm Pl})^{1/4}T_{\rm RH}^{1/2}$
%and then decreases as $T \propto a^{-3/8}$,
%
%In fact, until $t\ll  \Gamma_\phi^{-1}$ assuming $\rho_\phi
%\gg \rho_R$ the system approximately evolves as
%
%i.e.
%\beq
%T = \bigg[ \frac{9}{2 g \pi^3}\bigg]^{1/4} {a^{-3/8}} (1-a^{-5/2}) \Hub^{1/4} M_{\rm Pl}^{1/2} \Gamma_\phi^{1/4}\eeq
%This scaling
%continues until the time $t \sim \Gamma_\phi^{-1}$, when the
%radiation-dominated phase starts with temperature $T\sim
%T_{\rm RH}$, defined as
%%       \footnote{Approximating the
%%       statistical distributions with Boltzmann-Maxwell functions gives
%%        $\rho_R=3g_*T^4/\pi^2$ and $T_{\rm RH}=[\pi \Gamma_\phi^2 M_{\rm Pl}^2/
%%       (8g_*)]^{1/4}$.}
%\begin{equation}
%T_{\rm RH}=.
%\end{equation}

Let us now consider the evolution of the Higgs field throughout the thermal phase.
Because of thermal corrections, the Higgs potential receives an extra mass term $\frac12 m_T^2 h^2$,
where $m_T\sim gT$ and $g$ represents the relevant combination of coupling constants. This expression for the thermal mass holds up to field values $h\circa{<} 2\pi T$.
The thermal corrections to the potential can be approximated as~\cite{leptog}
\begin{equation}
V_T \approx \left(0.21-0.0071\log_{10}\frac{T}{\GeV}\right) T^2 \frac{h^2}{2} e^{-\frac{h^2}{(2\pi T)^2}}\ ,
\end{equation}
where we added an exponential cut-off at high values of $h$.
$V_T$ helps in stabilising the Higgs potential by shifting the instability to higher scales, in much the same way as the mass term due to the coupling $\xi_H$ does.

Figure~\ref{hendvsTRH} shows the maximum allowed value of $h_{\rm end}$ in order for the Higgs not to fall into its true vacuum at (or above) Planckian field values. A direct comparison with the right panel of fig.~\ref{Vpostinfl} shows that, for high enough reheating temperatures, thermal effects are indeed of extreme relevance.

\begin{figure}[t]
\begin{center}
$$\includegraphics[width=0.45\textwidth]{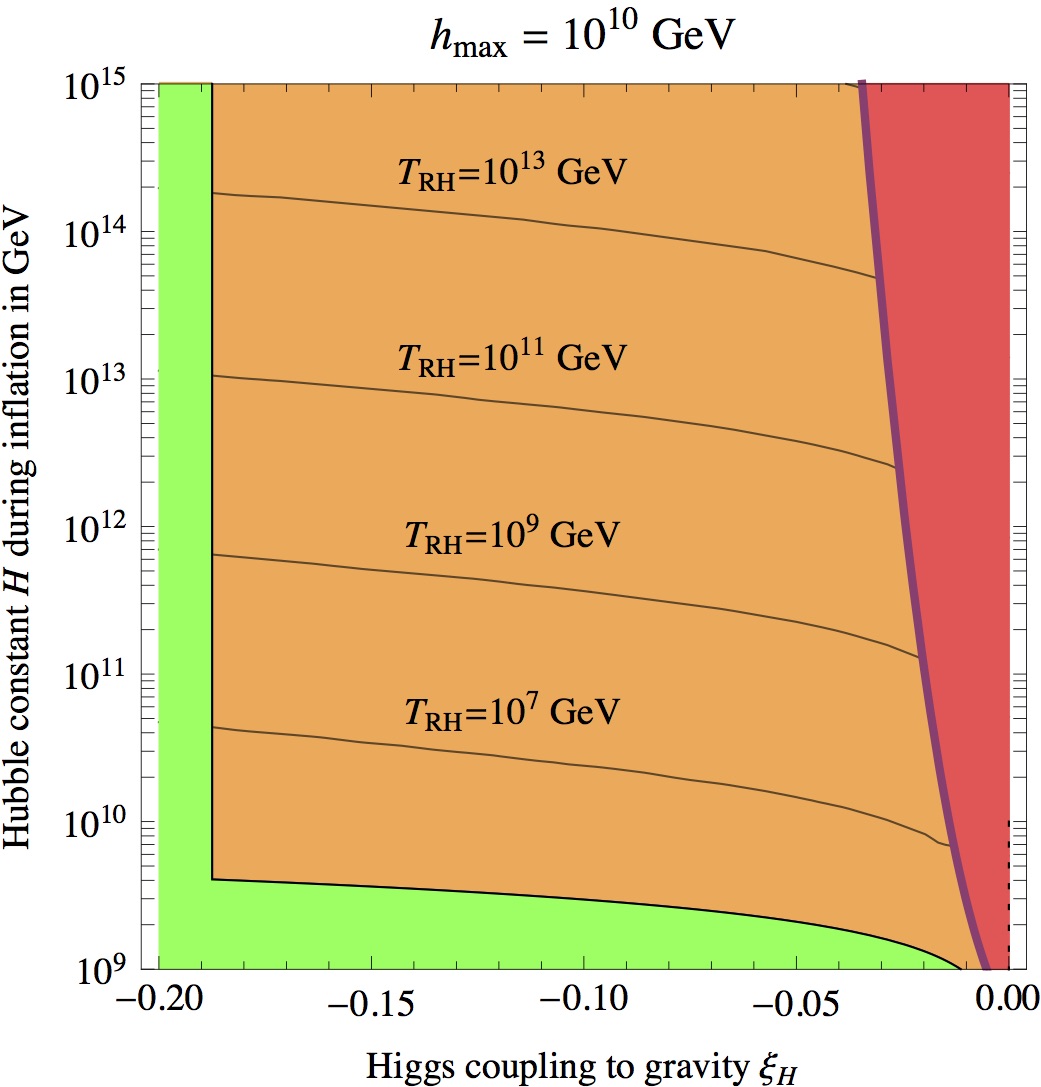}\qquad
\includegraphics[width=0.45\textwidth]{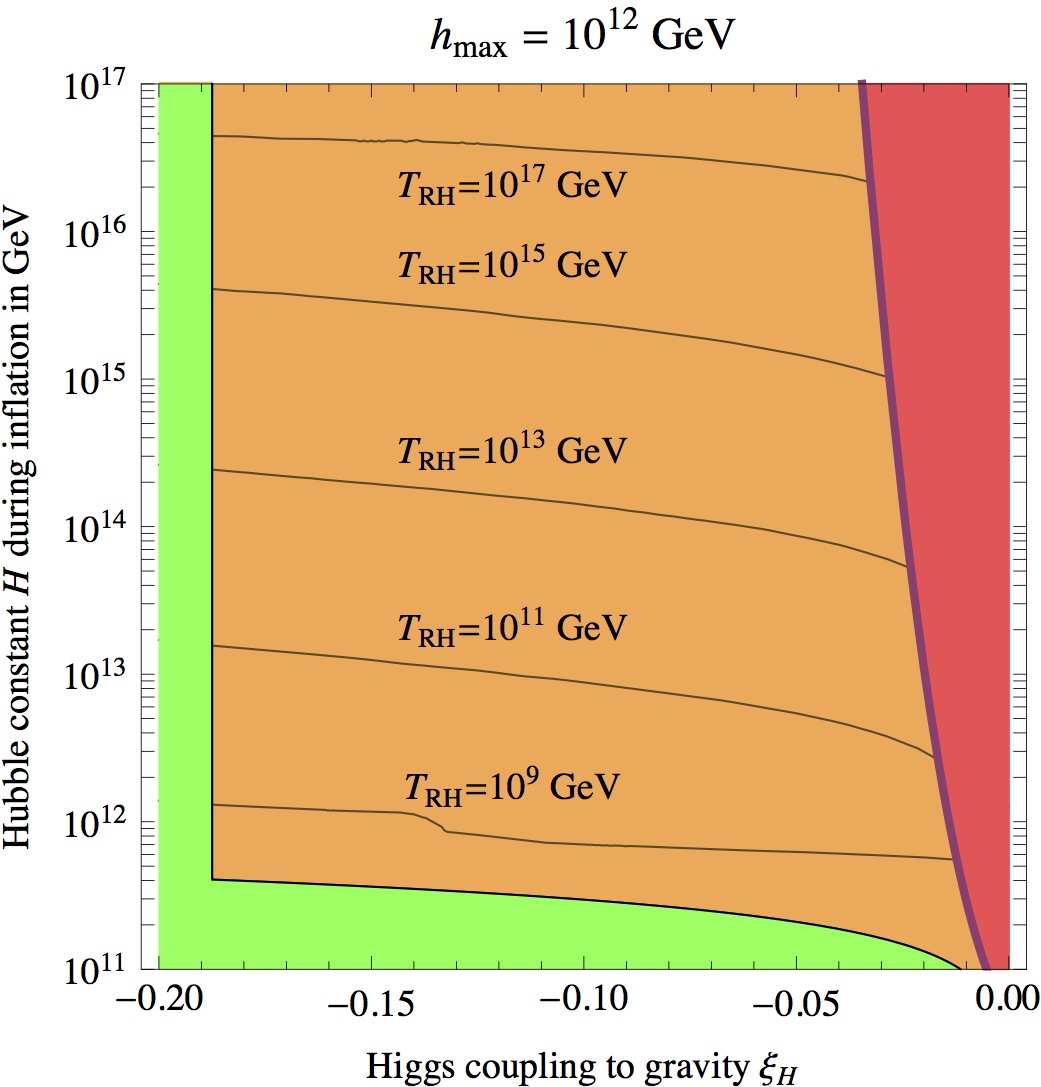}
$$
\caption{\em Minimal reheating temperature $T_{\rm RH}$ needed to 
prevent the fall of the Higgs down into its deep true vacuum, assuming two different values for the instability scale $h_{\rm max}$ of the Higgs potential. 
% and different $\xi_H$. Black lines show that, assuming $T_{\rm max}$ equal the average value at the end of inflation $\langle h^2\rangle ^{1/2}$, the temperature-induced mass term is enough to avoid the instability when the Higgs fluctuates around its instability scale.
\label{TmaxvsH}}
\end{center}
\end{figure}

It is not difficult to understand the behaviour of the maximum allowed value of $h_{\rm end}$ as a function of the Hubble rate. Let us consider for simplicity the case $\xi_H =0$ (left panel in fig.~\ref{hendvsTRH}).
From eq.~(\ref{eqafter}) one can see that in a time scale of the order of the Hubble time, when the scale factor changes
by order unity and the maximum temperature has been reached, the Higgs field changes by an amount (we neglect factors of order unity)
\begin{equation}
h-h_{\rm end}\simeq -\frac{1}{H^2}V'(h_{\rm end})\simeq -\frac{m_T^2(T_{\rm max}) h_{\rm end}}{H^2} - \frac{\lambda h^3_{\rm end}}{H^2},
\end{equation}
where we have approximated the zero temperature potential as $\lambda h^4/4$ and one has to remember we are considering the region
where $\lambda<0$. Therefore, we obtain the approximate expression, valid soon after inflation,
\be
h\simeq h_{\rm end}\left(1-\frac{T_{\rm max}^2}{H^2} -\lambda \frac{h^2_{\rm end}}{H^2}\right).
\label{supper}
\ee
For $h$ to roll towards the origin, a necessary condition is that the right-hand side of \eq{supper} is smaller than one, which implies $h_{\rm end} \circa{<}T_{\rm max}/|\lambda|^{1/2}$. This explains the approximate flatness of the curves in the right panel of fig.~\ref{hendvsTRH} for small $H$. For $H\gg T_{\rm max}$, the approximate scaling of the bound on $h_{\rm end}$ as $H^{-1/3}T_{\rm max}^{4/3}$ can be understood in the following way. Being the Hubble rate large, the term $(a/H^2)\partial V/\partial h$ in  eq.~(\ref{eqafter}) can be neglected up to the moment when 
the second time derivative term or the first time derivative term become of the order of the potential term. This means that the Higgs field
does not move much from its initial condition $h_{\rm end}$ up to the moment when 
\be
\frac{5}{2a}\frac{d h}{d a}\sim  \frac{a}{H^2}m_T^2(T)h
=\frac{a}{H^2}T_{\rm max}^2 a^{-3/4} h.
\ee
This implies that the Higgs field starts moving away from $h_{\rm end}$ when 
\be
a\sim a_* = \left(\frac{45\, H^2}{8\, T_{\rm max}^2}\right)^{4/9}.
%a=a_*\sim \left[45\ln(h_{\rm end}/ h_{\rm max}) H^2/8T_{\rm max}^2\right]^{4/9}.
\ee
 Imposing that at this value of $a$ the finite temperature term in the potential dominates over the negative quartic term gives
\be
h_{\rm end}\circa{<}\left( \frac{8}{45}\right)^{1/6} \frac{H^{-1/3}\, T_{\rm max}^{4/3}}{\sqrt{|\lambda|}}\, ,
%h_{\rm end}\circa{<}\left[\frac{45}{8}\ln(h_{\rm end}/ h_{\rm max})\right]^{-1/6}\frac{T_{\rm max}^{4/3}}{H^{1/3}},
\ee
which reproduces the right scaling shown in fig.~\ref{hendvsTRH}.

In fig.~\ref{hendvsTRH} we let $h_{\rm end}$ and $H$ vary independently. However, the inflationary dynamics correlates the two variables, assigning a certain probability to $h_{\rm end}$ for any given $H$. Although we have not used any relation between the two variables, in fig.~\ref{hendvsTRH} we have indicated in red the region in which the field $h$ has overwhelming probability to slide towards large values and thus the corresponding parameters essentially cannot be the outcome of inflation.  

\smallskip

Figure~\ref{TmaxvsH} shows the minimal value of $T_{\rm RH}$ for which the thermal corrections  prevent the fall of $h$ down to its deep minimum.
In other words, it shows how the limit on the Hubble constant $\Hub$ corresponding to the orange area of fig.~\ref{RHbounds} can be relaxed, depending on the reheating temperature $T_{\rm RH}$. For sufficiently large $T_{\rm RH}$, regions in which
the Higgs field fluctuates around its instability scale $h_{\rm max}$ can be recovered by post-inflationary thermal effects.
On the other hand, the thermal phase cannot save regions in which the Higgs fell down into its deep negative-energy minimum
during inflation. In the next section we will show that most of these regions eventually expand and hence 
a viable cosmology require that no such regions are produced during inflation. This excludes the red area in  fig.~\ref{RHbounds}.

%The basic conclusion is that the stronger bound on the Hubble constant $\Hub$
%obtained by demanding that inflation does not produce regions where 
%the Higgs field lies fluctuates around its instability scale $h_{\rm max}$
%(gray region in in fig.~\ref{RHbounds}) 
%is not needed, because the thermal phase can easily `burn' such bubbles.
%
%The thermal phase cannot `burn' the regions where the Higgs fell down to its deep negative-energy minimum
%during inflation;  in the next section we will show that such regions would expand and thereby 
%one must avoid their production during inflation, obtaining the red bound in  fig.~\ref{RHbounds}.
%
%

\begin{figure}[t]
$$
\includegraphics[width=70mm,height=70mm]{\figs/plotC2}\qquad
\includegraphics[width=70mm,height=70mm]{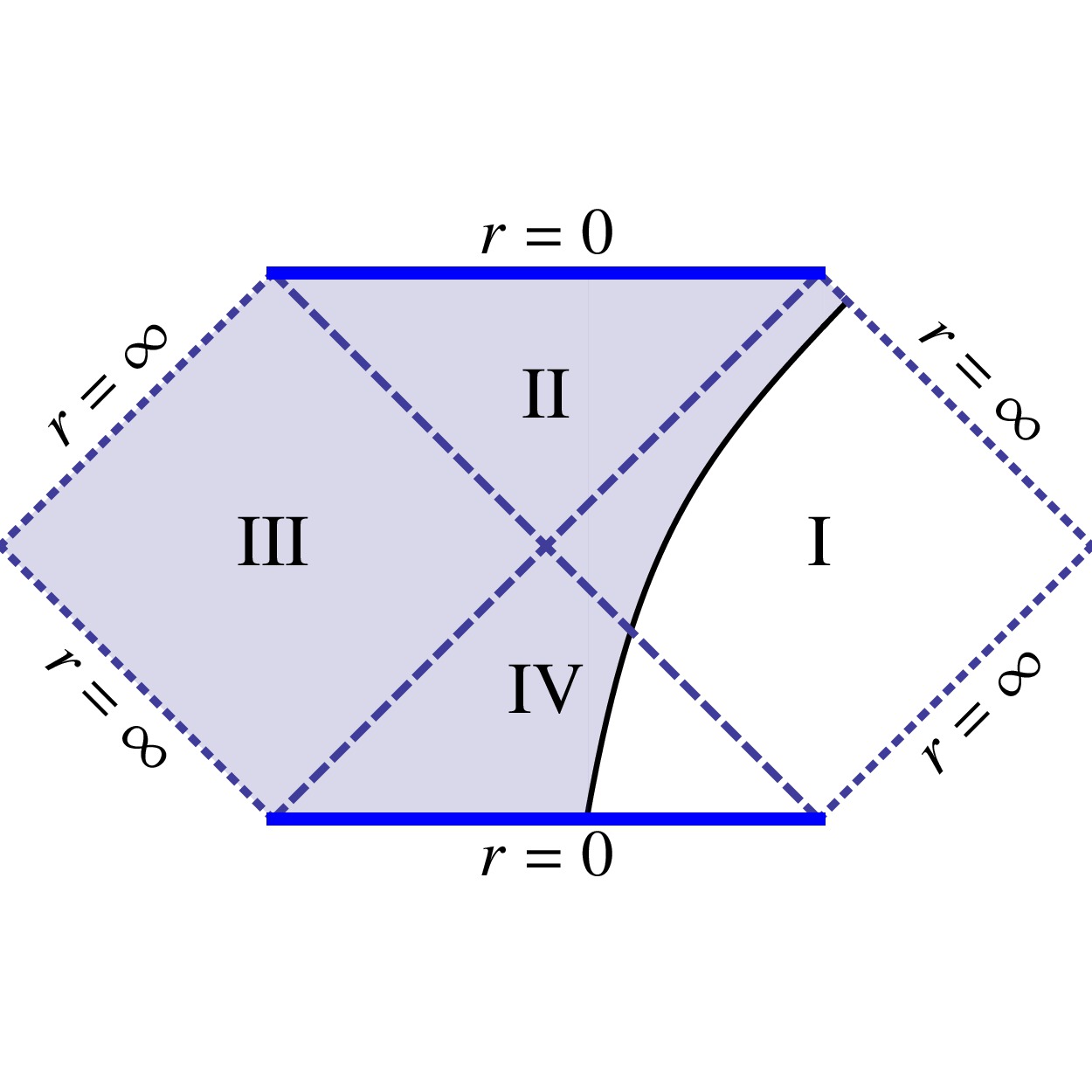}$$
\caption{{\bf Penrose diagram describing an AdS  bubble that expands in external Minkowski}.
Left: {\em The wall trajectory corresponding to line C of fig.~\ref{h1}, in AdS space}.
Right: {\em The wall trajectory corresponding to line C of fig.~\ref{h1},  in the Schwarzschild geometry.}
\label{C1}}
\end{figure}

\subsection{Bubble evolution in Minkowski spacetime}\label{bubblem}
The discussion of bubble evolution in an external Minkowski spacetime
is analogous to the de Sitter discussion of section~\ref{bubble}. 
A first difference is that the effective potential
of \eq{potds}, which dictates the evolution of the bubble,
is simplified when we set $\ell_{\rm out}=\infty$ ({\it i.e.}\ $\delta=0$).
A second key difference is that the external Minkowski space 
has no causal horizons: if, after inflation ends,
bubbles expand at the speed of light, they engulf the whole space.

\medskip

An important task is to determine whether Higgs bubbles expand or shrink.
The complete analysis is presented in appendix~\ref{AdS}, where all the possible
wall trajectories are determined. In summary, 
there are two scenarios in which bubbles do not take over the whole space:
either they start small enough so that they shrink, 
or they expand but remain hidden behind a black-hole horizon
(a possibility that corresponds to $\Delta<0$).
In the following we examine if either of these possibilities 
is realised for Higgs bubbles, making them benign.

\medskip

We first consider bubbles with positive 
\beq \Delta\equiv
 {1}/{\ell^2_{\rm in}}-(4\pi G\sigma)^2 >0,\eeq which 
 are bigger than their Schwarzschild radius and thereby can expand in the naive Newtonian way.
This is the case depicted in the Penrose diagram in fig.~\ref{C1}, to be compared with  fig.~\ref{W1} for external dS.
The black continuous curve denotes the trajectory of the wall:
the bubble starts small and expands indefinitely within the asymptotically 
flat spacetime. 
The total space is constructed by patching the part of the diagram on the right of the wall with the part of the
left diagram on the left of the wall. The shaded areas correspond to the
parts that must be eliminated in order to join the remaining parts along the wall trajectory.
From the point of view of an external observer, the bubble asymptotically expands at the speed of light and asymptotically reaches null infinity,
filling all space.

\begin{wrapfigure}{R}{0.5\textwidth}
  \vspace{-2ex}
    \begin{center}
\includegraphics[width=0.45\textwidth]{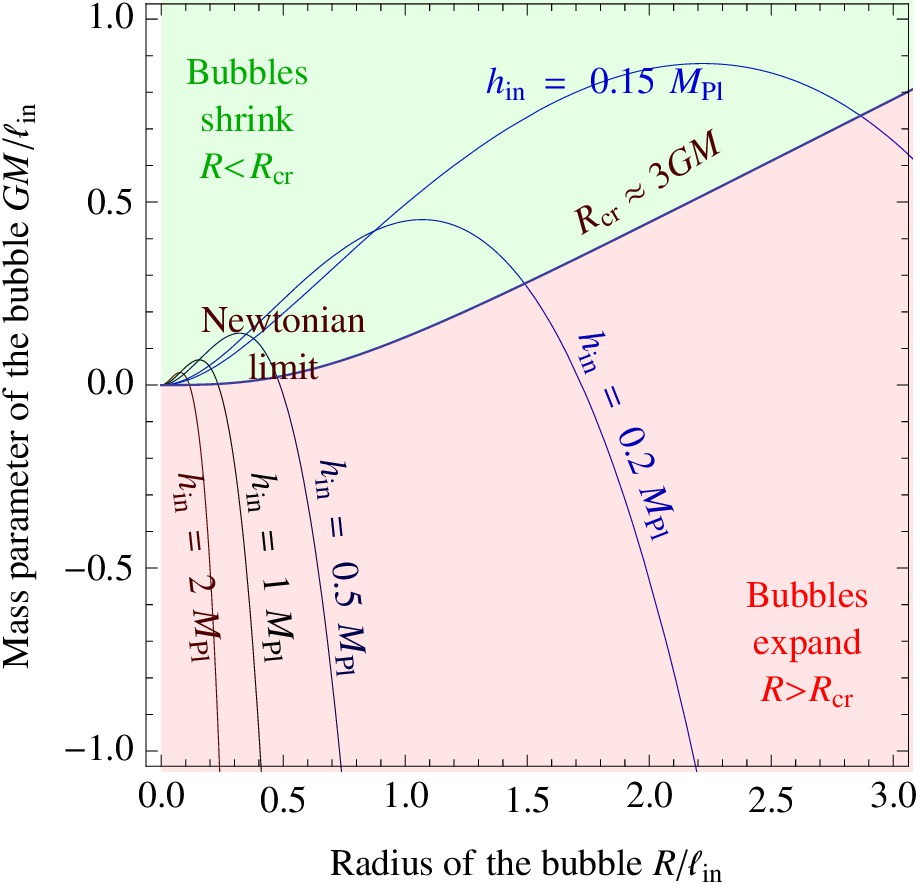}
  \end{center}
  \vspace{-2ex}
  \caption{\em\label{Hbubbles}  The boundary in the $(R,M)$ plane that separates expanding bubbles  from shrinking bubbles.
The curves are the estimated masses of Higgs bubbles, as functions of their radius,
 for different values of $h_{\rm in}$, the unknown Higgs field value at the deep minimum. 
  \label{estimate}}
\end{wrapfigure}

%Do these bubbles expand?  
%Appendix~\ref{AdS} presents a full study of the possible wall trajectories %$\tilde R(\tilde\tau)$.
Bubbles may also shrink because of their surface tension, if they are small enough ($R< R_{\rm cr}$), and start with a small wall velocity. 
The critical radius, separating the two types of evolution,
can be computed from the potential of \eq{potds}, but the resulting 
expression is not very illuminating. We present the complete discussion
in appendix \ref{AdS6}.
The result can be simplified by assuming $\dot R=0$, 
and the Newtonian limit $\kappa\ll 1$ (such that $\Delta >0$). In this case,
the critical radius is obtained by 
extremising the sum of the surface and volume energy ($4\pi R^2 \sigma-4 \pi R^3V_{\rm in}/3$) with respect to $R$, thus
finding $R_{\rm cr}  = 2 \sigma/V_{\rm in}$ and a bubble mass
$M = 16\pi\sigma^3/3V_{\rm in}^2$. 
The exact result, valid even beyond the Newtonian approximation,  
is shown in
fig.~\ref{estimate}.
In the ultra-relativistic limit the critical 
radius becomes $R_{\rm cr}=3GM$, 
slightly larger than the Schwarzschild radius $2GM$.

Figure~\ref{estimate} also shows the estimated $M(R)$ corresponding to Higgs bubbles for different values of $h_{\rm in}$, the unknown Higgs value at its deep minimum.
We use  $V_{\rm in}\sim |\lambda| h_{\rm in}^4$ and estimate 
the surface tension from the Newtonian expression
\beq \sigma \approx \int dr \bigg[\frac12 \bigg(\frac{\partial h}{\partial r}\bigg)^2+V(h) - V(h_{\rm in})\bigg]
\sim \frac{h_{\rm in}^2}{\Delta r} + \Delta r |\lambda| h_{\rm in}^4 \circa{>} \sqrt{|\lambda|} h_{\rm in}^3\eeq
minimised for a bubble thickness $\Delta r \sim 1/
( h_{\rm in}\sqrt{|\lambda}|)$.

The estimate for $\sigma$ is inserted in the full expression for $M$,
\eq{square}.
The meaning of negative values of $M$ is discussed in the appendix, in sections
\ref{negmas} and \ref{AdS6}: they correspond to bubbles for which the negative 
volume contribution to the total energy budget is dominant. 
For our present purposes, the conclusion is that
inflationary fluctuations  create a number of bubbles with a variety of values of $R$ and $\sigma$, including
bubbles with $R>R_{\rm cr}$, which expand. 
Bubbles produced by inflationary fluctuations tend to appear with characteristic size $R\sim 1/H = \ell_{\rm out}$
and with negligible $\dot R$. If the Higgs value is near the deep minimum of the potential, such bubbles are expected to have small or negative mass (because of the
negative volume contribution to the energy) and, therefore, expand.

%While this situation would prevent the expansion of bubbles, 
%in the realistic case where the wall is replaced by a continuous variation of the Higgs field
%with SM potential $V \simeq \lambda |H|^4$,
%the condition $\ex=-1$ is realised only if the variation of $\langle H\rangle$ between the two minima
%happens in a sub-Planckian length.
%Surely there are field configurations that do not satisfy this condition.
%Furthermore, even if this condition were initially satisfied, the bubble would evolve
%towards a smoother bubble with smaller surface tension.
%In conclusion, the mechanism that can prevent bubble expansions is not generic enough
%(indeed, quite the opposite).
%
%However most, if not all, bubbles are not of this type.
%In conclusion, AdS bubble expand into Minkowski, asymptotically at the speed of light.

%The form of the potential is very similar to that in
%fig.~\ref{h1}. There are now two horizons, which lie above the
%curve of the potential for $\ex=1$, or 
%are tangent to it at $z=1$ for $\ex=-1$. 
%The various trajectories correspond to solutions of constant 
%$E=-{\kappa^2}/(G^2M^2\rho^4)$, as depicted in fig.~\ref{h1}.
%The two types of behavior, characterized by $\ex=\pm 1$, correspond now
%to the gravitational self-energy $\kappa^2$ 
%of the bubble being smaller or larger than
%the total difference in energy density $1/\ell^2+1/\ell_{\rm out}^2$ between the dS and 
%AdS spacetimes.

Bubbles with sufficiently large surface tension $\sigma$
have $\Delta <0$ and  
 their expansion is energetically disfavoured  from a Newtonian point of view.
However, there exist 
expanding solutions for such bubbles 
within general relativity. They are discussed in appendix \ref{AdS} (see fig.~\ref{G1}). The crucial characteristic is that the expanding region is not accessible
to an observer in the asymptotically flat space. 
In other words, an  observer outside the bubble only sees a black-hole horizon,
that protects him from its expansion.
While such bubbles would be benign,  $\Delta <0$ represents an  extreme case:
for the quartic Higgs potential, the surface tension gives a Planck-suppressed correction, such that
\beq  \Delta \sim  \frac{|\lambda| }{M_{\rm Pl}^2}\bigg[  h_{\rm in}^4 -   \frac{h_{\rm in}^6}{M_{\rm Pl}^2} \bigg]\eeq
is negative only when the deep minimum of the SM potential is super-Planckian,
$h_{\rm in}\circa{>} M_{\rm Pl}$. It is then impossible to make firm predictions;
strong gravitational effects may induce various dangerous effects.
As long as the deep minimum is in the calculable
sub-Planckian region ($h_{\rm in}\circa{<} M_{\rm Pl}$),
bubbles have super-Planckian tension only for extreme Higgs field configurations,
{\it e.g.}\ if the variation of $h$ between the two minima happens 
within a sub-Planckian length.
Inflationary fluctuations tend to create bubbles with bigger thickness and smaller surface tension.
Furthermore, even if the condition $\Delta<0$ were initially satisfied, the bubble would evolve
towards a smoother configuration with smaller surface tension
by reconfiguring the Higgs field profile. 
%In summary, we conclude that the possibility $\Delta <0$ does not lead to stability, 
%because most inflationary  bubbles  have $\Delta >0$.

Our study has been carried out within the thin-wall limit, because this is
the only
setup for which an analytical treatment is possible. As we pointed out above,
the realistic situation is more likely to involve configurations with a
smooth
transition region from the interior AdS space to the false-vacuum exterior.
For these, the fundamental dynamics is mainly determined through the
interplay between
the negative energy density in the interior and the positive contribution
from the transition
region. We expect that our analysis captures the essential features of the
evolution of
such configurations as well.

\bigskip

The important conclusion that we draw from our study is the following.
{\em No robust general-relativistic effect prevents large-field Higgs bubbles from expanding
and engulfing all Minkowski space.} 
As a result, a viable cosmology requires that 
no expanding bubbles are present in our past light-cone. In other words,
the condition $p(h\to \infty)< e^{-3N}$ for $N\approx 60$ $e$-folds must hold and the red region in fig.~\ref{RHbounds} is excluded.

\begin{figure}[t]
\begin{center}
$$\includegraphics[width=0.7\textwidth]{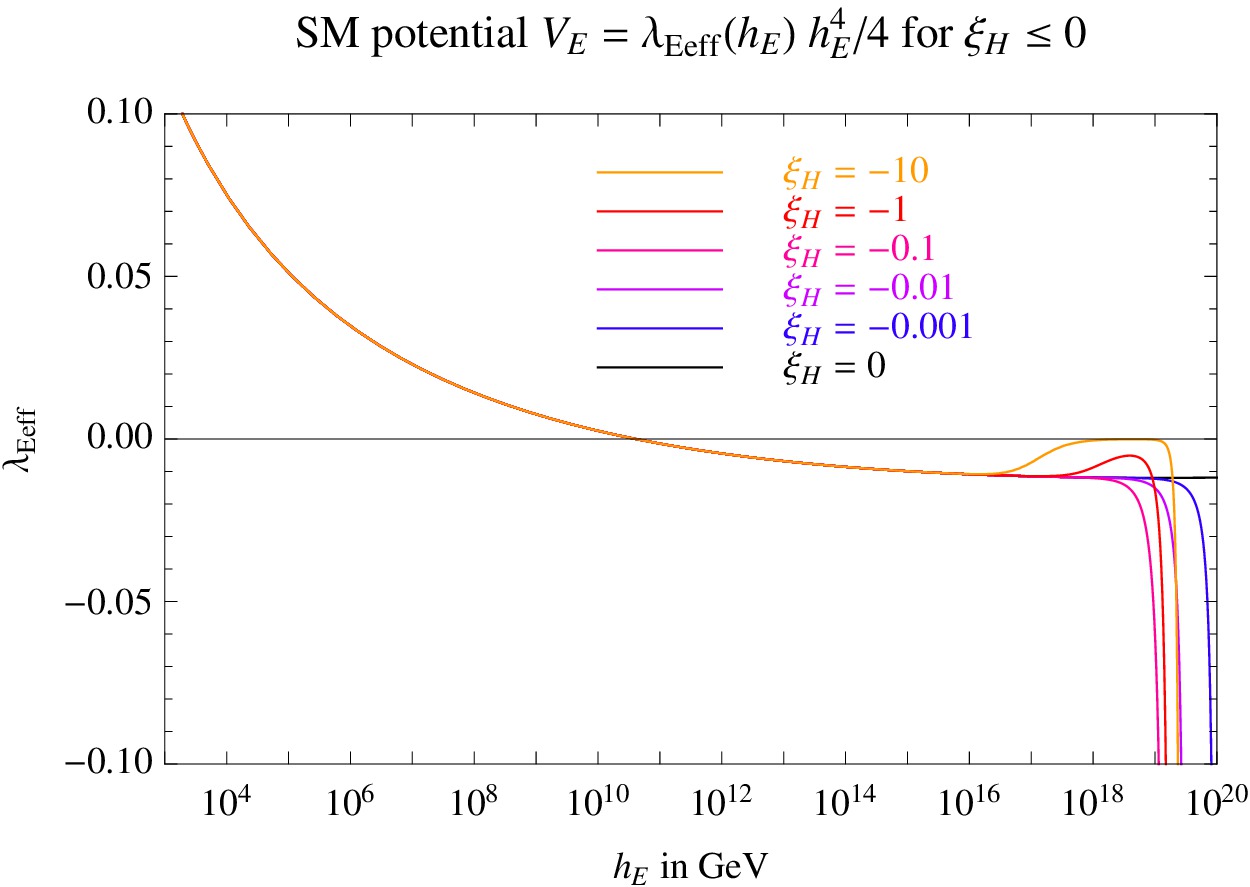}$$
\caption{\em The SM Higgs potential $V_E\equiv \lambda_{E\rm eff}(h_E) h_E^4/4$ in the presence of a negative  $\xi_H$ coupling,
written in the Einstein frame in terms of the canonical  Higgs field $h_E$.
\label{VSMxi}
}
\end{center}
\end{figure}

\subsection{The Higgs potential for $\xi_H\neq 0$ at zero temperature}
%We conclude our study of the Higgs evolution after inflation with a remark concerning the non-minimal gravitational coupling $\xi_H$. 
%As previously discussed, a coupling
%$\xi_H$ helps to stabilise the Higgs field during inflation. However, it could have an opposite effect in the classical potential at zero temperature and $H\approx 0$, in presence of a Higgs instability generated by SM interactions ($\lambda_{\rm eff}<0$).
We conclude our study of the Higgs evolution after inflation with a remark concerning the non-minimal gravitational coupling $\xi_H$.
As previously discussed, a coupling $\xi_H$ helps to stabilise the Higgs field during inflation.
However, long after inflation, at zero temperature and $H\approx 0$, it could have an opposite effect in the classical potential, in presence of a Higgs instability generated by SM interactions ($\lambda_{\rm eff}<0$).
In this setup an additional source of instability is generated by $\xi_H$ at Planckian values of the Higgs field, with no effect on our discussion of the Higgs dynamics during the inflationary and post-inflationary phase.

In order to investigate the phenomenon, we focus on the real component $h$ of the Higgs doublet
$\Hig = (0,h/\sqrt{2})$ and 
perform a Weyl rescaling to the Einstein frame
$g_{\mu\nu}^E = g_{\mu\nu}\times f$ with 
$f=1+ \xi_H h^2/\bar M_{\rm Pl}^2$. Then,
the  Einstein-Hilbert term becomes canonical and the action is
\beq \label{eq:LE}
\Lag_E =
 \sqrt{\det g_E} \bigg[   - \frac{\bar M_{\rm Pl}^2}{2} R_E
+ Z  \frac{(\partial_\mu h)^2}{2} 
 - V_E (h)
  \bigg]+\cdots\eeq
  \beq 
 Z = \frac{1}{f} + \bar M_{\rm Pl}^2  \frac{3 f^{\prime 2}}{2 f^2} ~ ,~~~~~
 V_E (h)= \frac{V(h)}{f^2}.
\label{eq:VHxi}
\eeq
We are studying the theory long after inflation, and therefore $V(h)= \lambda_{\rm eff}(h) h^4/4$.\footnote{During inflation $V$ contains an extra constant term $V_\phi$, which dominates the energy density, and $\Hub ^2 = V_\phi/(3\bar M_{\rm Pl}^2)$ is the Hubble constant during inflation.
By expanding $V_E$
at leading order in $h^2/\bar M_{\rm Pl}^2$ for $V_\phi\neq 0$,
we recover the Higgs mass term considered in the previous sections,
$m^2 = - 4\xi_H V_\phi /\bar M_{\rm Pl}^2=
-12 \xi_H \Hub ^2$.
The higher order terms were not relevant for our previous discussion.}
It is convenient to define a canonically normalised Einstein-frame
Higgs field $h_E$ through the equation $dh_E/dh = \sqrt{Z}$, where $Z$ is given in \eq{eq:VHxi}.
The field $h_E$ is such that
$h_E \simeq h$ for $h\ll\bar M_{\rm Pl}$ and $h_E\to \infty$ for $h\to \bar M_{\rm Pl}/\sqrt{-\xi_H}$ (hence $f\to 0$).

The Einstein-frame scalar potential becomes
\beq
V_E (h_E) =
\left. \frac{\lambda_{\rm eff}(h) h^4}{4(1+ \xi_H h^2/\bar M_{\rm Pl}^2)^2}\right|_{h=h(h_E)} ~.
\label{veer}
\eeq
In the limit of large $h_E$ (which corresponds to $h\to \bar M_{\rm Pl}/\sqrt{-\xi_H}$), the denominator in \eq{veer} nearly vanishes, while $\lambda_{\rm eff}$ is negative. In that regime of field configurations, we find $h(h_E) -\bar M_{\rm Pl}/\sqrt{-\xi_H}\propto \bar M_{\rm Pl}\exp(-\sqrt{2/3}h_E/\bar M_{\rm Pl})$ and the potential becomes
\be
\frac{V_E (h_E)}{\bar M_{\rm Pl}^4} \propto \lambda_{\rm eff} \exp \left(\sqrt{\frac83}\frac {h_E}{\bar M_{\rm Pl}}\right) , ~~~~{\rm for}~h_E \to \infty \, .
\label{VExp}
\ee
Here $\lambda_{\rm eff}$ is evaluated at $h=\bar M_{\rm Pl}/\sqrt{-\xi_H})$ and is negative. The exponential behaviour in eq.~(\ref{VExp}) contributes to amplify the source of instability already present. The effect is shown in fig.~\ref{VSMxi}, where we plot the effective coupling $\lambda_{E\rm eff}$, defined in analogy with previous effective quartic couplings by rewriting the potential in the Einstein frame as $V_E (h_E) \equiv \lambda_{E\rm eff}(h_E) h_E^4/4$. Negative values of $\xi_H$ appear to trigger a deep instability at Planckian field values $h\sim \bar M_{\rm Pl}/\sqrt{-\xi_H}$. However, for nonzero $\xi_H$, the ultraviolet cutoff of the SM
is no longer $\bar M_{\rm Pl}$, but $\bar M_{\rm Pl}/|\xi_H|$ \cite{uvcut}. Hence, the instability just described takes place 
above that cutoff, where one is losing control of the theory~\cite{Branch}. It has been pointed out that small primordial black holes
can seed Higgs-vacuum decay
and enhance its rate~\cite{gregory}. However, this process crucially depends on the number density 
of black holes at a given epoch.

\section{The quantum gravity prediction for the Higgs mass}\label{QG}
In this section we explore how a speculative conjecture that has been put forward in the context of quantum mechanical completions of gravity can lead to a sharp correlated prediction for the Higgs and top-quark masses. The intriguing result is that this prediction agrees quite well with the measured values of these masses. The reasoning is essentially based on two points.
\begin{enumerate}
\item  The empirical observation that we live in an accelerating universe.  

\item The theoretical conjecture that quantum gravity is ill-defined in de Sitter space~\cite{Witten:2001kn,Goheer:2002vf}.    
\end{enumerate}
The difficulties with dS quantum gravity have been argued from various perspectives~\cite{Witten:2001kn,Goheer:2002vf}. We summarise here in a very schematic way some of the arguments against a stable dS space, reviewed in~\cite{ArkaniHamed:2008ym}, extending them in light of some more recent developments. There is no positive conserved energy in dS (and, as a consequence, there cannot be unbroken supersymmetry). There is no classical compactification of ten- or eleven-dimensional supergravity to dS space, and stable dS space cannot be obtained  from any string or M-theory.  {Even in the general setting of quantum gravity, beyond the particular UV completion offered by string theory, other problems arise. It has been suggested that the quantum Hilbert space in dS is of finite dimension, limiting the variations of complex constructions.} Given that  the Gibbons-Hawking temperature sets a minimum temperature,  the finite dimensionality of the Hilbert space sets a maximum time scale, the so-called recurrence time~\cite{Goheer:2002vf}. In particular, this leads to the problem of the so-called Boltzmann brains~\cite{Page:2006nt}, and it has been suggested that its resolution calls for an unstable universe~\cite{Page:2006nt}. More generally, a  rigorous definition of the Hilbert space in dS seems to be problematic~\cite{Witten:2001kn}. In quantum gravity, it is difficult to define precisely local observables and one can rely only on asymptotic quantities, such as the $S$-matrix in Minkowski space and the boundary correlators in AdS. However, in dS, where asymptotic states fall behind the horizon, no such precisely defined observables seem to be present. 

In addition to these problems that have been known for some time, it has been found in~\cite{Dubovsky:2008rf,Dubovsky:2011uy,Lewandowski:2013aka} that there is a sharp universal bound on how much reheating volume slow-roll inflation is capable to create without being eternal. This is given by $e^{S_{\rm dS}/2}$, where $S_{\rm dS}$ is the entropy of the would be de Sitter space with Hubble rate evaluated at the time of reheating. Larger overall expansion is possible only by making space infinite. This generalises at the quantum level the bound on the duration of inflation found in~\cite{ArkaniHamed:2007ky}, and determines the universality of such a bound under the number of fields involved, higher derivative corrections, number of space time dimensions and slow roll parameters. The same phenomenon, {\it i.e.} not being able to produce arbitrarily large finite volumes, is shared by false vacuum inflation. These results seem to suggest that there are bounds on the kind of global spacetime structures that local quantum field theory can generate. This is related to, and somewhat supporting, the argument against de Sitter space in nature that we follow here.
%, extending it to any number of dimensions and number of fields, and to any higher dimension operator in the theory of inflation and gravity,. This phenomenon is shared by false-vacuum inflation, where arbitrarily large, but finite, volumes seem to be forbidden (see below). There seem therefore to be some definite discontinuity between finite and infinite spaces. 

\bigskip

Though none of these arguments  raise to the level of a proof of the inconsistency of dS space, they clearly give an idea of the conceptual difficulties that arise when considering quantum mechanics in dS space.  
%\xxx{``````LS: maybe this is enough? The paper has changed, and now we are not making a whole paper out of this section. Furthermore, I do not see much of the point of repeating too much the section of~\cite{ArkaniHamed:2008ym}, as very little new on that front has happened. In fact, I added the one on the entropy bound because it was a new thing.''""}

It could well be that the problems of asymptotic dS space are circumvented by Planckian dynamics,
which can for example open channels for vacuum tunnelling to a `landscape' of other minima with zero or negative vacuum energy.
This is certainly a possibility. However, it is interesting to note that, even without any special hypothesis about the gravitational sector, the SM Higgs offers an easy way out to the problem. A solution is automatically found if the present dS space is only metastable.

As soon as the decay rate per unit space time volume $\Gamma$ of the false vacuum  is non zero, the asymptotic space is not dS. True vacuum bubbles are nucleated and expand at the speed of light. There are two critical values of the decay rate~\cite{Guth:1982pn}, both valued around $H^4_\Lambda$, with $H_\Lambda$ being the Hubble rate of the would-be de Sitter region.  

If $\Gamma$ is larger than the largest critical point, bubbles percolate, fill the whole space, inflation ends globally, and the asymptotic spacetime is the one of the true vacuum (if this is AdS, the instability grows and leads to a singularity in about one Hubble time). The value of this critical point is $ \Gamma_2/H_\Lambda^4=9/4\pi\simeq 0.71$~\cite{ArkaniHamed:2008ym}. 

The precise value of the second critical point is  unknown, but is bounded to be in the range~\cite{Guth:1982pn} $1.1 \times 10^{-6}\leq \Gamma_1/H_\Lambda^4\leq 0.24$. If the decay rate lies between these two critical points, then bubbles percolate and long chains of bubbles form, connecting arbitrarily distant point in the otherwise dS space. Finally, when the decay rate is even slower than the second critical point, then bubbles do not percolate, and the asymptotic spacetime is the one called `false vacuum eternal inflation'. In this phase,  there is an infinite amount of space that keeps inflating, but each single point decays at some time into the true vacuum through the nucleation of a bubble.\footnote{Bubbles continuously form and collide an infinite number of times 
among each other but they do so  in relatively smaller and smaller regions: there are points in the inflationary space for which the probability that they are connected by a stream of  bubbles is zero. When bubbles collide in our past light cone, a very sharply defined disk shape is impressed in the CMB~\cite{Kleban:2011yc}. The optimal analysis  to search for such a signal in the WMAP data has been recently performed, with no evidence found~\cite{Osborne:2013hea}.}
%When the decay rate per unit space-time volume is much smaller than $H^4$, the asymptotic space time is not dS, but false vacuum eternal inflation~\cite{Guth:1982pn}.  

\bigskip

On the other hand,
there is no question that AdS and Minkowski are well-defined spaces from the point of view of quantum gravity, although we have a non-perturbative formulation of quantum gravity only for AdS. 
%Even if an AdS bubble leads to a singularity, quantum gravity has no issues with singularities: it is rather us that have difficulties understanding them. The opposite might instead be true for de Sitter space.
  So, it is possible that quantum gravity allows for an eternally inflating false-vacuum  space-time, but not for dS space. We will show that the opportunity of circumventing dS space through the Higgs field is given to nature only for a narrow range of Higgs boson masses. Interestingly, it seems that nature did not miss the opportunity because the measured Higgs boson mass lies exactly within this range.

%The spirit of this paper is to opt for simplicity and minimality in the properties of the theory and we use a sharp Occam razor to cut off anything beyond the SM up to scales of about $10^{12}$--$10^{14}$~GeV or more, where new dynamics may possibly exist, in association with inflation, neutrino masses, axion, unification. 
%In this spirit, the naturalness problems of the cosmological constant and the Higgs vev may be resolved by the statistical properties of the multiverse. 

Let us for a moment accept the quantum-gravity arguments against stable dS space, and let us assume that our universe, which we observe to be today in a dS phase, escapes the problems through a future decay of the Higgs vacuum. This implies the bound  on the Higgs boson mass~\cite{instab}
\beq \frac{M_h}{\GeV} < 129.6 + 2.0 \left(\frac{M_t}{\GeV}-\Mtexp\right) -0.5\left( \asdiff \right) \pm 0.3 
%\pm 0.6_{\rm non-pert} 
\ .
\label{eq1}
\eeq

The vacuum decay rate induced by the SM Higgs instability,
exponentially suppressed by the action of its bounce solution, $S \approx 8\pi^2/(3|\lambda|) \sim 10^3$,
is always faster, and typically much faster, than what is needed to avoid the dS problems.
Indeed, for comparison, the de Sitter entropy is $S_{\rm dS}\sim \pi \mp^2/H_\Lambda^2  \sim 10^{120}$.
In general, the action of any Coleman-de Luccia instanton out of a false vacuum with positive energy density is never larger than  $S_{\rm dS}$, no matter how high we make the false vacuum barrier  (see the discussion of~\cite{ArkaniHamed:2008ym} for a review).\footnote{One can check this by taking the limit in which the wall energy density goes to infinity in eq.~(3.16) of the paper by Coleman and de Luccia~\cite{deluccia}.} Therefore, when the Coleman-de Luccia instanton is present and can be reliably computed, the lifetime is bound by the Poincar\'e recurrence time of de Sitter space, of order $e^{S_{\rm dS}}/H$, up to logarithmic factors. It is expected that this lower bound on the decay rate is a property shared by all theories with a de Sitter false vacuum within field theory and perturbative gravity.

\medskip

There are also lower bounds on the Higgs mass. A first bound is obtained by requiring that the tunnelling rate away from the EW breaking state with small and positive cosmological constant is not faster than the age of the universe. 
% This can be viewed just as an observational constraint, since we evidently do not live in an AdS space with large negative cosmological constant. 
%But, even without using any bias from our point of view of observers, one can argue that AdS is ruled out because it leads to eventual collapse of space-time. {\it To be done: consider case with matter and radiation; study cosmological evolution.} 
In the current universe, the volume of our past light cone at the current time $T_U$ is
\beq
{\rm Vol}_4(T_U)=0.08 H_\Lambda^4\ ,
\eeq
where $H_\Lambda^2=\Lambda^4/3\bar M_{\rm Pl}^2$ is the Hubble rate produced by the observed vacuum energy
$\Lambda^4$.
Imposing that the probability $p=e^{- {\rm Vol}_4(T_U) \, \Gamma}$ 
that our universe experienced vacuum decay in the past is small enough,
implies an upper bound on the vacuum decay density rate $\Gamma$
\beq
\Gamma<\frac{1}{ {\rm Vol}_4(T_U)} \log \left(\frac{1}{p}\right)
\eeq
and, within the SM, a lower bound on the Higgs mass~\cite{instab}
\beq 
\frac{M_h}{\GeV}
> 111 + 2.8  \left(\frac{M_t}{\GeV}-\Mtexp \right)
-0.9 \left( \frac{\alpha_s(M_Z)-0.1184}{0.0007}\right) \pm 1  .
\label{eq2}
\eeq

\begin{figure}[t]
$$\includegraphics[width=0.83\textwidth]{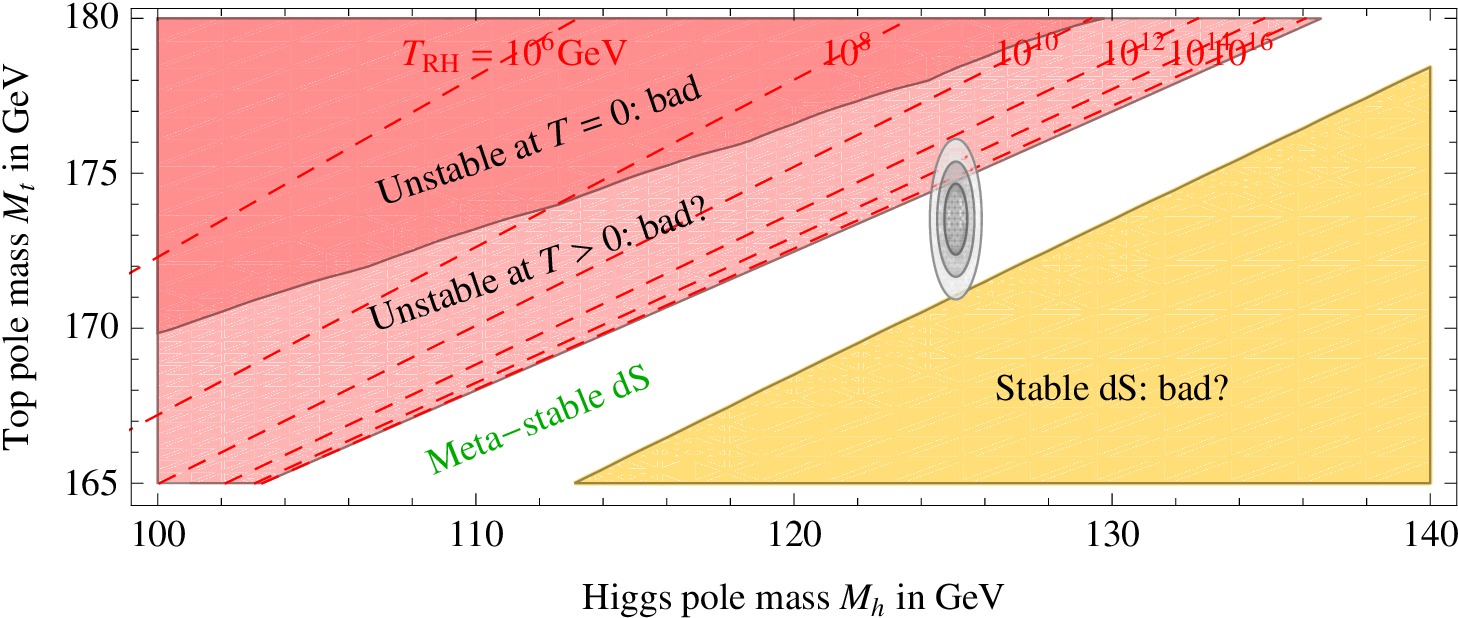}$$  % instead of plot2
$$\includegraphics[width=0.83\textwidth]{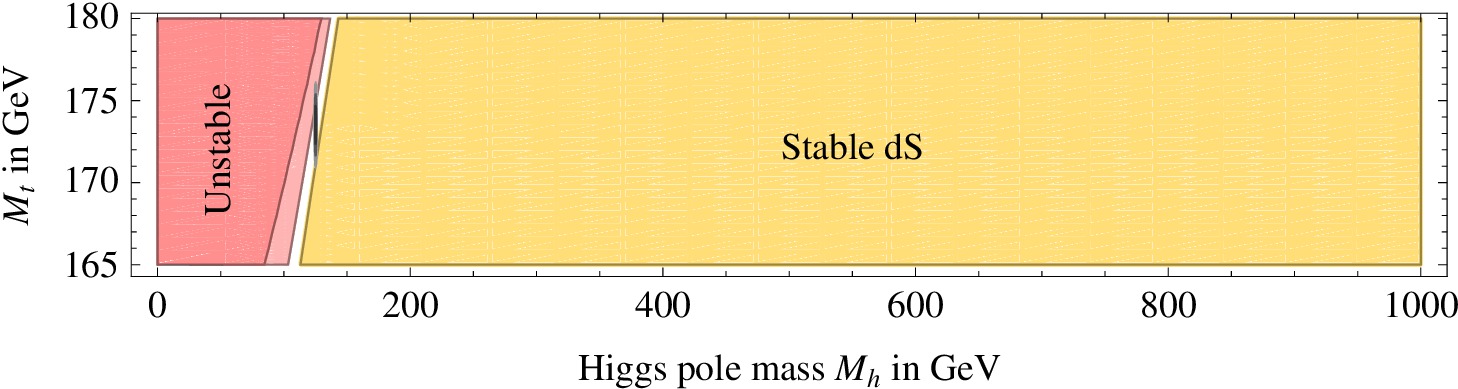}$$
\caption{\em 
The allowed meta-stability window of the Higgs mass.  The ellipse indicates the measured values of $M_h$ and $M_t$.
The orange region is excluded by assuming that ``stable dS" is unacceptable. 
The red region is excluded by vacuum decay at zero temperature.
The pale-red region is excluded by the requirement that the universe must have been hot in the past
(the dashed red curves show boundaries for different values of the reheating temperature).
The bottom panel shows the same result in the full range of a-priori possible Higgs masses, in order to emphasise the smallness of the surviving meta-stability region.}
%\caption{\em  {From the point of view taken in this paper, the natural window for the Higgs mass would have been everywhere up to the perturbative unitarity bound of $WW$ scattering, at about $800\GeV$. Then, the allowed window in which the Higgs vacuum is unstable, but has not yet decayed, in particular in the case of a hot universe, is remarkably, and suggestively, small: about 1\% of the region.  We take this plot as a suggestion that nature, for quantum gravity reasons, might wish to forbid a de Sitter space ever to be realized in our universe. } %Same as before, but from a PR point of view: ``Wow, how small the allowed window is...!
%}
\label{fig1} \label{fig2} 
\end{figure}

A stronger lower bound is obtained from the requirement that the universe underwent a hot phase. There are good reasons to believe that the universe has been very hot at an early epoch. Indeed, processes such as inflation and leptogenesis suggest that the primordial universe reached high temperatures.  We have seen in section~\ref{postI}
how a large reheating temperature helps in forcing the Higgs to its weak scale meta-stable minimum.
However, such high temperatures could have prematurely destabilised the Higgs metastable vacuum. The requirement that this did not happen implies
\beq 
\frac{M_h}{\GeV}
> 124.2 - \frac{190}{\log^2_{10} \frac{T_{\rm RH}}{\GeV}}
 + 2.0  \left( \frac{M_t}{\GeV}-\Mtexp \right)
-0.6\left(  \frac{\alpha_s(M_Z)-0.1184}{0.0007}\right) \pm 1.
\label{eq3}
\eeq
Equations~(\ref{eq1}) and (\ref{eq3}) define a fairly narrow range of possible Higgs masses (see fig.~\ref{fig1}).
Loosely speaking, one might claim that quantum-gravity favours
\beq {122\GeV <  M_h < 129.4\GeV\qquad \hbox{for $M_t = \Mtexp\GeV$}}
.\eeq
Given that the Higgs mass is now  precisely measured, one can better use $M_h$ as input and predict the top mass in the range
\beq
\bbox{171\GeV < M_t < 175 \GeV.}
\eeq
The coincidence that the Higgs and top masses are within the predicted range can be viewed as an indirect indication that nature took the opportunity offered by the Higgs to avoid the problem of an asymptotic dS space.

%Here we could add a discussion on the impact of scanning the top quark mass. At this stage, this does not seem very motivated because: {\it (i)} It does not add any new qualitative features (there is always a correlation between top and Higgs masses); {\it (ii)} The minimal hypothesis is to have the Higgs potential parameters scan while the other SM parameters stay put.

%At first sight, it looks that thermal effects give a stronger bound because $T_{\rm RH}\sim \sqrt{H \mp}$, larger than $H$.
%However, we may consider the possibility that $H$ during inflation is different than the value at the end of inflation. 
%At any rate, the problem is interesting and maybe we can find something new. 
%One aspect that has not been discussed in the literature is the impact of a ``thermal" mass of the Higgs induced by the Gibbons-Hawking temperature during inflation.  

\section{Conclusions}\label{concl}
Assuming that the SM holds up to large energies, we
studied under which conditions the cosmological evolution does not disrupt the electroweak vacuum,
in spite of the presence of an instability of the SM effective Higgs potential $V(h)$ at field values $h > h_{\rm max}$.

\smallskip

As a preliminary step, in section~\ref{Vxi} we clarified the gauge-dependence of the effective potential.
The Nielsen identities show that the gauge-dependence of the effective action corresponds to different
ways of parameterising the same physics in field space: the physical content of the effective action is gauge-independent.
We have shown how this implies that the classical equation of motion (as well as the related Langevin and Fokker-Planck equations used later)
are gauge-independent, because the gauge-dependence of the effective potential is compensated by the gauge-dependence of the
kinetic term.  Furthermore, we showed how, in the basis in which the kinetic term is canonical, the full effective potential becomes gauge-independent in the limit in which only the leading-log corrections are retained (which, for our purposes, is a very good approximation, see fig.~\ref{fig:VSMxi}).
For the present best-fit values of the SM parameters one has
$ h_{\rm max} \approx 5\times 10^{10}\GeV$, but $ h_{\rm max}$
can  vary by orders of magnitude if the top mass $M_t$ is varied within its uncertainty band.

\medskip

Next, in section~\ref{HI} we studied Higgs fluctuations during inflation. In our study, we also took into account the effect of a Higgs mass $m^2$ induced, during inflation, either by a mixed quartic coupling between the Higgs and the inflaton, or by a non-minimal Higgs coupling to gravity $\xi_H$. Not being radiatively stable, such a coupling is expected to be generally present and leads to   $m^2 =  -12\xi_H H^2$, where $H$ is the Hubble constant during inflation, given by 
$ H\approx  8\times 10^{13}\GeV\sqrt{{r}/{0.1}}$.
%, we can relate the ratio $H/ h_{\rm max} $, which is the crucial parameter in our analysis, to the observable tensor-to-scalar power ratio $r$ according to $H/ h_{\rm max}\approx 1600 \sqrt{{r}/{0.1}}$.
Present cosmological data constrain $r \circa{<} 0.1$, but future measurements will have greater sensitivity.
  
If $m^2 < 9 H^2/4$ the Higgs undergoes inflationary quantum fluctuations, which we computed
via a Langevin equation that bypasses the need of imposing appropriate boundary conditions encountered in the Fokker-Planck equation used in previous works.
We find that the parameter space in the plane $H/ h_{\rm max} $ vs $\xi_H$ splits into 3 regions (see fig.~\ref{RHbounds}):
\begin{itemize}
\item[$\color{green}\bullet$] `green' region, where the Higgs remains below its  instability scale at the end of inflation, and thus inflationary fluctuations do not destabilise the electroweak vacuum.
\item[$\color{orange}\bullet$] `orange' region, where the Higgs can probe field values above the instability scale ($|h|> h_{\rm max}$), but quantum fluctuations dominate over classical evolution and prevent the Higgs from falling into its true AdS minimum; the ultimate fate of the Higgs is determined by post-inflationary dynamics.
\item[$\color{red}\bullet$] `red' region, where the Higgs fluctuates above the instability scale and falls down into its true minimum, presumably ending inflation in that patch of space.
\end{itemize}
In section~\ref{sec4} we followed the evolution of the Higgs field through the reheating process, in order to assess the viability of parameters corresponding to the `orange' region. Thermal effects can rescue the Higgs field, letting it slide towards the origin of the SM potential, if the reheating temperature after inflation $T_{\rm RH}$ is sufficiently large. We derived upper bounds on $H/ h_{\rm max} $, for given $T_{\rm RH}$, as shown in fig.~\ref{TmaxvsH}. The result is that thermal effects can easily make the `orange' region cosmologically acceptable.

\smallskip

On the contrary, we found that the `red' region is problematic.
By approximating the large-field Higgs patches as spherical bubbles with small thickness, we could perform
a general relativistic computation in order to determine whether such bubbles shrink or expand.
The computation addresses several relevant counter-intuitive phenomena.
While we identified mechanisms that can make some of the bubbles innocuous
(small bubbles with low wall velocity shrink, bubbles with large tension expand hidden behind a black-hole horizon),
we find that inflation produces Higgs `bubbles' that expand, at least as long as they are in the computable sub-Planckian regime.
During inflation these bubbles are not lethal, as they remain behind a de Sitter horizon and are diluted by space expansion.
However, after inflation they keep on growing at the speed of light, eventually swallowing all space.
Therefore, we must require that inflationary fluctuations do not produce
any of these regions in our past light-cone.

\medskip

This leads us to our final result: the `red' region of fig.~\ref{RHbounds} is excluded.
If $|\xi_H |< 0.01$ one needs a Hubble constant smaller than $0.045~ h_{\rm max}$.
This constraint gets weaker (stronger) for negative (positive) $\xi_H$: {\it e.g.}\ $H < 10^4~ h_{\rm max}$ for $\xi_H \approx - 0.03$.
A small negative $\xi_H$ however leads to a new, super-Planckian instability of the SM potential in the Einstein frame, see fig.~\ref{VSMxi}.
In a similar way, a direct coupling of the inflaton to the Higgs could also relax the limits on $H$ but, contrary to the case of $\xi_H$, it does not lead to any instabilities at large field values.

\bigskip

Finally,  in section~\ref{QG} we explore a new speculative idea.
Assuming that the present acceleration of the universe is due to a small cosmological constant, 
and accepting the conjecture that quantum gravity is ill-defined in a de Sitter space, we argue that vacuum decay is a necessary way out for the universe.
We show that vacuum decay triggered by the Higgs instability is fast enough to resolve this conceptual problem.

Basically the SM phase diagram in the $(M_t , M_h)$ plane is reinterpreted:
the instability region remains `bad', the stability region becomes `bad', 
and the only `good' region is the narrow meta-stability strip of parameter space.
%the cosmological Higgstory, as studied in the first part of our paper, 
%can happily end in the meta-stable electroweak minimum.
As discussed in section~\ref{postI}
a large enough reheating temperature may play an important role in the universe, and the requirement that thermal effects do not induce an excessively fast vacuum decay  
provides a further restriction in the  Higgs and top masses, as shown in fig.~\ref{fig1}.
One could view this restriction as a remarkably precise post-diction for the Higgs or top masses.

\small

\subsubsection*{Acknowledgments}
We thank M. Garny and T. Konstandin for very useful discussions and Gino Isidori and Joan Elias-Mir\'o for participating in the early stages of this work.
J.R.E. thanks CERN for hospitality and partial financial support.
This work was supported by the ESF grant MTT8. The work of J.R.E. has
been supported by the Spanish Ministry MEC under grants FPA2013-44773-P, FPA2012-32828; by the Generalitat grant 2014-SGR-1450 and by the Severo Ochoa excellence program of MINECO (grant SO-2012-0234).
 L.S.~is supported by the DOE Early Career Award DE-FG02-12ER41854, by the National Science Foundation under PHY-1068380, and by
 the European Commission under the ERC Advanced Grant BSMOXFORD 228169.
The work of N.T. has been co-financed by the European Union (European Social Fund ESF) and Greek national 
funds through the Operational Program ``Education and Lifelong Learning" of the National Strategic Reference 
Framework (NSRF) - Research Funding Program: ``THALIS. Investing in the society of knowledge through the 
European Social Fund''.

\appendix

\section{Evolution of  bubbles}\label{AdS}

In this appendix we study the evolution of a region of true vacuum with 
negative vacuum energy density, which lies 
within the false-vacuum asymptotically flat or de Sitter space.
The basic question is whether this region, which we call a bubble (assuming
spherical symmetry), expands or contracts. 
For an outside observer, the presence of the bubble has a gravitational effect
equivalent to the presence of a central mass. As a result, the exterior metric 
is of the Schwarzschild or Schwarzschild-de Sitter (SdS) type.  
We study an idealised configuration with constant vacuum energy density in the
interior and exterior of the bubble, as well as constant surface tension. 
The study of a realistic bubble, corresponding to a space-time dependent Higgs configuration, is not possible analytically. However, we believe that our treatment
captures the main aspects of the problem, 
determined essentially by the difference in the
local energy density of the Higgs field on either side of the bubble wall.

We employ the  thin-wall approximation and
parameterise the wall and the inner and outer space as described in section~\ref{bubble}. The interior of the bubble is assumed to
be a part of anti-de Sitter spacetime, described by the metric (\ref{ads-metric}).
The exterior of the bubble is described by the metric (\ref{schw-metric}). 
The case 
$V_{\rm out} \not= 0$ corresponds to the SdS 
spacetime, while 
the case $V_{\rm out}=0$ to 
an exterior Schwarzschild metric.
The metric on the wall is given by eq.~(\ref{wall-metric}).

\subsection{Matching the geometries} \label{AdS1}

The metric must be continuous over the whole space. This means that 
\begin{eqnarray}
f_{\rm in}(R) \, \dot{\eta} &=& \ex_1\left(\dot{R}^2+f_{\rm in}(R) \right)^{1/2},
\label{doteta} \\
f_{\rm out}(R) \, \dot{t}&=&\ex_2\left(\dot{R}^2+f_{\rm out}(R) \right)^{1/2},
\label{dott} \end{eqnarray}
where $\ex_1=\pm 1$, $\ex_2=\pm 1$ are possible sign choices
and a dot denotes a derivative with respect to $\tau$.
Since $f_{\rm in}\geq 1$, the value of $\ex_1$ determines the relative flow of the two timelike
coordinates $\eta$ and $\tau$. It is natural to make the choice $\ex_1=1$, which is also the only consistent choice (see below). 
We consider only this value in the following.
The relation between $t$ and $\tau$ is more complicated because $f_{\rm out}$ can be negative. We follow the convention of~\cite{blau}, according to which the
flow of proper time is such that future-directed world lines correspond to a growing Kruskal-Szekeres coordinate $V$ (so that $\dot{V}>0$).

\medskip

The matching of the two regions can be done following~\cite{blau}.
The four-velocity of a point on the wall is 
$U^\mu_{\rm in}=(\dot{\eta},\dot{R},\vec{0})$ and 
$U^\mu_{\rm out}=(\dot{t},\dot{R},\vec{0})$
in each of the frames.
A  spacelike vector $\xi^\mu$ perpendicular to the wall must be orthogonal to $U^\mu$. 
In order to determine it uniquely, we have to specify whether it points towards the interior or the exterior.
For spaces with horizons, such as the exterior space, 
we adopt the convention of~\cite{blau}. We assume that 
$\xi^\mu$ points towards increasing values of the Kruskal-Szekeres coordinate $U$. We also assume that the exterior
lies on the `right' of the wall in the Penrose diagram. 
 For the AdS space in the interior, we assume that $\xi^\mu$ points towards increasing values of the global coordinate $r$. With these conventions $\xi^\mu$ 
points from the interior towards the exterior.
It is given by 
\begin{eqnarray}
\xi^\mu_{\rm in}&=&\left(\frac{\dot{R}}{f_{\rm in}},f_{\rm in} \, \dot{\eta},\vec{0}\right)
=\left(\frac{\dot{R}}{f_{\rm in}},(f_{\rm in}+\dot{R}^2)^{1/2},\vec{0}\right)
\label{perp1} \\
\xi^\mu_{\rm out}&=&\left(\frac{\dot{R}}{f_{\rm out}},f_{\rm out}\, \dot{t},\vec{0}\right)
=\left(\frac{\dot{R}}{f_{\rm out}},\ex_2(f_{\rm out}+\dot{R}^2)^{1/2},\vec{0}\right),
\label{perp2} \end{eqnarray}
in each of the frames. It has been normalized to $-1$.

\bigskip

The junction conditions connect the discontinuity in the extrinsic curvature to the surface tension:
\be
(K_{\rm out})^i_j-(K_{\rm in})^i_{~j}=-4\pi\sigma G\delta^i_{~j}.
\label{junction} \ee
We  match the $\theta\theta$ component of the extrinsic curvature
(the other components give equivalent relations), which is 
$K_{\theta\theta}= \xi^\mu\partial_\mu r^2/2$.
Evaluated on either side of the wall, it is given by
\begin{eqnarray}
(K_{\rm in})_{\theta\theta}&=&(f_{\rm in}+\dot{R}^2)^{1/2}R\equiv \beta_{\rm in} R,
\label{extr1} \\
(K_{\rm out})_{\theta\theta}&=&\ex_2(f_{\rm out}+\dot{R}^2)^{1/2}R\equiv \beta_{\rm out} R.
\label{ext2} \end{eqnarray}
Thereby, the  $\theta\theta$ matching condition is
\be
\ex_2(f_{\rm out}+\dot{R}^2)^{1/2}-(f_{\rm in}+\dot{R}^2)^{1/2}=\beta_{\rm out}-\beta_{\rm in}=-4\pi G\sigma  R.
\label{junction12} \ee

\subsection{Bubbles in asymptotically flat spacetime} \label{AdS2}

We consider first the case $V_{\rm out}=0$, which corresponds to 
an exterior Schwarzschild metric. We shall discuss later the case 
$V_{\rm out} \not= 0$, corresponding to a SdS 
spacetime.

The square of eq.~(\ref{junction12}) can be put in the form:
\be
2GM=\left(\kappa^2-\frac{1}{\ell^2_{\rm in}} \right)R^3+2\ex_2 \kappa R^2
\left(1-\frac{2GM}{R}+\dot{R}^2 \right)^{1/2}.
\label{square1} \ee
For large $R$ 
and non-relativistic wall velocity, the last parenthesis becomes equal to 1.
The resulting expression indicates that the mass $M$ of a large-radius bubble  
is dominated by a volume contribution 
proportional to $\kappa^2-1/\ell^2_{\rm in}$. 
The total volume effect can be negative or positive, depending on the value of
\beq 
\ex \equiv  {\rm sign} \left( \frac{1}{\ell^2_{\rm in}}-\kappa^2 \right).
\label{epss}  
\eeq
As a result, it is possible for the total mass $M$ to become negative.

Solving eq.~(\ref{junction12}) for $M$ one finds a result that,
for $\ex_2=1$, has a simple Newtonian interpretation:
\be
2GM=-\left(\frac{1}{\ell^2_{\rm in}}+\kappa^2 \right)R^3+2\kappa R^2
\left(1+\frac{R^2}{\ell^2_{\rm in}}+\dot{R}^2 \right)^{1/2},
\label{square} \ee
with $\kappa\equiv 4\pi G \sigma$. 
For small $R$,
the mass $M$ attributed to the bubble of AdS 
by an outside observer contains a volume term 
proportional to $-1/\ell^2_{\rm in}-\kappa^2$. The contribution $-1/\ell^2_{\rm in}$ corresponds to the 
vacuum energy density, while $-\kappa^2$ reproduces correctly 
the gravitational self-energy of the wall. The second term in eq.~(\ref{square})
can be expanded for small $R$ and $\dot{R}$. One recovers the surface energy of the bubble, with nonrelativistic correction, and the surface-volume binding energy \cite{Guth:1982pn}.
The leading term for small $R$ is the positive surface energy $\sim \kx R^2$, which
indicates that small bubbles tend to collapse in order to minimise their energy. 
The case $\ex_2=-1$ does not lead to solutions with a simple Newtonian 
interpretation, even though
it contains acceptable configurations for the global geometry.
%The complete analysis for both signs of $\ex_2$ is presented 
%in detail in the following subsection.

By squaring eq.~(\ref{junction12}) a second time, 
we can express the `kinetic energy' $\dot R^2$
in terms of a conserved `energy' $E$ and an effective `potential energy'.
We express the result as the equation
for the one-dimensional motion of a particle in a `potential' $V$
\be
\left(\frac{d\Rt}{d\ttau} \right)^2+V(\Rt)=E,
\label{eom} \ee
where
\be
V(\Rt)=-\left( \frac{\ex_m+\ex \Rt^3}{\Rt^2}\right)^2-\ex_m\frac{\gamma^2}{\Rt},
\qquad
E=-\frac{\kappa^2}{G^2M^2\rho^4}
\label{energ}\label{pot} \ee
and
\be
\epsilon_m={\rm sign}(M).
\label{massign} \ee
The dimensionless  `coordinate' variable $\Rt$ and the `time' variable $\tilde\tau$ are defined as
\beq
\Rt= \rho R,\qquad
\label{zz} 
\ttau=\frac{2\kappa}{\gamma^2}\tau.
\label{tautil}
\eeq
The parameter
$\rho$, defined as 
\beq
\rho^3 = \frac{1}{2G|M|}\left| \frac{1}{\ell^2_{\rm in}}-\kappa^2\right|,
\label{rrho} \eeq
sets a characteristic inverse length-scale, while
$\gamma$ parameterises the surface-energy term in $V$:
\beq
\gamma=\frac{2\kappa}{\left| {\ell^{-2}_{\rm in}}-\kappa^2\right|^{1/2}},\qquad
\hbox{i.e.}\qquad
\kappa^2=\frac{1}{\ell^2_{\rm in}}\frac{\gamma^2}{\gamma^2+4\ex}.
\label{kapgam} 
\label{gamm}
\eeq

The form of the solutions of eq.\ (\ref{eom}) can be revealed more easily through the following observations:
\begin{itemize}
\item
The sign $\ex_2$ disappeared when performing the second squaring, so that
eq.~(\ref{eom}) describes the solutions of eq.~(\ref{junction12}) with both
values of $\ex_2$. 
We can rewrite eq.~(\ref{junction12}) in terms of the new parameters as
\be
\beta_{\rm in}=\beta_{\rm out}+4\pi G \sigma R = \frac{G|M|\rho^2}{\kappa}\frac{1}{\Rt^2}\left(\ex_m+\ex \Rt^3+\frac{\gamma^2}{2}\Rt^3 \right),
\label{cond1} \ee
where we have used eq.~(\ref{square}). For positive-mass bubbles 
($\ex_m=1$) we have $\beta_{\rm in}>0$. This is obvious for 
$\ex=1$. It also holds for $\ex=-1$ , because 
$\gamma^2>4 $ in this case. We conclude that 
the only consistent value for $\ex_1$ for positive-mass bubbles 
is $\ex_1=1$ (the value we assumed).
\item
We can also write 
\be
\beta_{\rm out}=\frac{G|M|\rho^2}{\kappa}\frac{1}{\Rt^2} \left(\ex_m+\ex \Rt^3 \right),
\label{cond2} \ee
from which it is apparent that, for positive-mass bubbles, 
$\beta_{\rm out}$ is positive and $\ex_2=1$
for $\ex=1$, while $\beta_{\rm out}$ changes sign at $\Rt=1$ for $\ex=-1$. 

\item
For negative-mass bubbles ($M<0$ or $\ex_m=-1$) the variable $t$ is always timelike.
The value of $\ex_2$ determines the relative flow of $t$ and $\tau$. It is natural
to make the choice $\ex_2=1$ in this case. The possibility $\ex_2=-1$ does not
lead to a physical solution, as we discuss in subsection \ref{negmas}.

\item
It is apparent from eqs.~(\ref{energ}), (\ref{rrho}) that, for fixed $\ell_{\rm in}$ and $\kappa$, the total energy $E$ is a function of $M$. As a result, the nature of
the various
solutions of eq.~(\ref{eom}) is directly related to the mass of the bubble. 
%For $\ex=-1$, the mass $M_S$ for which $E(M_S)=-\gamma^2=V(\Rt=1)$ is given by the relation
%\be
%M_S=\mb \left(\frac{\gamma^2}{4}-1 \right)^{1/2},
%\label{mss} \ee
%with $\mb=l/(2G)$.

\item
The `potential' is maximal at $\Rt=\Rt_{\rm max}$, given by
\be
2\Rt_{\rm max}^3=\ex_m\left(\ex +\frac{\gamma^2}{2}\right) +\sqrt{\left(\ex+\frac{\gamma^2}{2}\right)^2+8}.
\label{zmm} \ee
%It can be checked easily that $\Rt_{\rm max}>1$.
The value of the `potential' at its maximum is
\be
V(\Rt_{\rm max})=-3 \frac{\Rt^6_{\rm max}-1}{\Rt_{\rm max}^4}.
\label{potmax} \ee
For positive-mass bubbles ($\ex_m=1$) we have $\Rt_{\rm max}>1$.
%The `energy' $E$ equals the maximal `potential energy' for the critical value %$M=M_{\rm cr}$ given by
%The critical mass $\Rt_{cr}$\as{ \Rt is not a mass?}
% for which $E(M_{cr})=V(\Rt_{\rm max})$ is 
%\be
%M_{\rm cr}=\mb \left(\ex+\frac{\gamma^2}{4} \right)^{1/2} %\gamma^3\frac{\Rt^6_m}{3^{3/2}(\Rt^6_m-1)^{3/2}}.
%\label{mcr} \ee
\item
The Schwarzschild radius of a bubble with positive mass $M$ is $r_{H}=2GM$,
which, in terms of the variable $\Rt$, becomes
\be
E=-\frac{\gamma^2}{\Rt_{H}}.
\label{hori} \ee
This relation determines the location of the horizon on a solution of eq.
(\ref{eom}) with given $E$. 
Making use of the definition (\ref{pot}) of the `potential', we can write
\be
E
%=-\frac{\gamma^2}{\Rt_{H}}
=V(\Rt_{H})+\left(\frac{1+\ex\, \Rt_{H}^3}{\Rt^2_{H}}   \right)^2.
\label{horiz} \ee
For $\ex=-1$ the curve $-\gamma^2/\Rt_{H}$, depicting the location of the
horizon, is tangent to the curve $V=V(\Rt)$ at $\Rt=1$. 
For $\ex=1$ the curve for the horizon is always located above the curve for the `potential'.

\end{itemize}

\begin{figure}[!t]
\centering
$$
\includegraphics[width=0.3\textwidth]{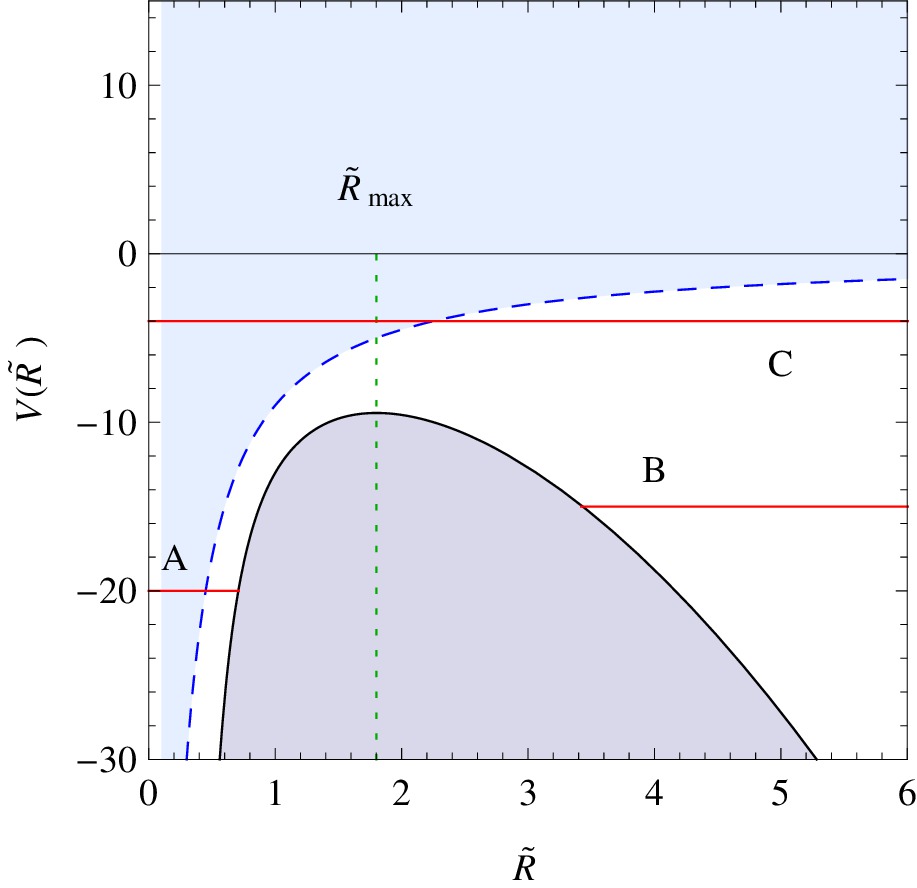}\qquad 
\includegraphics[width=0.3\textwidth]{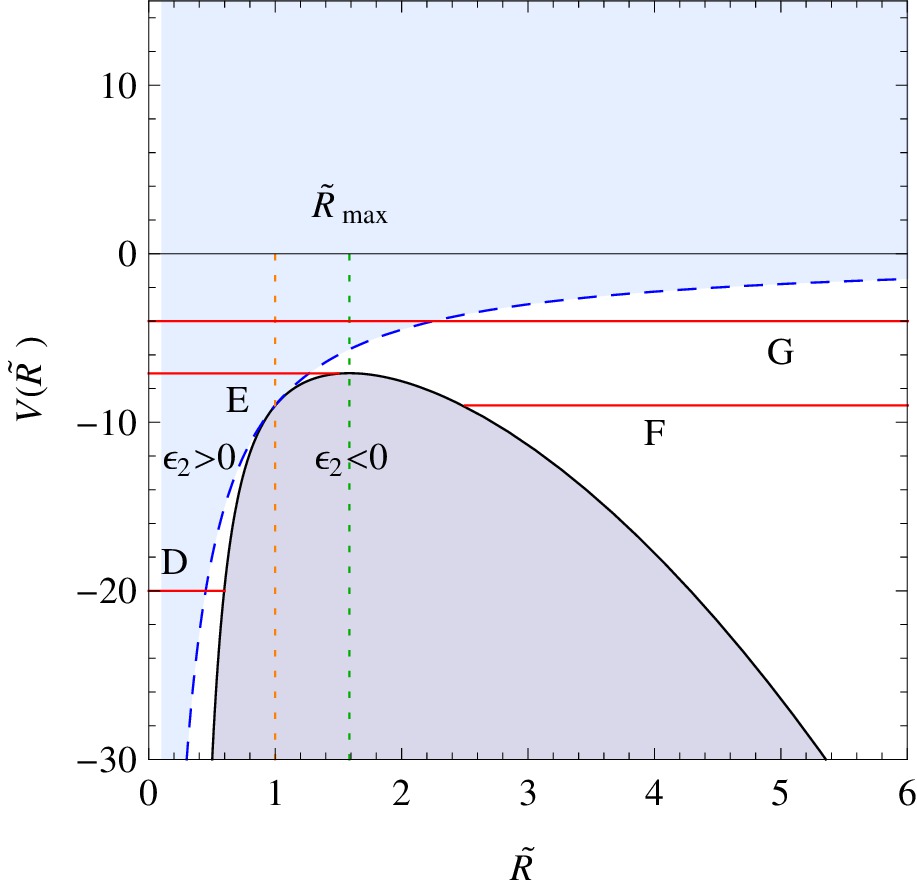}\qquad \includegraphics[width=0.3\textwidth]{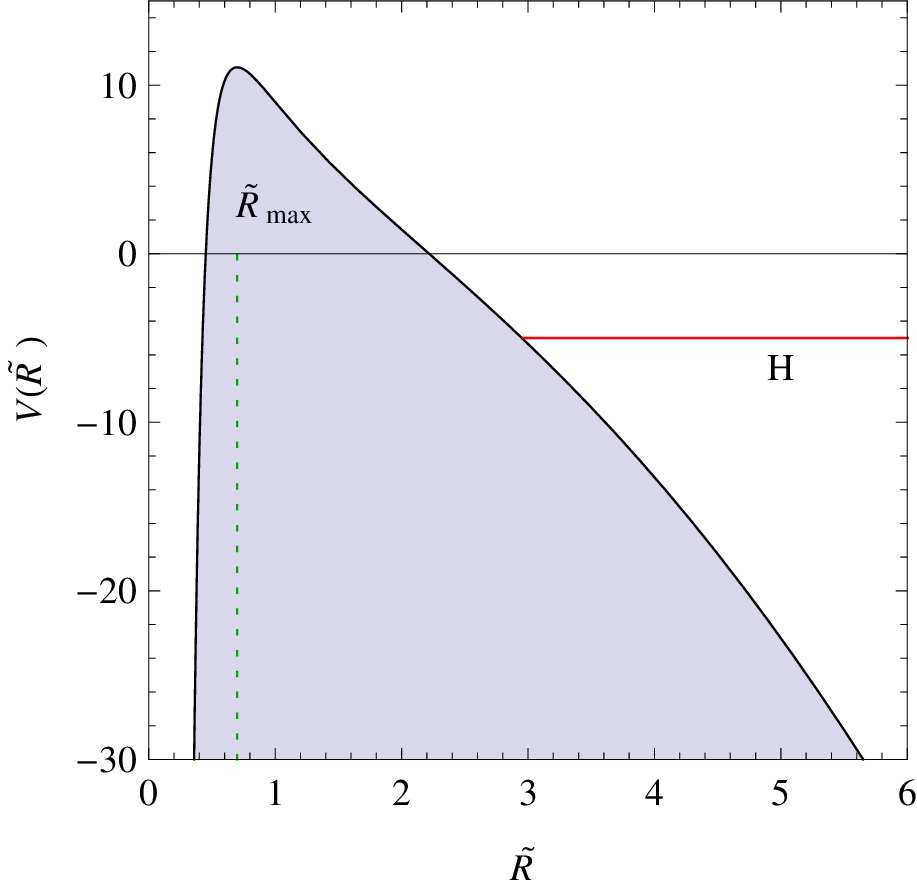}$$
\caption{\em The `potential' of eq.~(\ref{pot}) for 
$\gamma=3$ and $\ex=1$, $\ex_m=1$ (left), for $\ex=-1$, $\ex_m=1$ (middle), and for
$\ex=1$, $\ex_m=-1$ (right).}
\label{h1}
\end{figure}

The above features are depicted graphically in fig.~\ref{h1}. 
 The solid black curve depicts the `potential' $V(\Rt)$, which has a maximum at $\Rt=\Rt_{\rm max}$. 
The dashed blue curve indicates the location of the horizon. 
For $\ex=1$ (i.e. $1/\ell^2_{\rm in}>\kappa^2$) the curve for the 
horizon is always located above the `potential'.  
For $\ex=-1$  (i.e. $1/\ell^2_{\rm in}<\kappa^2$) the curve for the horizon is tangent to the `potential' at $\Rt=1$. The function $\beta_{\rm out}$ changes sign at this point. In the centre
plot we have separated with a red vertical dashed line the regions in which 
$\ex$ (and $\beta_{\rm out}$) has opposite signs.   
%For points below the 
%dashed blue line, which correspond to the region of space outside the horizon, the %derivative $\dot{t}$ also turns from positive to negative as
%$\Rt$ increases through this line.

\smallskip

The various types of trajectories can be deduced from these plots.
We plot a few lines with constant $E$ that
stop when $E=V$: at this point $\dot R=0$ and the motion of the wall is reversed.
There are various types of trajectories for which the bubble expands indefinitely. 
If $\ex=-1$, such evolution can be obtained only for $\ex_2=-1$.

\subsection{Evolution of positive-mass bubbles} \label{AdS3}

We consider first the case $M>0$, or, equivalently, $\ex_m=1$. 
The evolution of the wall is best depicted using Penrose diagrams. 
The diagrams for the most characteristic types of wall evolution are presented in figs.~\ref{A2}, \ref{C1}, \ref{G1}. 
Each figure contains a pair of diagrams.
In each pair, the left diagram 
depicts AdS space, which has the simple structure of a cylinder, with the line marked $r=0$ corresponding to its centre and the line
$r=\infty$ to conformal infinity. The right diagram represents the 
complete Schwarzschild geometry, which includes two singularities, marked $r=0$, and the corresponding horizons. 
The black continuous curve in each diagram denotes the trajectory of the wall. 
Thick black lines denote singularities, dashed lines horizons and dotted lines denote conformal infinities.
The total space is constructed by patching the part of the left diagram on the left of the wall with the part of the
right diagram on the right of the wall. The shaded areas correspond to the
parts that must be eliminated in order to join the remaining parts along the
wall trajectory.

The crucial relation for the fate of space is between the gravitational
self-energy of the wall $\kappa^2$ and the vacuum energy $-1/\ell^2_{\rm in}$. Naively, 
one expects that, if $1/\ell^2_{\rm in} > \kappa^2$ (i.e. $\ex=1$), large bubbles will grow indefinitely
because the system gains energy in the process. In the opposite case with
$1/\ell^2_{\rm in} < \kappa^2$ (i.e. $\ex=-1$), the bubbles will shrink for similar 
energetic reasons.
These simple expectations, which are based on Newtonian intuition, are only
partly fulfilled in the complete analysis. More complicated scenarios are
realised as well.

\subsection*{Case $\bma{\ex=1}$}

\subsubsection*{Small  bubbles with small initial wall velocity do not expand}

\begin{figure}[t]
$$
\includegraphics[width=70mm,height=70mm]{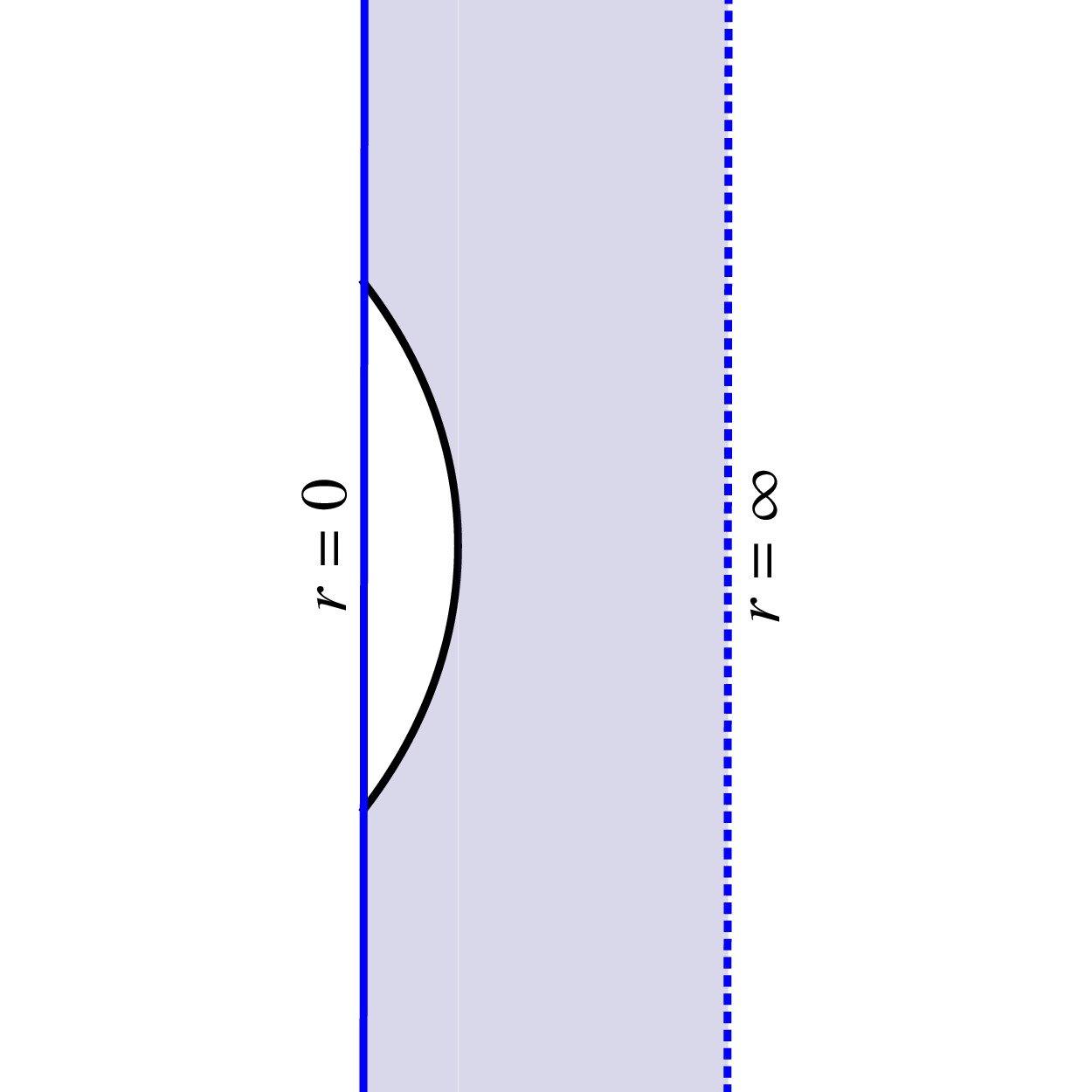}\qquad
\includegraphics[width=70mm,height=70mm]{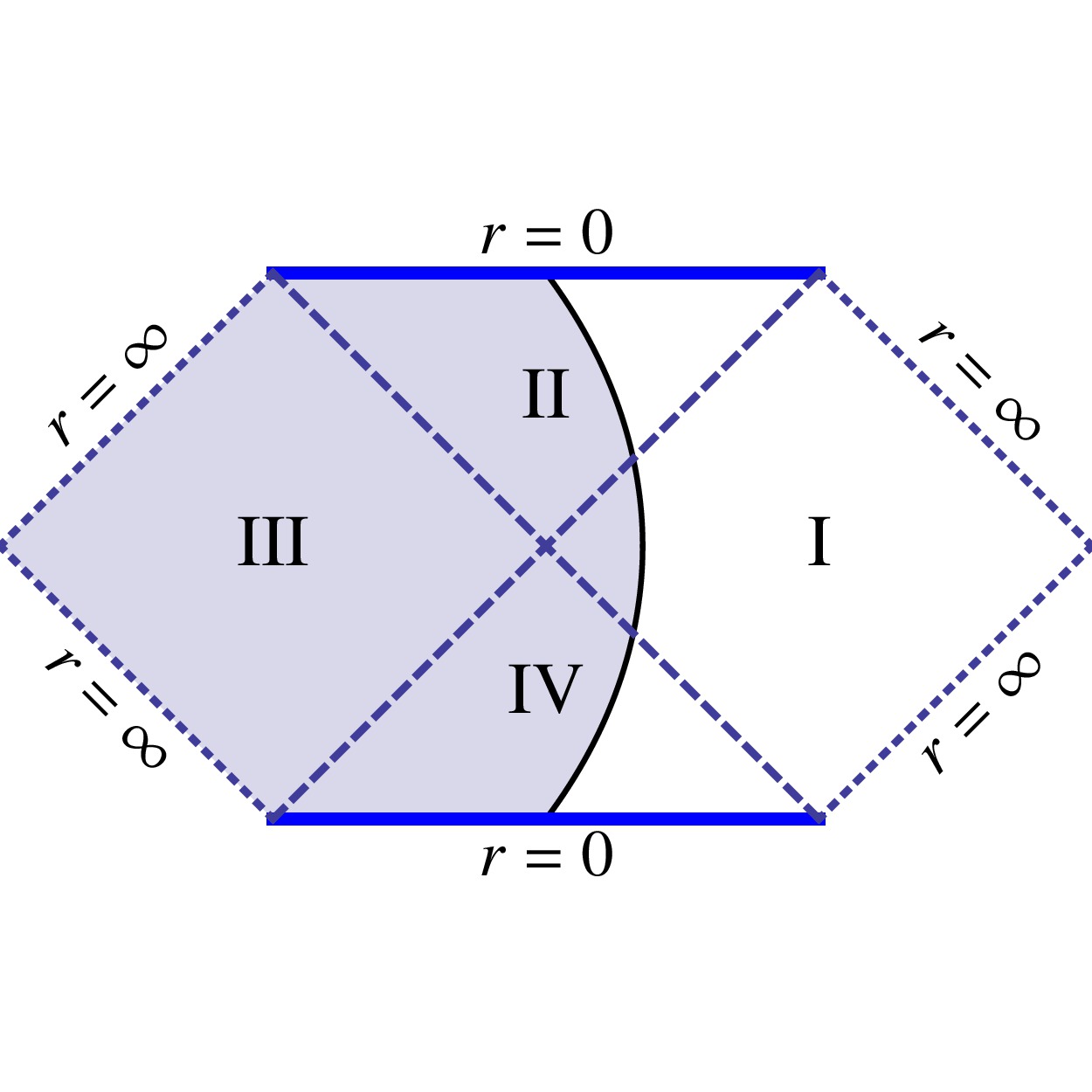}$$
\caption{{\bf Small  bubbles with small initial wall velocity do not expand}.
Left: {\em The wall trajectory corresponding to line A of fig.~\ref{h1}, in AdS space}.
Right: {\em The wall trajectory corresponding to line A of fig.~\ref{h1},  in the Schwarzschild geometry.}
\label{A2}}
\end{figure}

\begin{figure}[t]
$$
\includegraphics[width=70mm,height=70mm]{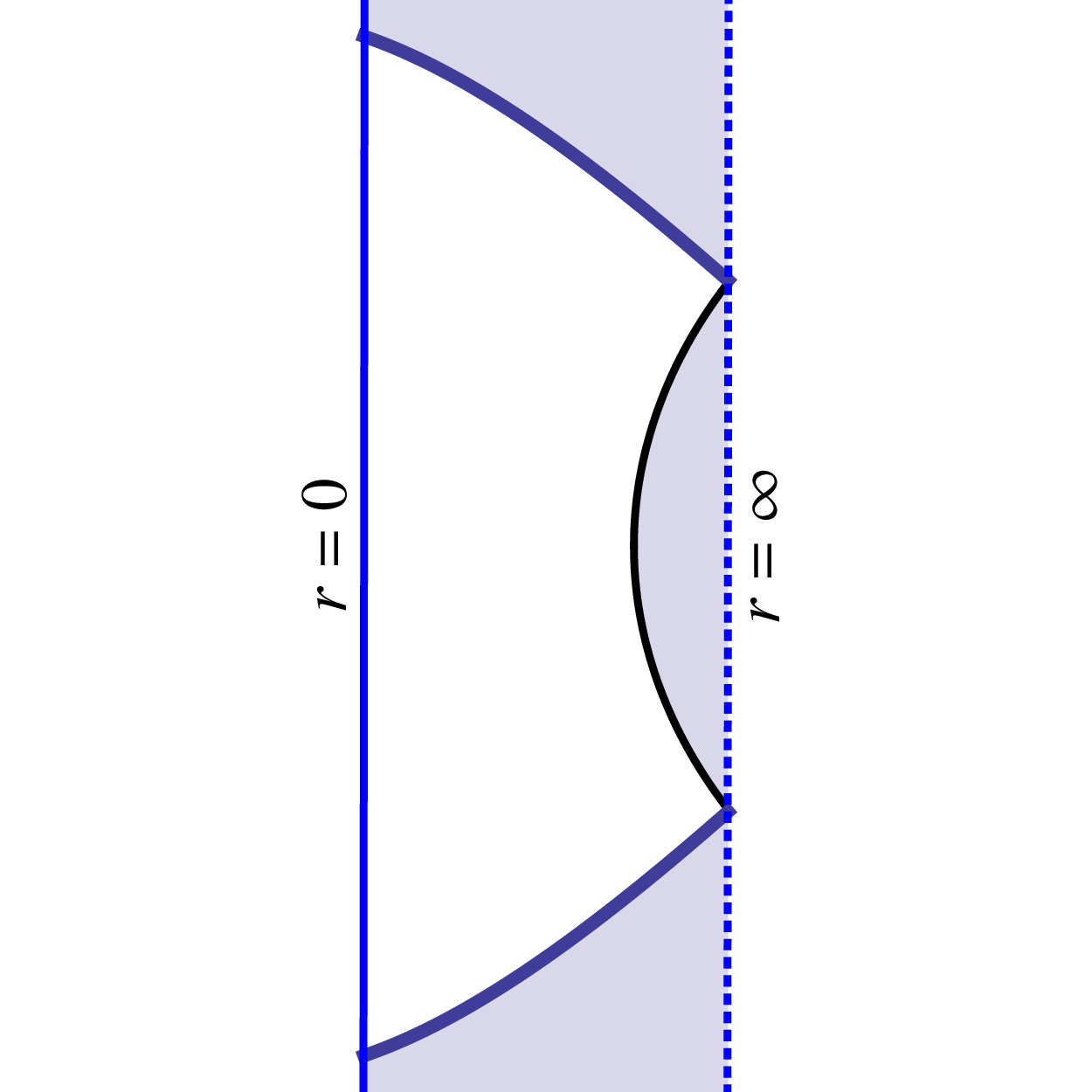}\qquad
\includegraphics[width=70mm,height=70mm]{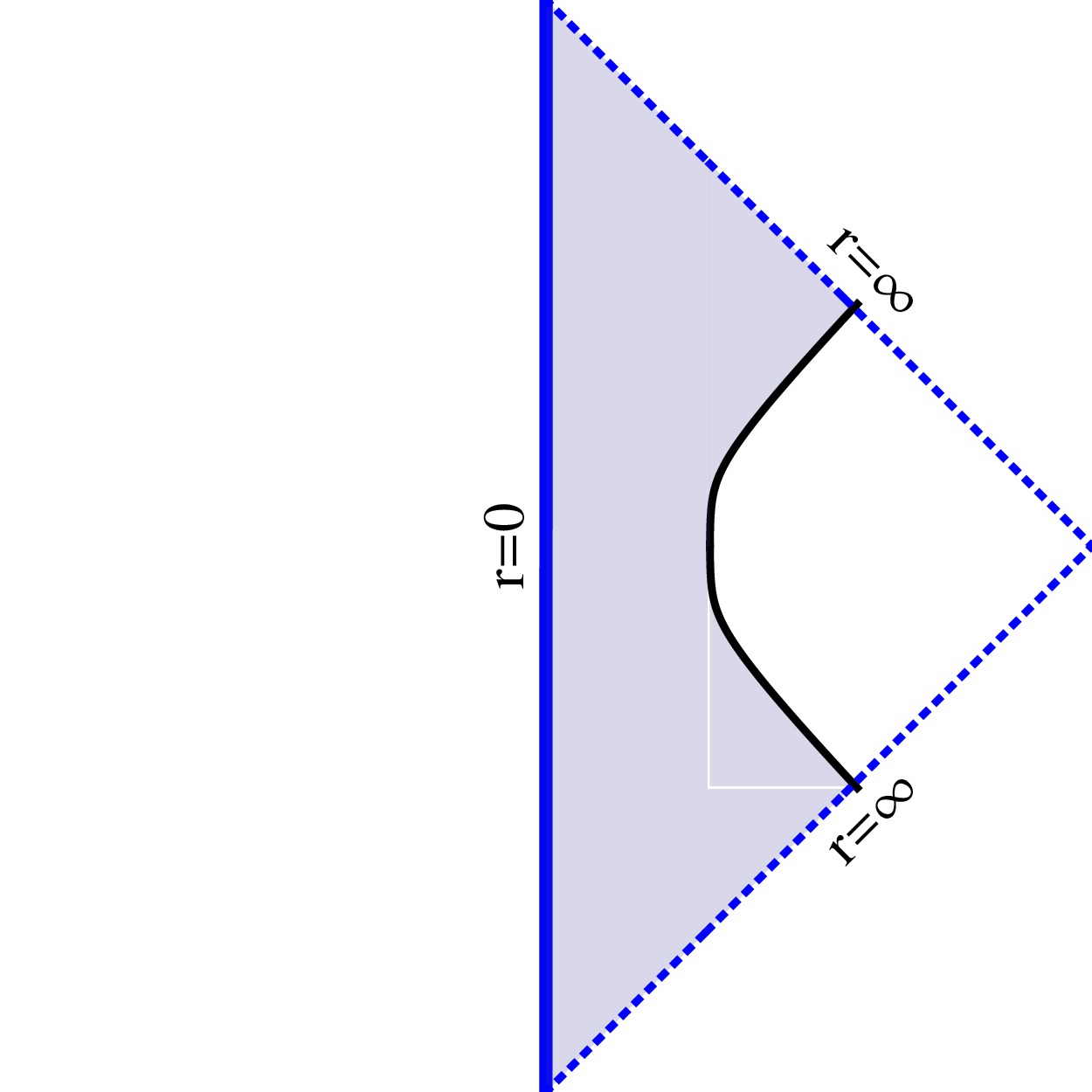}$$
\caption{{\bf The evolution of negative-mass bubbles}.
Left: {\em The wall trajectory corresponding to line H of fig.~\ref{h1}, in AdS space}.
Right: {\em The wall trajectory corresponding to line H of fig.~\ref{h1},  in the negative-mass Schwarzschild geometry.}
\label{H1}}
\end{figure}

Line A of fig.~\ref{h1} describes the evolution of a bubble whose volume 
energy receives its largest contribution from 
the negative vacuum energy density ($1/\ell^2_{\rm in}>\kappa^2$). However,
the surface contribution to the energy, arising from the wall tension, is the 
dominant factor and tends to make the bubble shrink. 
The bubble has small initial wall velocity, which prevents it from 
evolving to a size sufficiently large for the volume contribution to the
energy to dominate. As a result the surface tension wins: the bubble reaches a maximum size and subsequently collapses 
falling within its own horizon. 

The space corresponding to this solution is
depicted in fig.~\ref{A2}. It results from eliminating the shaded areas in
each of the two Penrose diagrams and patching the remaining parts along the
wall trajectory. 

\subsubsection*{Small  bubbles with large initial wall velocity expand}

Line C of fig.~\ref{h1} corresponds 
to a bubble with similar characteristics as in the previous case, but with
much larger wall `kinetic energy'. This is is apparent by the fact that the
total energy is less negative. We can consider a bubble that starts very small
(with almost vanishing $r$ or $\Rt$). Even though the surface contribution to the
`potential energy' dominates, the initial velocity is sufficiently large for the
bubble to expand. Eventually the bubble develops a size for which the volume
contribution to the `potential energy' becomes dominant over the surface
contribution. From this point on, the bubble expands indefinitely, with its
wall approaching asymptotically the speed of light.

The corresponding evolution of space is depicted in fig.~\ref{C1}. 
After a finite time $\eta$ the wall reaches the boundary of AdS space. 
As the AdS boundary is timelike, there is 
a Cauchy horizon, beyond which the spacetime cannot be determined without additional boundary conditions. It is expected that, within the full theory beyond the thin-wall limit, 
a spacelike singularity develops before the Cauchy horizon \cite{freivogel}.  This is depicted by a thick blue line
in the left diagram of fig.~\ref{C1}. From a mathematical point of view, the solution also describes the reverse process.

\subsubsection*{Large bubbles expand}

Line B of fig.~\ref{h1} 
describes the evolution of a bubble so large 
that its surface tension is irrelevant.
The bubble starts with infinite radius, shrinks
to finite size and then re-expands. There are two singularities in the 
Penrose diagram of AdS space, starting from the points at which
the wall trajectory reaches the boundary \cite{freivogel}. The whole trajectory lies with the region I of the Schwarzschild geometry.

\subsection*{Case $\bma{\ex=-1}$}

\subsubsection*{Small  bubbles with small initial wall velocity do not expand}

Line D of fig.~\ref{h1} describes evolution very similar to that 
for line A. The contribution from the surface tension dominates the
`potential energy', while the `kinetic energy' is small. The bubble 
expands up to a certain size, and subsequently recollapses. The space
is described by Penrose diagrams very similar to those of fig.~\ref{A2}.

Line E of fig.~\ref{h1} describes a similar scenario, but now
the extrinsic curvature $\beta_{\rm out}$ (or, equivalently $\ex_2$) 
changes sign during the evolution. 
This implies that the wall trajectory  crosses  regions IV, III and II of the Schwarzschild geometry instead
of the regions IV, I, II (see fig.~\ref{A2}).

\begin{figure}[t]
$$
\includegraphics[width=70mm,height=70mm]{\figs/plotC2}\qquad
\includegraphics[width=70mm,height=70mm]{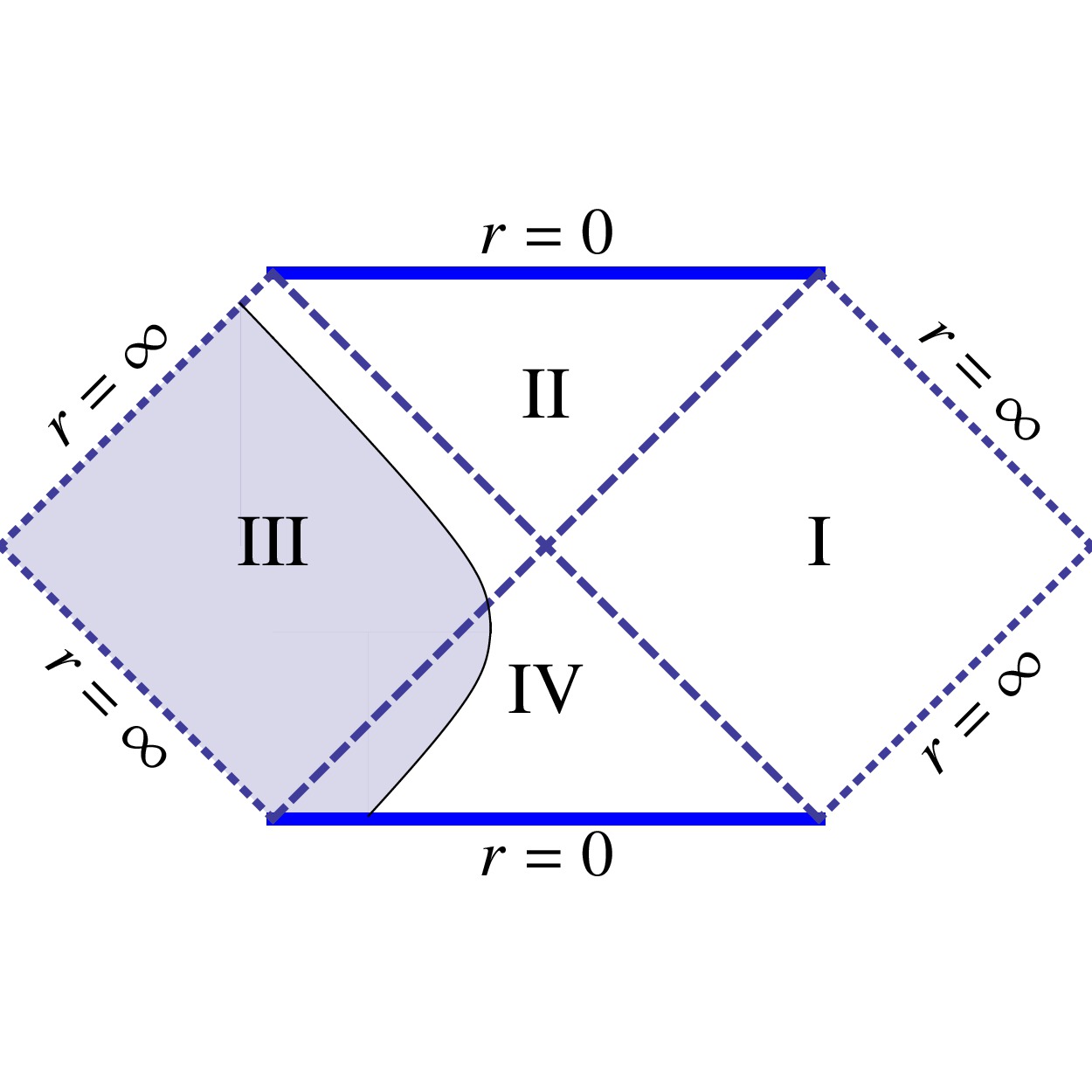}$$
\caption{{\bf Large bubbles with $1/\ell^2_{\rm in}<\kappa^2$
expand behind the horizon}.
Left: {\em The wall trajectory corresponding to Line G of fig.~\ref{h1}, in AdS space}.
Right: {\em The wall trajectory corresponding to Line G of fig.~\ref{h1},  in the Schwarzschild geometry.}
\label{G1}}
\end{figure}

\subsubsection*{Small  bubbles with large initial wall velocity expand behind the horizon}

As we have seen already, the case $\ex=-1$ may lead to 
evolution that cannot be deduced through a purely Newtonian approach. 
For $1/\ell^2_{\rm in}<\kappa^2$ the wall self-energy dominates the negative vacuum energy, so that the growth of the bubble seems energetically unfavourable. 
The Newtonian intuition suggests that such bubbles cannot expand.
However, there is a relativistic solution  described by line G of fig.~\ref{h1}. The corresponding space evolution is depicted in fig.~\ref{G1}. 
The crucial difference with respect to the case $\ex=1$, 
depicted in fig.~\ref{C1}, is that the wall trajectory is located within
the regions IV and III, instead of the regions IV and I of the 
Schwarzschild space-time: in simpler words the bubble expands inside its Schwarzschild radius.
The extrinsic curvature $\beta_{\rm out} R$ of
eq.~(\ref{ext2}) changes sign along the trajectory G of the wall, while it stays positive for the trajectory C, as can be seen in 
fig.~\ref{h1} \cite{blau}. 
Asymptotic regions of flat space-time survive:
the growth of the AdS region and its singularity
are hidden behind the horizon and do not affect an observer located in region I.\footnote{A pictorial representation of an 
analogous  situation  for a dS bubble is 
given in fig.~13 of~\cite{blau}, 
in which case the AdS singularity is absent.}

\subsubsection*{Large bubbles expand behind the horizon}

Line F of fig.~\ref{h1} describes a large bubble that initially shrinks, 
reaches a minimal size and subsequently expands. The whole evolution
lies entirely within the region III of the Schwarzschild space-time
and is hidden
behind a horizon for an observer located in region~I.

\subsection{Evolution of negative-mass bubbles} \label{negmas}

We next turn to the solutions with negative mass $M$, or, equivalently, $\ex_m=-1$.
The metric (\ref{schw-metric}), with $V_{\rm out}=0$, has a naked timelike singularity
at $r=0$ in this case. However, this metric is relevant only for the bubble exterior,
while the interior is described by the AdS metric (\ref{ads-metric}). As long as the
bubble expands and the wall moves to increasing values of $R$, the global geometry
is free of singularities.

The form of the `potential' for $1/\ell^2>\kappa^2$, $M<0$, depicted in the right
plot of fig.~\ref{h1}, allows for such a solution. 
For $\ex=1$, $\ex_m=-1$, the `potential' has a positive maximal value. On the
other hand, the `energy' $E$ of eq.~(\ref{pot}) is always negative. This allows for
only one possible type of solutions, the one corresponding to line H of 
fig.~\ref{h1}. It represents a bubble that starts with infinite radius, shrinks to 
a finite value of $R$, and subsequently re-expands. Its mass is negative, because
the radius is always sufficiently large for the 
negative volume contribution to the energy content to be dominant. 
The Penrose diagram for this solution is depicted in fig.~\ref{H1}.
There are two singularities in the 
Penrose diagram of AdS space, starting from the points at which
the wall trajectory reaches the boundary.
The Schwarzschild metric with negative mass has a naked singularity at $r=0$, 
depicted by
the vertical solid line in the right plot of fig.~\ref{H1}.
However, this singularity is irrelevant for our problem because 
it is
eliminated when the white areas of the two plots are joined along the wall
trajectory. 
%We have marked the right plot as `I' in order to indicate the correspondence with
%the similarly marked area in the Schwarzschild geometry of fig.~\ref{C1}.

It must be pointed out that it is not possible to construct negative-mass 
solutions corresponding
to horizontal lines extending from $\Rt=0$ to the `potential' in fig.~\ref{h1}.
As can be seen from eq.~(\ref{cond2}) such lines would require $\beta_{\rm out}<0$, or, 
equivalently $\ex_2=-1$. As we have already remarked, this choice would require
the timelike coordinates $\tau$ and $t$ to increase in opposite directions. 
For positive mass, the Schwarzschild geometry has sufficient structure to
permit solutions with both signs of $\beta_{\rm out}$,
such as the one corresponding to line G of fig.~\ref{h1}, which 
is depicted in fig.~\ref{G1}. However, for negative mass, the Penrose diagram 
cannot be extended beyond that depicted in fig.~\ref{H1}, unless completely 
disjointed regions are introduced. For this reason, the only meaningful solution
is the one of fig.~\ref{H1}.

\subsection{The AdS `crunch'} \label{AdS5}

%\caption{\em }

As we have seen, the particular structure of AdS implies that the evolution of 
the bubble must lead to a singularity. 
This is apparent in fig.~\ref{C1}:
the form of spacetime,
after the finite time $\eta$ at which the wall reaches the timelike boundary,
cannot be determined without additional boundary conditions. 
It is expected that, for a physical system that realises an approximation of the 
idealised bubble evolution that we consider, 
a spacelike singularity must develop in the interior of the
bubble \cite{freivogel}.  

\begin{wrapfigure}{R}{0.4\textwidth}
  \vspace{-2ex}
    \begin{center}
   \includegraphics[height=0.4\textheight]{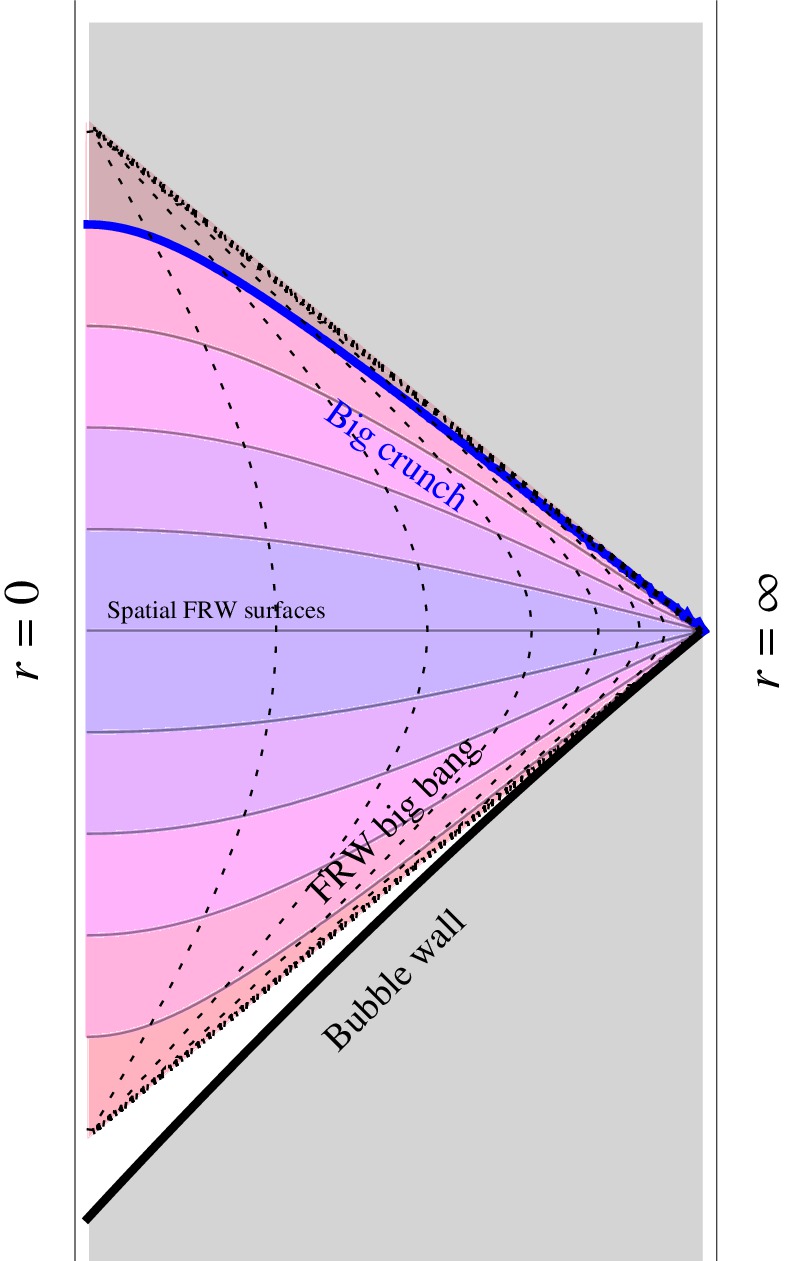}
  \end{center}
  \vspace{-3ex}
  \caption{\em The AdS interior of the bubble in conformal coordinates,
 showing the crunch and (in color) a patch in FRW coordinates.  
  \label{wallcrunch}}
    \vspace{-3ex}
\end{wrapfigure}

For the problem at hand, the nature of this singularity can be understood through
the picture of the AdS `crunch' presented in~\cite{deluccia}.
As shown there, a part of AdS space can be viewed as an open Friedmann-Robertson-Walker (FRW) universe with negative energy density. 
The coordinate change~\cite{chimento}
\begin{eqnarray}
r&=&\ell_{\rm in} \sin\frac{\that}{\ell_{\rm in}}\,\sinh \psi
\nonumber \\
\cos \frac{\that}{\ell_{\rm in}}&=& \left(1+\frac{r^2}{\ell_{\rm in}^2} \right)^{1/2}\, \cos\frac{\eta}{\ell_{\rm in}}
\label{coordchange} \end{eqnarray}
puts the AdS metric (\ref{ads-metric}) in the form
\be
ds^2=-d\that^2+\ell_{\rm in}^2\sin^2\frac{\that}{\ell_{\rm in}}\,
\left(d\psi^2+\sinh^2\psi \, d\Omega^2_2 \right).
\label{crunch-metric} \ee
This metric describes an homogeneous FRW universe that is born with a 
big `bang' at $\that=0$
and collapses in a big `crunch' at $\that=\ell_{\rm in}\pi$. 
The coordinates $\that,\psi$ do not cover the whole AdS space, but only
a triangular patch of its Penrose diagram. This is bounded by the dot-dashed null 
lines in fig.~\ref{wallcrunch}. 
Even though these lines represent only a coordinate singularity in the idealized
picture of a pure AdS bubble interior, the big `crunch' becomes a true physical
singularity in the presence of a fluctuating Higgs field, as argued in~\cite{deluccia}.
From continuity it is apparent that the Higgs fluctuations become very large in the
neighbourhood of the null line.

The connection of the AdS `crunch' to the bubble evolution can be obtained by 
establishing the relative position of the FRW `triangle' and the wall 
trajectory on the Penrose diagram. As the FRW observer views a homogeneous universe,
the AdS patch to which he has access must be located sufficiently deep inside
the bubble for the Higgs field to have a constant value. At late times, the wall
moves with approximately the speed of light. It is expected that 
the wall trajectory and the lower side of the FRW `triangle' will
converge asymptotically
as the AdS boundary is approached, as depicted in fig.~\ref{wallcrunch}. 
The singularity developing below the Cauchy horizon appears on a spacelike curve
emanating from the point at which the wall reaches the AdS boundary. 
The homogeneity of space viewed by the FRW observer indicates that this singularity 
should correspond to a constant-$\that$ surface. In fig.~\ref{wallcrunch} we 
depict the slicing of the FRW `triangle' with such surfaces. The thick solid line  
represents the possible location of the `crunch', very close to the upper
side of the FRW `triangle'.

The most important consequence of the above picture is that the `crunch' never 
reaches the bubble wall. This is apparent in fig.~\ref{wallcrunch}, as the 
black solid line, representing the wall trajectory, and the blue solid line,
representing the `crunch', never cross. They seem to merge on the AdS boundary.
However, this is an illusion created by the Penrose diagram. The bubble wall 
lies always slightly outside the FRW `triangle', as its speed never becomes
exactly equal to that of light. 

A final observation relevant for the asymptotic wall expansion concerns the
corresponding time scales in the various frames. 
Let us consider a very large bubble with $\epsilon_1=\epsilon_2=1$, see eqs.~(\ref{doteta})--(\ref{dott}), expanding almost at the speed of light, 
such that $R\gg \ell_{\rm in}, 2GM$ and $\dot{R}\gg 1$. The evolution of the wall 
in terms of the three time coordinates of the systems (\ref{wall-metric}), (\ref{ads-metric}), (\ref{schw-metric}) is given by 
\be
R=R_0 e^{c_1\tau}=\frac{R_0}{1-c_2\eta}=R_0+t,
\label{timess} \ee
with $c_1\simeq 1/(2\ell^2_{\rm in}\kappa)$, $c_2\simeq R_0/\ell^2_{\rm in}$ for $1/\ell^2_{\rm in}\gg\kappa^2$. 
It is apparent that the wall reaches the AdS boundary within a finite amount of
the time coordinate $\eta$, while it requires an infinite amount of time $\tau$ or
$t$. In particular, an observer in the asymptotically flat spacetime infinitely far 
from the
bubble is reached by the wall only after an infinite amount of time $t$.

\subsection{Critical bubbles} \label{AdS6}
For given $\ell_{\rm in}$ and $\kappa$, corresponding to given interior vacuum energy 
and surface tension, there is a critical bubble radius $\Rcr$.
Bubbles that start with negligible wall velocity and $R>\Rcr$ follow trajectories
with increasing $R$, while bubbles that start with $R<\Rcr$ have diminishing $R$
and eventually collapse to a black hole. The critical radius corresponds to 
the maximum of the potential of fig.~\ref{h1}. One can imagine a horizontal line,
tangent to the top of the potential. The right part of the line is the limiting
case of lines starting on the potential and 
describing expanding bubbles, while its left part is the limiting
case of similar lines describing collapsing bubbles. 
As the `energy' $E$ is negative, while the potential has a maximum with a positive
value for $\ex_m=-1$ (see the third plot of fig.~\ref{h1}), 
it is obvious that there are no critical bubbles with negative mass. 

\begin{figure}[t]
$$
\includegraphics[width=0.3\textwidth]{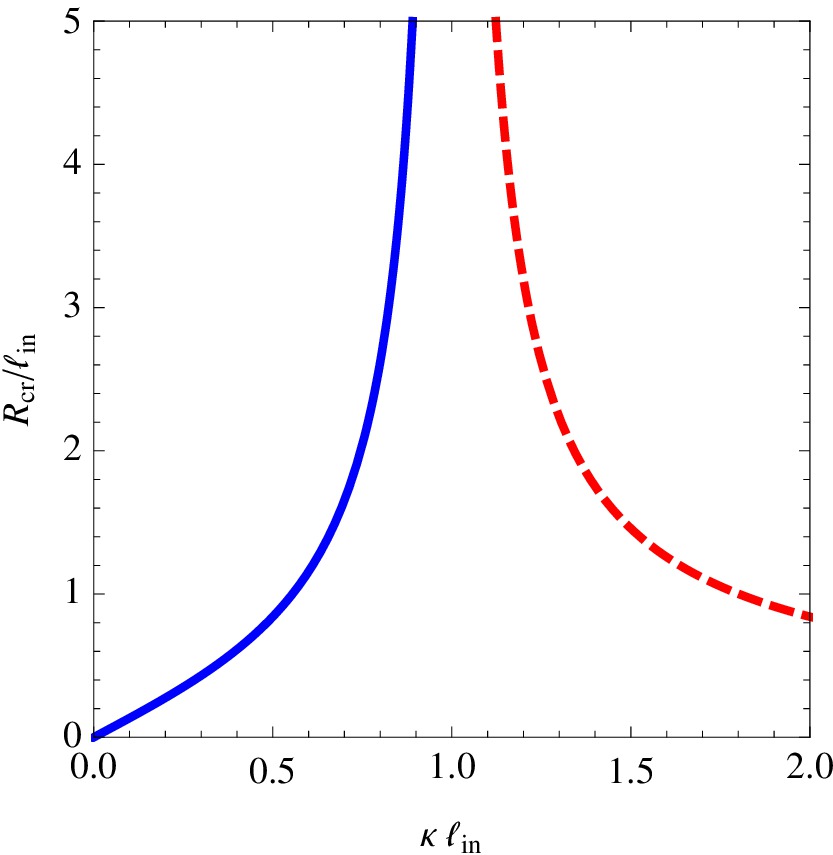}\qquad
\includegraphics[width=0.3\textwidth]{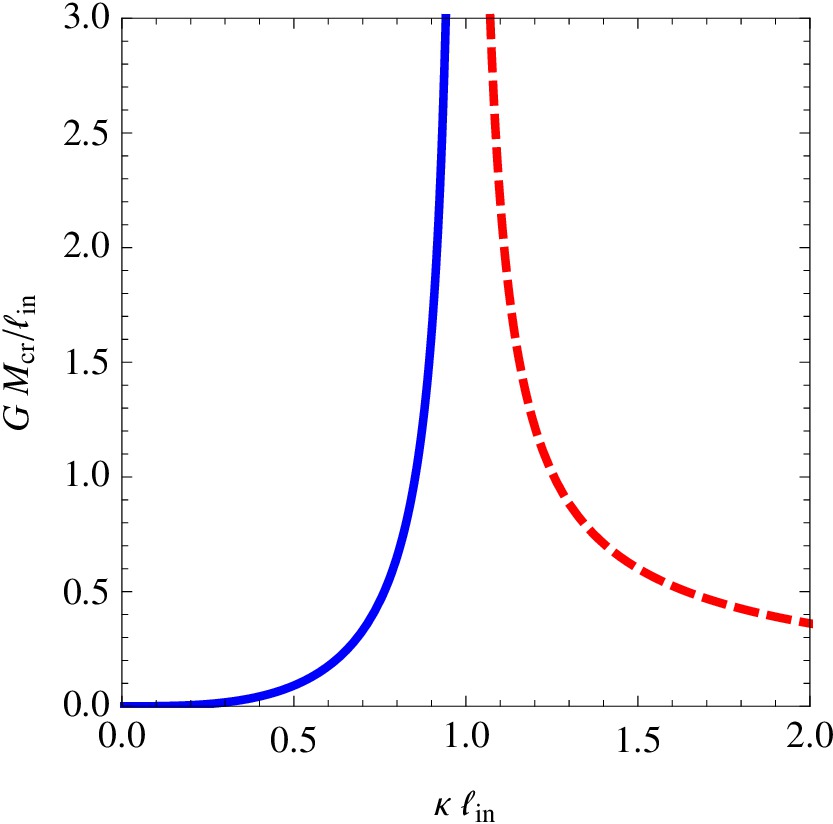}\qquad
\includegraphics[width=0.3\textwidth]{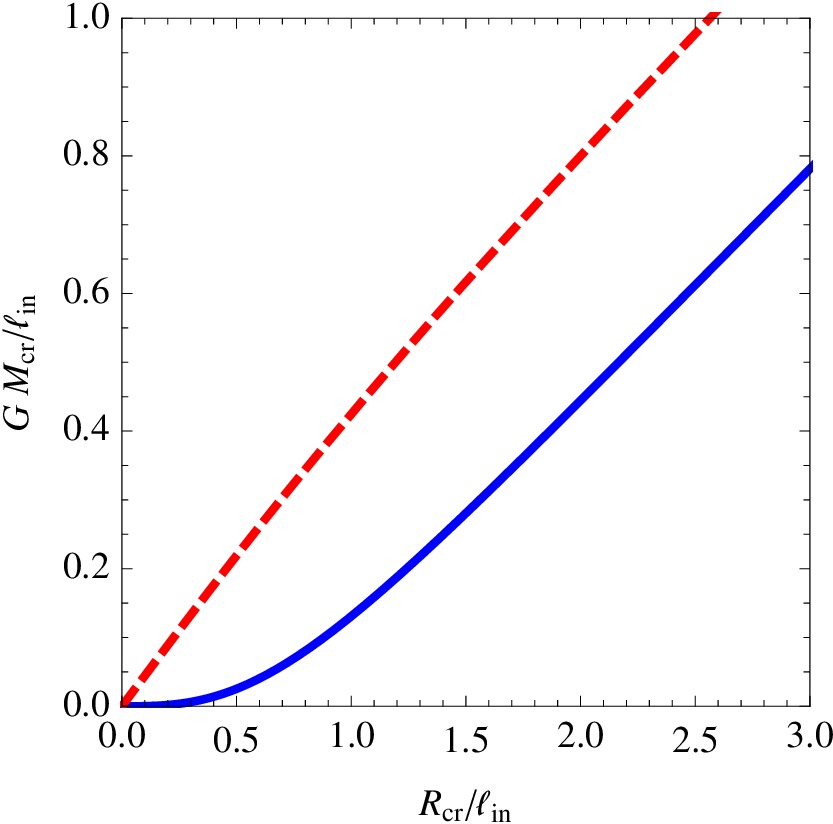}
$$
\caption{
{\em The radius and mass of critical bubbles as a function of $\kappa\ell_{\rm in}$ for $\Delta >0$ (blue curves)
and $\Delta<0$ (red dashed).
}
\label{crit}}
\end{figure}

The maximum of the potential and its value at this point are given by 
eqs.~(\ref{zmm})--(\ref{potmax}). The value of the `energy' can be obtained
from eq.~(\ref{eom}) with $d\Rt/d\ttau=0$ and $\Rt=\Rt_{\rm max}$. 
Expressions for $\Rcr$ and the corresponding mass $M_{\rm cr}$ 
can then be obtained by combining 
eqs.~(\ref{pot})--(\ref{gamm}). As these expressions 
are not very illuminating we do not
present them explicitly. The quantities $\Rcr/\ell_{\rm in}$, $G\Mcr/\ell_{\rm in}$ are functions of
the dimensionless combination $\kappa\ell_{\rm in}$. In fig.~\ref{crit}
we present the functions $\Rcr(\kappa\ell_{\rm in})/\ell_{\rm in}$, $G\Mcr(\kappa \ell_{\rm in})/\ell_{\rm in}$ and
$[G\Mcr/\ell_{\rm in}](\Rcr/\ell_{\rm in})$.

The critical bubbles have certain characteristics:
\begin{itemize}
\item
Their radius is always larger than the Schwarzschild radius. This can be
deduced from fig.~\ref{h1}, in which it is apparent that the location of the maximum 
of the potential is always outside the horizon. 
\item
There are two branches of critical bubbles, corresponding to
$1/\ell_{\rm in}^2>\kappa^2$ or $\ex=1$ (solid lines), and $1/\ell_{\rm in}^2<\kappa^2$ or $\ex=-1$ (dashed lines). 
\item
The radius diverges for $\kappa\to 1/\ell_{\rm in}$, as the effective energy density in
the interior of the bubble vanishes in this limit. 
\item
The branch with $\ex=1$ 
reproduces correctly for
$\kappa\to 0$ the Newtonian limit of nonrelativistic bubbles with
$\Rcr=4\ell_{\rm in}^2\kappa/3$ and $G \Mcr=16\ell_{\rm in}^4\kappa^3/27$. 
\item
The branch with $\ex=-1$ is not visible
to an observer located in region I of the Penrose diagram. 
\item
For given vacuum-energy scale $\ell_{\rm in}$ and critical-bubble radius $\Rcr$, 
the bubbles with $\ex=-1$ are more massive than the ones with $\ex=1$. 
(Note that the two types of bubbles also have different surface tension $\kappa$.)
\end{itemize}

The most interesting solutions are those that 
describe bubbles visible to an observer in the
asymptotically flat region. These are bubbles for which a Newtonian
limit exists within their parameter range. Their mass-to-radius relation
is depicted in the third plot of fig.~\ref{crit}. The critical bubbles correspond
to the solid line. The parameter range above this line corresponds 
to collapsing bubbles, while the range below to expanding bubbles. Expanding
bubbles can have negative mass, so their parameter range includes the region 
below the positive $R$-axis.

\subsection{Bubbles in asymptotically de Sitter spacetime} \label{AdS7}

The evolution of an AdS bubble within an asymptotically dS spacetime can be analysed
in complete analogy to the previous discussion for an asymptotically
flat spacetime.   
The metric of eq.~(\ref{schw-metric}) now contains the function
$f_{\rm out}(r)=1-r^2/\ell_{\rm out}^2-{2GM}/{r}$. There are two horizons,
corresponding to the zeros of $f_{\rm out}(r)$.
The Penrose diagram of the Schwarzschild-de Sitter (SdS) 
spacetime is depicted in the right part of 
fig.~\ref{W1}. It is a combination of the 
diagrams for the Schwarzschild and dS spacetimes \cite{gibbons}. 
Thick blue lines denote curvature singularities, the dashed lines horizons and the dotted lines conformal infinities.
The two thin vertical lines at the ends of the
diagram indicate that the pattern is repeated indefinitely on either side.

\begin{wrapfigure}{R}{0.5\textwidth}
   \vspace{-3ex}
     \begin{center}
    \includegraphics[height=0.3\textheight]{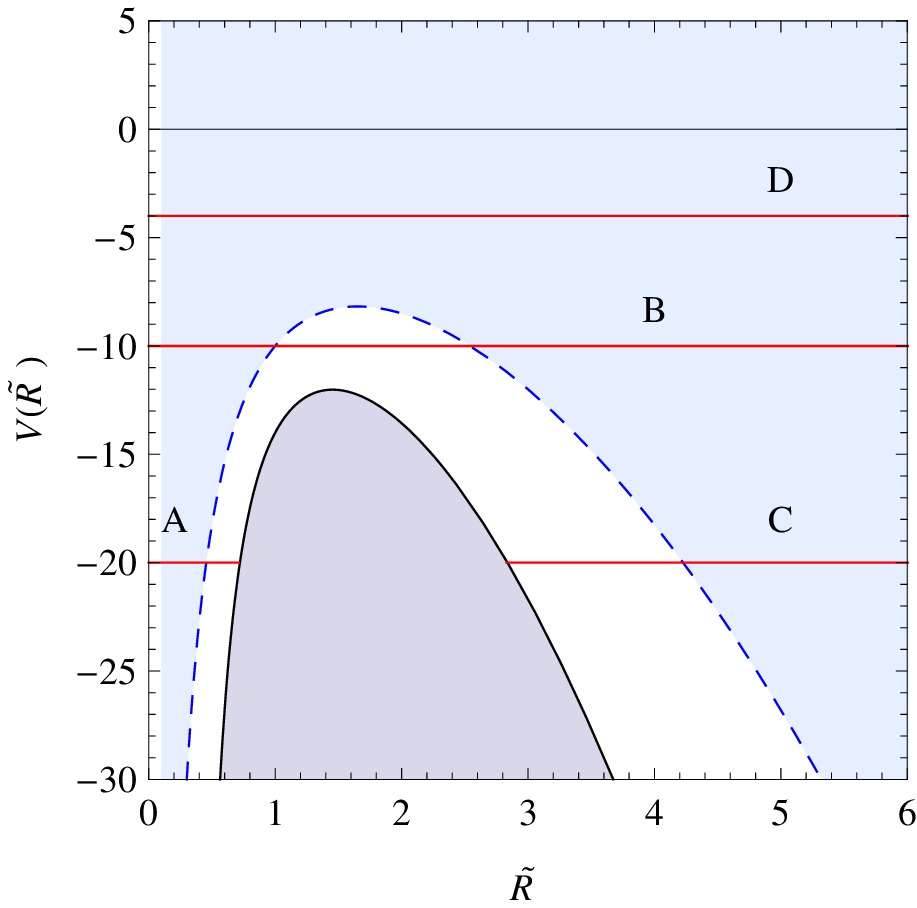}
   \end{center}
   \vspace{-3ex}
   \caption{\em 
   The `potential' of eq.~(\ref{potdsn}) with
     $\gamma=3$, $\delta=1$, $\ex=1$,
   for AdS bubbles in Schwarzschild-de Sitter space. 
   \label{dspot}}
 \end{wrapfigure}

The matching across the 
domain wall, located at $R(\tau)$, proceeds as before. We do not analyse the many
possible cases, as the analysis is a straightforward generalisation of
the discussion in the previous subsections. 
We focus instead on the novel aspects of the SdS case. For a positive mass $M$,
the motion of the wall is again determined 
by eq.~(\ref{eom}{}), with $\Rt=\rho R$. However, the `potential' now has the form
\be
V(\Rt)=-\left( \frac{1+\ex \Rt^3}{\Rt^2}\right)^2-\frac{\gamma^2}{\Rt}-\delta^2 \Rt^2,
\label{potdsn}\ee
with
\be
\rho^3 = \frac{1}{2GM}\left| \Delta \right|,
\qquad
\ex \equiv  {\rm sign}\, \Delta,
\qquad
\gamma=\frac{2\kappa}{
\sqrt{\left|\Delta \right|}},
\qquad
\delta^2=\frac{4\kappa^2}{\ell^2_{\rm out}\Delta^2},
\label{rrhodss} \ee
where
\be
\Delta=\frac{1}{\ell^2_{\rm in}}+\frac{1}{\ell^2_{\rm out}}-\kappa^2.
\label{ddelta} \ee
The form of the `potential' is very similar to that in
fig.~\ref{h1}. The horizon corresponds
to a value $\Rt_H$ such that 
\be
E=-\frac{\gamma^2}{\Rt_{H}}-\delta^2\Rt^2_{H}.
\label{horisds} \ee
It is again determined by eq.~(\ref{horiz}), but now has a different shape.
In fig.~\ref{dspot} we depict the `potential' (solid black line) and the horizon 
(dashed blue line)
for $\gamma=3$, $\delta=1$, $\ex=1$.

The various trajectories correspond to solutions of constant 
$E=-{\kappa^2}/(G^2M^2\rho^4)$, as depicted in fig.~\ref{dspot}.
The two types of behaviour, characterised by $\ex=\pm 1$, correspond now
to the gravitational self-energy $\kappa^2$ 
of the bubble being smaller or larger than
the total difference in energy density $1/\ell^2_{\rm in}+1/\ell^2_{\rm out}$ between the dS and 
AdS spacetimes. 
We do not analyse the form of all the possible trajectories, as
they are similar to those discussed earlier. In fig.~\ref{dspot} we have 
depicted a few characteristic cases for $\ex=1$.

Line A corresponds to a bubble that starts below the inner horizon,
crosses it, reaches a maximal radius and collapses falling again behind the horizon. 

Line C corresponds to a large bubble that stars with infinite radius, moves
within the outer horizon, reaches a minimal radius and then re-expands moving again
outside the outer horizon.
One may consider also the scenario in which the bubble is spontaneously created
with vanishing wall velocity at a certain radius and expands, with the wall moving
outside the outer horizon. In this scenario, line C is covered only once. 

Line B corresponds to a bubble that starts with a very small radius and expands 
indefinitely, with
its wall crossing the inner and outer horizons successively.
Its speed asymptotically approaches the speed of light. 
The form of the wall trajectory on the Penrose diagram is depicted in fig.~\ref{W1}.
The total space is constructed by patching the white regions of the two plots
in fig.~\ref{W1}. 

Line D corresponds to the evolution of a bubble that does not cross any horizons.
The reason is that `energies' that approach zero correspond to increasing values 
of the mass parameter $M$. For sufficiently large $M$, the metric function 
$f_{\rm out}(r)$ does not vanish at any $r$, but stays always negative. The 
space has a naked spacelike singularity at $r=0$.  
However, this part of spacetime is eliminated and replaced by the interior of the
AdS bubble. 

\medskip

There are many other possibilities for $\ex=-1$ or for negative bubble mass. 
These can
be analysed in complete analogy to the trajectories depicted in the second
and third plot of fig.~\ref{h1}.
They correspond to collapsing bubbles or bubbles expanding behind 
horizons, which are not visible to an observer located in the asymptotic de 
Sitter space.

The crucial question pertinent to the scenario of Higgs fluctuations during
inflation is whether the expanding AdS bubbles can completely eliminate the
surrounding dS space and thus terminate inflation. 
It is apparent from fig.~\ref{W1} that asymptotically the wall trajectory reaches
spacelike infinity. The wall location separates two spacelike regions: one of
them is replaced by the interior of the AdS bubble, while the other remains 
part of an external dS spacetime.  
Asymptotically the total spacetime 
contains large AdS bubbles within large 
dS regions. This scenario is in contrast to the case of 
asymptotically flat spacetime, in which the wall asymptotically
reaches null infinity and the whole space is
engulfed by the AdS bubbles. In other words, the inflationary growth 
guarantees that, even when the size of the AdS regions grows with the speed of light,
the external regions grow even faster, so that they survive at late times.

\end{document}